\documentclass[12pt]{article}
\pdfoutput=1

\usepackage{amssymb,bm}

\usepackage{xcolor}
\usepackage[colorlinks]{hyperref}
\hypersetup{
  colorlinks=true,
  linkcolor=purple,
  filecolor=red,
  citecolor=blue,      
  urlcolor=violet
}

\usepackage{scicite}
\usepackage{graphicx}
\usepackage{times}
\usepackage[normalem]{ulem}
\usepackage{soul}
\usepackage{multirow}
\usepackage{array}
\usepackage{booktabs}
\usepackage{lineno}
\usepackage{overpic}
\usepackage{amsmath}
\usepackage{threeparttable}

\usepackage{subfig}
\usepackage[font=footnotesize]{caption}

\graphicspath{images/}

\usepackage[symbol]{footmisc}

\graphicspath{{\subfix{./images/}}}

\newenvironment{sciabstract}{%
\begin{quote} \bf}
{\end{quote}}

\title{An energetic dirty fireball detected in soft X-rays}

\author{}
\date{}

\begin{document}

\baselineskip 20pt


\maketitle



\footnotetext[1]{Corresponding authors:  X.-Y. Wang (xywang@nju.edu.cn), B. Zhang (bing.zhang@unlv.edu), \\
\indent\indent P. G. Jonker (p.jonker@astro.ru.nl), Y. Liu (liuyuan@nao.cas.cn), W. Yuan (wmy@nao.cas.cn)}
\footnotetext[2]{These authors contributed equally to this work.}

{\noindent C.-Y. Dai$^{1,2}$†, J. Quirola-V\'asquez$^{3}$†, Y.-H. Wang$^{4,5}$†, H.-L. Li$^6$, J. Yang$^{1,2}$,  X.-L. Chen$^{7}$, A.-L. Wang$^{8,9}$, H. Sun$^6$,  X.-Y. Wang$^{1,2}$*, B. Zhang$^{10,4}$*, P. G. Jonker$^3$*, Y. Liu$^6$*, W. Yuan$^{6,11}$*, D. Xu$^{6,12}$, Z.-G. Dai$^{13}$, M.~E. Ravasio$^{3,14}$, L. Piro$^{15}$,  P. O'Brien$^{16}$, D. Stern$^{17}$, H.-M. Zhang$^{18}$, Y.-P. Yang$^7$, T. An$^{8,11}$, Y.-L. Qiu$^6$, L.-P. Xin$^{6}$, W.-X. Li$^6$,   R.-Y. Liu$^{1,2}$,   X.-F. Wu$^{19}$, C.-Y. Wang$^{20}$, D.-M. Wei$^{19}$, Y.-F. Huang$^{1,2}$,  F.~E. Bauer$^{21}$, W.-H. Lei$^{22}$, B.-B. Zhang$^{1,2}$, N.-C. Sun$^{6,11,23}$, H. Gao$^{23,24}$, V. S. Dhillon$^{25,26}$, J. An$^{6}$,   C.-H. Bai$^{27}$,   A. Martin-Carrillo$^{28}$, H.-Q. Cheng$^6$, J.~A. Chacon Chavez$^{14}$, Y. Chen$^9$, G.-W. Du$^7$, J.~N.~D. van Dalen$^3$,  A. Esamdin$^{27}$, Y.-Z. Fan$^{19}$, X. Gao$^{27}$, F. Harrison$^{29}$, J.-W. Hu$^6$, M.-Q. Huang$^{13}$, S.-M. Jia$^9$, A.~J. Levan$^{3,30}$, C.-K. Li$^9$, D.-Y. Li$^6$, E.-W. Liang$^{18}$, S. Littlefair$^{25}$, X.-W. Liu$^7$, Z.-Y. Liu$^{13}$, Z.-X. Ling$^{6,11}$, D.~B. Malesani$^{31,32}$, H.-W. Pan$^6$,  A. Rodriguez$^{29}$, A. Rossi$^{33}$, D. Mata S\'anchez$^{26,34}$, J. Sánchez-Sierras$^3$,  X.-J. Sun$^{35}$, M.~A.~P. Torres$^{26,34}$, A.~P.~C. van Hoof$^3$, X.-F. Wang$^{36}$,  Q.-Y. Wu$^6$, X.-P. Xu$^{6,11}$, Y.-F. Xu$^{6,11}$, Y.-W. Yu$^{37}$, C. Zhang$^6$, M.-H. Zhang$^6$, S.-N. Zhang$^9$, Y. Zhang$^{27}$, Y.-H. Zhang$^{38}$, Z.-P. Zhu$^6$\\

\noindent \normalsize{$^{1}$School of Astronomy and Space Science, Nanjing University, Nanjing 210093, China.}\\
\normalsize{$^{2}$Key Laboratory of Modern Astronomy and Astrophysics (Nanjing University), Ministry of Education, China.}\\
\normalsize{$^{3}$Department of Astrophysics/IMAPP, Radboud University, P.O. Box 9010, 6500 GL, Nijmegen, The Netherlands.}\\
\normalsize{$^{4}$Nevada Center for Astrophysics and Department of Physics and Astronomy, University of Nevada, Las Vegas, NV 89154, USA.}\\
\normalsize{$^{5}$Department of Astronomy, University of Wisconsin, Madison, WI 53706, USA.}\\
\normalsize{$^{6}$National Astronomical Observatories, Chinese Academy of Sciences, Beijing 100101, China.}\\
\normalsize{$^{7}$South-Western Institute for Astronomy Research, Yunnan University, Kunming, Yunnan 650504, People's Republic of China.}\\
\normalsize{$^{8}$Shanghai Astronomical Observatory, Chinese Academy of Sciences (CAS), 80 Nandan Road, Shanghai 200030, China.}\\
\normalsize{$^{9}$Key Laboratory of Particle Astrophysics, Institute of High Energy Physics, Chinese Academy of Sciences, Beijing 100049, China.}\\
\normalsize{$^{10}$Department of Physics, University of Hong Kong, Pokfulam Road, Hong Kong 999077, China.}\\
\normalsize{$^{11}$School of Astronomy and Space Science, University of Chinese Academy of Sciences, Chinese Academy of Sciences, Beijing 100049, China.}\\
\normalsize{$^{12}$Altay Astronomical Observatory, Altay, Xinjiang 836500, China.}\\
\normalsize{$^{13}$Department of Astronomy, School of Physical Sciences, University of Science and Technology of China, Hefei 230026, China.}\\
\normalsize{$^{14}${Instituto de Astrof{\'{\i}}sica, Facultad de F{\'{i}}sica and Centro de Astroingenier{\'{\i}}a, Facultad de F{\'{i}}sica, Pontificia Universidad Cat{\'{o}}lica de Chile}, UC, Campus San Joaquín, Av. Vicuña Mackenna 4860, Macul, Santiago, 7820436, RM, Chile.}\\
\normalsize{$^{15}$INAF—Istituto di Astrofisica e Planetologia Spaziali, Rome, Italy.}\\
\normalsize{$^{16}$School of Physics and Astronomy, The University of Leicester, University Road, Leicester LE1 7RH, UK.}\\
\normalsize{$^{17}$Jet Propulsion Laboratory, California Institute of Technology, 4800 Oak Grove Drive, Mail Stop 264-789, Pasadena, CA 91109, USA.}\\
\normalsize{$^{18}$Guangxi Key Laboratory for Relativistic Astrophysics, School of Physical Science and Technology, Guangxi University, Nanning 530004, Guangxi, China.}\\
\normalsize{$^{19}$Purple Mountain Observatory, Chinese Academy of Sciences, Nanjing 210023, China.}\\
\normalsize{$^{20}$Department of Astronomy, Tsinghua University, Beĳing 100084, China.}\\
\normalsize{$^{21}$Instituto de Alta Investigaci{\'{o}}n, Universidad de Tarapac{\'{a}}, Casilla 7D, Arica, Chile.}\\
\normalsize{$^{22}$Department of Astronomy, School of Physics, Huazhong University of Science and Technology, Wuhan, 430074, People’s Republic of China.}\\
\normalsize{$^{23}$Institute for Frontier in Astronomy and Astrophysics, Beijing Normal University, Beijing 102206, China.}\\
\normalsize{$^{24}$School of Physics and Astronomy, Beijing Normal University, Beijing 100875, China.}\\
\normalsize{$^{25}$Astrophysics Research Cluster, School of Mathematical and Physical
Sciences, University of Sheffield, Sheffield, S3 7RH, UK.}\\
\normalsize{$^{26}$Instituto de Astrofı\'isica de Canarias, IAC, E-38205, La Laguna, Tenerife, Spain.}\\
\normalsize{$^{27}$Xinjiang Astronomical Observatories, Chinese Academy of Sciences, Urumqi 830011, China.}\\
\normalsize{$^{28}$School of Physics and Centre for Space Research, University College Dublin, Belfield, Dublin 4, Ireland.}\\
\normalsize{$^{29}$Department of Astronomy, California Institute of Technology, 1200 East California Blvd, Pasadena, CA 91125, USA.}\\
\normalsize{$^{30}$Department of Physics, University of Warwick, Coventry, CV4 7AL, UK.}\\
\normalsize{$^{31}$Cosmic Dawn Center (DAWN), Denmark.}\\
\normalsize{$^{32}$Niels Bohr Institute, University of Copenhagen, Jagtvej 128, Copenhagen, 2200, Denmark.}\\
\normalsize{$^{33}$Osservatorio di Astrofisica e Scienza dello Spazio, INAF, Via Piero Gobetti 93/3, Bologna, 40129, Italy.}\\
\normalsize{$^{34}$Departamento de Astrof\'isica, Univ. de La Laguna, E-38206 La Laguna, Tenerife, Spain.}\\
\normalsize{$^{35}$Shanghai Institute of Technical Physics, Chinese Academy of Sciences, Shanghai, 200083, China.}\\
\normalsize{$^{36}$Physics Department, Tsinghua University, Beijing, 100084, China.}\\
\normalsize{$^{37}$Institute of Astrophysics, Central China Normal University, Wuhan 430079, People's Republic of China.}\\
\normalsize{$^{38}$Innovation Academy for Microsatellites, Chinese Academy of Sciences, Shanghai 201210, China.}\\
}

\begin{sciabstract}
The collapse of massive stars drives explosions that power relativistic fireballs.  If only a small amount of matter is entrained, such clean fireballs can expand with Lorentz factors $\Gamma> 100$,  accounting for gamma-ray bursts (GRBs). 
It has been hypothesized that energetic explosions with more baryon contamination, dubbed ``dirty fireballs'', may exist in nature, but they have not been observed.
Here we report the observation of an extragalactic fast X-ray transient, EP241113a, detected by Einstein Probe. Compared to GRBs, it has a similar isotropic energy of $1.4\times 10^{51}$ erg,  but significantly lower spectral peak energy. 
Theoretical modeling of its early X-ray afterglow suggests a
relativistic jet with a low Lorentz factor of $\Gamma \sim 20$ aligned close to the line-of-sight, signifying the prototype of a dirty fireball.  


\end{sciabstract}

{\bf One sentence summary}: Einstein Probe detected an extragalactic fast X-ray transient with an energy comparable to those of GRBs but a much lower spectral peak energy, signifying the prototype of the long-anticipated dirty fireballs originating from the collapse of massive stars.


\section{Introduction}
The collapse of massive stars drives explosions, some of which launch energetic ``clean fireballs'' with a small amount of matter entrained in the ejecta\cite{Woosley2006}. These explosions, known as gamma-ray bursts (GRBs) in observations since the 1970s\cite{GRB1973,Zhang2018},  are found to have ultra-relativistic jets with large Lorentz factors $\Gamma > 100$ due to small baryon loading\cite{Lithwick2001,Racusin2011}. 
The population of long-duration  GRBs has a broad distribution in isotropic energy and spectral hardness, with X-ray-rich GRBs and X-ray flashes extending to the less energetic, softer regime\cite{Campana2006,Sakamoto2008}. So far, known low-$\Gamma$ explosions are typically low-luminosity events\cite{Campana2006,Sun2025}.  Observationally, these explosions show a characteristic peak in their spectral energy distribution (SED) in the  $E^2{\rm d}N/{\rm d}E$  representation, whose rest-frame peak energy  $E_{\rm p}(1+z)$ ($z$ is the redshift) is positively correlated with the isotropic emission energy,$E_{\gamma, \rm iso}$, known as the``Amati relation''\cite{Amati2002}.
On the other hand, it has been speculated that energetic dirty fireballs, explosions with energy similar to that of GRBs but with much smaller Lorentz factors and  softer spectra, may exist in nature\cite{Paczynski1998,Dermer1999,Huang2002,Rhoads2003,Zhangwq2004}.  Such explosions have not been detected before.
There have been efforts to search for the late-stage  afterglow emission of such dirty fireballs, and some extragalactic optical transients were
suggested to be the candidates of  optical afterglows \cite{Cenko2013,Ho2022}. However, the prompt burst of the explosions, which carries the smoking-gun signatures of the dirty fireball,  remains undetected. The Einstein Probe (EP) mission\cite{Yuan2025}, with a combination of a large field of view and unprecedented sensitivity in soft X-rays, is beginning to unveil the nature of the mysterious extragalactic fast X-ray transients (EFXTs) population. So far, the identified  EFXTs are either GRBs at high redshift\cite{Liu2025}, or nearby, low-luminosity events similar to X-ray flashes\cite{Sun2025}.

\section{Observations of EP241113a}
EP241113a was discovered by the Wide-field X-ray Telescope (WXT) onboard the EP satellite in the
0.5–4 keV band (Fig.~\ref{fig_WXT_data}a) at  2024-11-13T19:12:53 (UTC). {Fermi}/GBM had full spatial and temporal coverage of  EP241113a, but no significant gamma-ray source was detected\cite{GCN.38238}. The flash of X-ray emission captured by the WXT is characterized by multiple pulses.  The transient has a duration of $T_{90}=204\pm 24 {\, \rm s}$ (the time during which the central
90\% of the fluence is observed) with a time-averaged unabsorbed X-ray
flux of $1.1^{+0.5}_{-0.2}\times10^{-9}{\rm \, erg\, cm^{-2} \,s^{-1}}$ in the $0.5-4.0$ keV band (the quoted errors correspond to the 68\% confidence level unless specified otherwise).  The integrated spectrum within $T_{90}$ can be  fitted
by an absorbed power law with a photon index $\Gamma_X=2.7_{-0.4}^{+0.7}$ ($dN/dE\propto E^{-\Gamma_X}$) and an intrinsic absorption with an equivalent neutral hydrogen column density $N_{\rm int}=3.2^{+1.6}_{-0.9} \times 10^{22} \, \rm cm^{-2}$ in excess of that of our own Galaxy\cite{supplementary} (see Fig.~\ref{fig_WXT_data}b and  Table~\ref{tab_spectrum_fitting}). 
This apparently steep spectral shape (an index of $-0.7$ in $E^2{\rm d}N/{\rm d}E$ spectrum) implies that the SED of EP241113a should peak somewhere within or below the 0.5-4 keV WXT band, to avoid energy diverging toward low energies.  
The spectrum also appears to be softer than the low-energy part of GRBs below $E_{\rm p}$, which has a typical index of  $\Gamma_X=1.08^{+0.43}_{-0.44}$\cite{Fermi-catalog}.
We try to constrain the spectral peak energy by fitting the WXT spectrum with models having an energy break. We use three models,  a Band function\cite{Band1993}, a smoothly broken power law model and a cutoff power law model, which are  commonly adopted to describe the broad-band spectra of GRBs and X-ray flashes. In the fittings, an upper limit is set to the low-energy photon index as $\Gamma_X < 2$ to allow a peak in the $E^2{\rm d}N/{\rm d}E$ spectrum, which is also reasonable in light of the values of GRBs.
For the Band function and the smoothly broken power law models, the peak energies cannot be
well determined, but an upper limit can be derived with $E_{\rm p,UL} = 1.9 \, \rm{keV}$ and $E_{\rm p,UL} = 1.6 \, \rm{keV}$ (at the $95\%$ confidence level), respectively (see Fig.~\ref{fig_ep_curved_model}\cite{supplementary}). For the cutoff power-law model, the best-fit peak energy is  $E_{\rm p} = 1.3^{+0.3}_{-0.6}\,\rm{keV}$ at the 68\% confidence level with an upper limit of $E_{\rm p,UL} = 2.4\,\rm{keV}$ at the 95\% confidence level (see Fig.~\ref{fig_ep_curved_model}).  Based on the Bayesian Information Criterion (BIC) comparison, the Band and smoothly broken power-law models can be ruled out statistically when compared to  the power-law model\cite{supplementary}. Therefore, a  conservative upper limit on the spectral peak energy, $E_{\rm p,UL} = 2.4 \, \rm{keV}$,  can be derived for EP241113a. 
This peak energy is significantly lower than those of typical GRBs \cite{Band1993, Kaneko2006, Nava2011} and GRB-associated EFXTs (e.g., EP240315a; \cite{Liu2025}). 
A comparison between the soft X-ray emission and the upper limit imposed by GBM\cite{supplementary} is shown in  Fig.~\ref{fig_WXT_data}d, also suggesting a soft spectrum in the keV to MeV range.  

An autonomous observation of EP241113a was performed with the EP Follow-up X-ray Telescope (EP-FXT) about 2 minutes later, which detected an uncatalogued X-ray source at R.A. = 131.9964 deg, Dec = 52.3815 deg (J2000) with an uncertainty of about 10 arcsec (90\% C.L.~statistical and systematic), consistent positionally with  the WXT transient within the uncertainties\cite{GCN.38211} (see  Fig.~\ref{fig_WXT_FXT_image}). During the 5000-second exposure, a significant decline is observed in the X-ray flux before it transitions into a plateau phase (Fig.~\ref{fig_WXT_luminosity}). Further observations of the event with EP-FXT were carried out later and the X-ray afterglow was detected until roughly 30 days after the trigger.   Overall, the light curve shows a steep decay at the end of the prompt emission and then a plateau before entering into a final ``normal'' decay. The temporal slopes ($\alpha$ in the power-law form $F_{\nu}\propto t^{-\alpha}$) for these three phases are $3.05^{+0.47}_{-0.41}$,  
$-0.16_{-0.24}^{+0.24}$, and $0.95_{-0.18}^{+0.18}$, respectively (see Fig.~\ref{fig_WXT_luminosity}b  and Ref.\cite{supplementary}). {Such a light curve,  seen in an EFXT for the first time,} is  similar to the canonical X-ray afterglow light curve observed in a large fraction of Swift-observed GRBs\cite{Zhang2006,Nousek2006}, hinting at a similar  origin of the emission from a relativistic jet (see below). 

The Large Binocular Telescope (LBT) observed the event in the $r'$ band   1.52 days after the burst, revealing a very faint source within the EP-FXT error circle (Fig.~\ref{fig_host}), which was not present in the Legacy Survey $r$-band images\cite{GCN.38233}. The source has an  AB magnitude of $r'$= 23.35$\pm$0.15. The Visible Telescope (VT)  onboard the Chinese-French Space Variable Objects Monitor mission (SVOM\cite{Wei2016}) performed optical observations of this event at five epochs, revealing a decaying optical afterglow at the same position\cite{supplementary}. The optical counterpart was further observed with the Low-Resolution Imaging Spectrometer (LRIS\cite{Oke1995}) on the Keck~I 10~m telescope on 2024 Nov.~30,  16.76 days after the trigger. The LRIS spectrum at the position of EP241113a shows a low signal-to-noise trace, likely corresponding to the host galaxy of the transient. A single, weak emission line is detected at $9420$\AA \,  (Fig.~\ref{fig_host}). Among the possible interpretations, the [O II] line at $z = 1.53$ is favored due to the lack of any other lines, which would otherwise be detected if the line were [O III]  or H$\alpha$\cite{supplementary}. A GTC HiPERCAM $ugriz$ observation obtained 54.2~days after the burst reveals a faint and extended source with a magnitude of $r=24.97\pm0.30$~AB~mag (Fig.~\ref{fig_host}), which we interpret as the host galaxy. Through modeling of the spectral energy distribution (SED) of the host galaxy, we obtain a photometric redshift of $z=1.58_{-0.16}^{+0.19}$\cite{supplementary}, which is fully consistent with the LRIS spectroscopic redshift. The host galaxy is estimated to have a stellar mass of $\rm M\sim10^{10} M_\odot$ and a star-formation rate of  $\rm \sim 12 M_\odot~yr^{-1}$, and the best-fitting SED is fully consistent with that measured through the Keck-LRIS spectra.

\section{Deviation from the Amati relation}
At a redshift of  $z = 1.53$, the peak isotropic X-ray luminosity and total isotropic energy of EP241113a in the 0.5–4 keV band are $L_{\rm p}=(5.1\pm 0.7) \times 10^{49} \, \rm erg \, s^{-1}$  and $E_{X, \rm iso} = 1.4^{+0.7}_{-0.2} \times 10^{51} \, \rm erg$, respectively.  Considering the upper limit on the
peak energy ($E_{\rm p} \le 2.4$ keV), we find that this event is  clearly an outlier in the “Amati relation”\cite{Amati2002} (see Fig.~\ref{fig_Amati}), which applies to  long GRBs, X-ray rich GRBs, X-ray flashes and low-luminosity GRBs. This is attributed to its extremely low peak energy and much higher energetics compared to other soft EFXTs. {In contrast, two of the EP-detected EFXTs, EP240315a and EP240801a \cite{Liu2025, Jiang2025}, follow the Amati relation. The low-luminosity event EP240414a also deviates from the Amati relation\cite{Sun2025}, but its energy is much lower than classic GRBs. EP241113a is unique in that its energy is comparable to those of classical GRBs, whereas its peak energy is much lower.}   Hence, EP241113a unveils a new parameter space of cosmic explosions.

\section{Jet properties inferred from the steep decay and plateau}
{The similarity of the X-ray light curves between EP241113a  and GRBs indicates a jet origin of EP241113a.  To investigate its jet properties, we conduct detailed theoretical modeling.} The steep decay phase, commonly observed in Swift GRBs\cite{Tagliaferri2005,ZhangBB2007}, has been interpreted as the high-latitude  prompt emission when central engine activity ceases\cite{Kumar2000,Zhang2006,ZhangBB2009}.  For a uniform jet, the decay of X-ray emission is predicted to follow
$F_\nu \propto t^{-\alpha} \sim t^{-(2+\beta)}$, where $\beta$ is the power law slope of the flux density spectrum (i.e., $F_\nu\propto \nu^{-\beta}$)\cite{Kumar2000}. The true GRB jets likely structured (e.g., a uniform core within $\theta_{\rm c}$ surrounded by a structured wing with decreasing energy per solid angle at larger viewing angles)\cite{Zhang2002,Rossi2002}. 
For a structured jet, the decay is preserved to be steep as long as the viewing angle ($\theta_{\rm v}$) is within the uniform core  of the structured jet, but becomes shallower if the viewing angle is outside  the core (see Fig.~\ref{fig_HLE_diff_thetav} and Ref.\cite{supplementary}). 
The observed slope of steep decay, $\alpha = 3.05^{+0.47}_{-0.41}$, is in good agreement with with the prediction for an on-axis viewing angle, $\alpha = 2+\beta = 1+\Gamma_X$,
where $\Gamma_X=2.1\pm0.1$ is the photon index during the steep decay phase. Therefore, an off-axis jet in EP241113a can be ruled out and the data indicates a jet seen on axis with an intrinsically soft high-energy emission spectrum\cite{supplementary}.

\begin{figure}[htbp]
    \centering
    \begin{overpic}[width=0.46\textwidth]{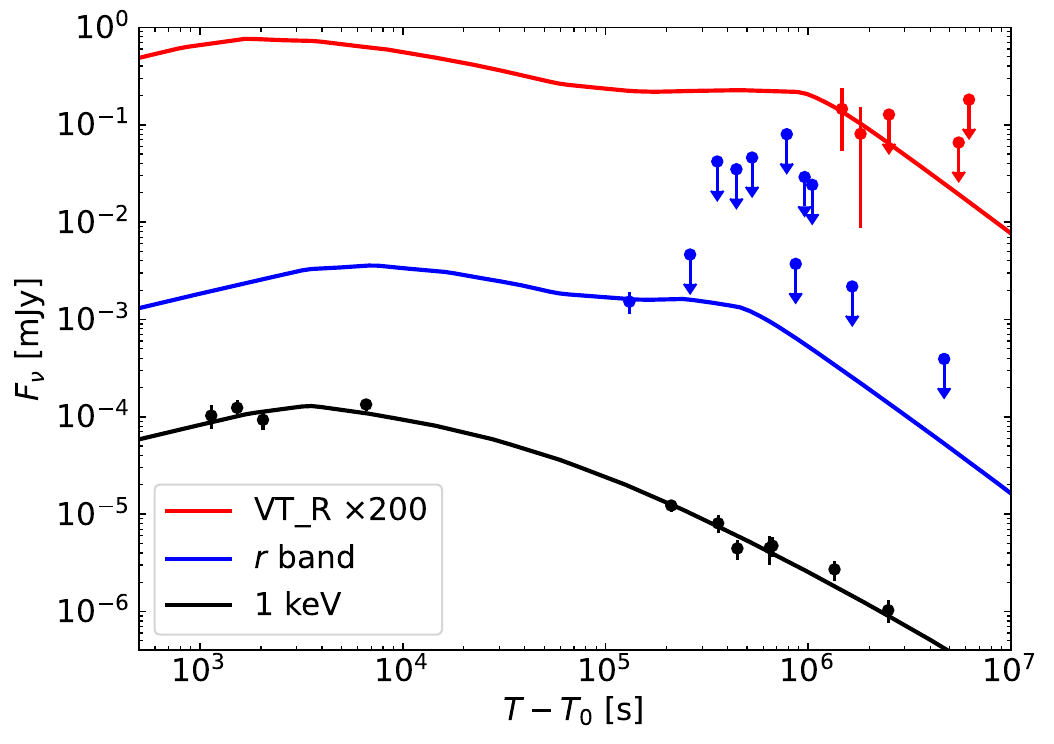}
        \put(2,73){\textbf{(a)}} 
    \end{overpic}
    \hspace{0.05\textwidth} 
    \begin{overpic}[width=0.46\textwidth]{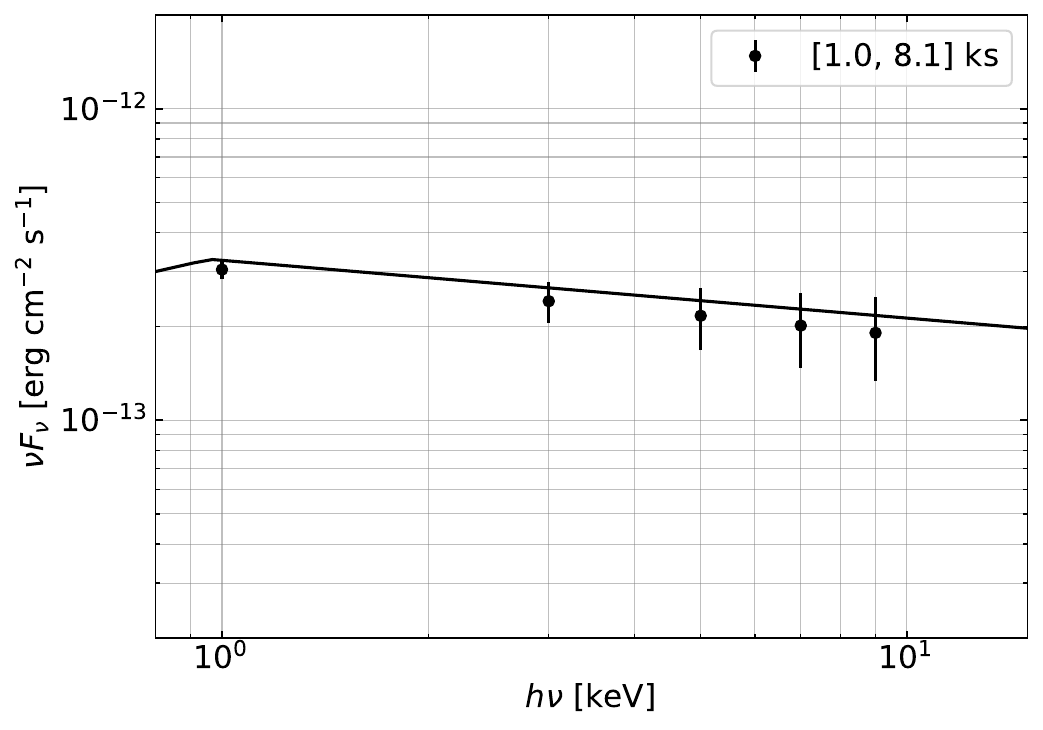}
        \put(2,73){\textbf{(b)}} 
    \end{overpic}

    \caption{\textbf{Modeling of the multi-band afterglow of EP241113a.} 
   (a) Modeling the multi-band light curve as synchrotron emission from a forward shock expanding in a wind environment, assuming a power-law structured jet with $k_{\rm c}=2$. The best-fit parameters, derived from the MCMC samples with the minimum reduced $\chi^2$, are as follows: $E_{\rm k} = 1.0 \times 10^{53} \, \rm erg$, $\Gamma_c = 23$, $\theta_{\rm c} = 0.48$, $p = 2.4$, $\epsilon_e = 1.3\times 10^{-2}$, $\epsilon_B = 3.5\times 10^{-2}$, $A_* = 1.5 \times 10^{-2}$, and $\xi = 4.8\times 10^{-3}$ (for parameter definitions, see Ref.\cite{supplementary}). The marginalized distributions of the parameters are shown in  Fig.~\ref{fig_af_contour} and the  Table~\ref{tab_afterglow_mcmc_results}. Note that, for clarity, the comparison between the upper limits of the other optical  data and radio data with the model is displayed in  Fig.~\ref{fig_af_up}. (b) Modeling the spectral data of EP241113a during $1.0$–$8.1 \, \rm ks$ with the forward shock synchrotron emission,  using the same  parameters as in panel (a). The spectral data are derived using the best-fit absorbed power-law model during the corresponding time interval.}
    \label{fig_afterglow_fitting}
\end{figure}

The plateau phase of GRBs is usually interpreted as a result of energy injection from the central engine\cite{Dai1998,Zhang2006}. 
However, the plateau light curve of  EP241113a is extremely flat, which requires $q\sim -0.3$ for injection of energy in the form of $L\propto t^{-q}$ in both a constant-density medium and a wind medium\cite{supplementary}. Such a small value of $q$ is unusual, since the quickest energy injection  predicts $q=0$, as expected from the  initial spin down of a millisecond pulsar. In fact, the plateau seen in most GRBs requires $q>0$\cite{Zhang2006}. 
A more natural scenario is that the jet is freely expanding in a wind medium and a nearly flat light curve can be produced without energy injection\cite{Shen2012,Dereli-Bégué2022}. The  density of the wind medium is described by $\rho= A R^{-2}$\cite{Dai1998-2,Chevalier1999}, where $R$ is the radius from the progenitor and $A=\dot{M}/4\pi v_w=5\times 10^{11} A_* \,{\rm g \, cm^{-1}}$ with $\dot{M}$  the mass loss rate and $v_w$  the wind velocity (a reference value of $A_*=1$ corresponds to $\dot{M}=10^{-5}M_\odot \, \rm yr^{-1}$ and $v_w=10^3 \, {\rm km \, s^{-1}}$). For this model, the break time from the plateau to the normal decay corresponds to the epoch when the jet starts to significantly decelerate (i.e., the deceleration time $t_{\rm dec}$). 
This leads to an estimate of the initial Lorentz factor of the jet\cite{supplementary}:
\begin{equation}
    \Gamma \simeq 16 E_{\rm k, 53}^{\frac{1}{4}} {A_{*}}^{-\frac{1}{4}} \left( \frac{t_{\rm dec}}{10^4 \, \rm s} \right)^{-\frac{1}{4}},
    \label{equation1}
\end{equation}
where $E_{\rm k, 53}$ is the isotropic-equivalent kinetic energy of the jet in units of $10^{53}{\, \rm erg}$. Detailed modeling of the afterglow using Markov Chain Monte Carlo (MCMC) method  gives $\Gamma\simeq 20$ (see Fig.~\ref{fig_afterglow_fitting} and  Fig.~\ref{fig_af_contour} in Ref.\cite{supplementary}), which is  consistent with the above analytical estimate (i.e., Eq.~\ref{equation1}) for the preferred parameter values in the MCMC fitting: $E_{\rm k}\simeq 1.0\times 10^{53} {\, \rm erg}$, $A_{*}\simeq 1.5\times 10^{-2}$, and $t_{\rm dec}\simeq 10^5 {\, \rm s}$. In addition,  a robust constraint on $\Gamma \sim 20$ can also be obtained from the flux of the X-ray plateau, independent of $E_{\rm k}$ and  $t_{\rm dec}$\cite{supplementary}.  This suggests that this soft X-ray explosion indeed has a much smaller Lorentz factor than those of typical GRBs. 
Adopting such a low-$\Gamma$ jet as a prior, the steep decay of EP241113a can be interpreted as the high-latitude emission of a jet viewed  with $\theta_{\rm v}<\theta_{\rm c}$ (see Fig.~\ref{fig_HLE_fix} and Ref.\cite{supplementary}). 

\section{Conclusions and implications}
All the data and modeling of EP241113a point toward a new type of cosmic explosion never detected before and are consistent with a long-expected dirty fireball\cite{Paczynski1998, Dermer1999, Huang2002, Rhoads2003}, i.e., an energetic explosion with a Lorentz factor much lower than those of GRBs. This implies that massive star core-collapse events can produce a variety of explosions with diverse properties. In particular, besides making traditional highly relativistic and energetic GRBs and their softer and fainter cousins, some explosions can produce energetic but slower, dirty jets contaminated by more baryons.

{There are a few possible physical mechanisms to make dirty fireballs. The core collapse of massive stars probably produces a hyperaccreting black hole  system surrounded by a neutrino-dominated accretion flow (NDAF)\cite{Woosley1999,Narayan2001}. 
A bipolar jet can be powered either by neutrino-anti-neutrino annihilation from the NDAF or, if the engine is highly magnetized, by the Blandford-Znajek mechanism\cite{BZ1997} that taps the spin energy of the black hole. The former is known to launch much dirtier jets than the latter\cite{Lei2013}. 
Therefore, EP241113a could be powered by a jet driven by neutrino annihilation with an unmagnetized central engine. 
Secondly, it is possible that the central engine is
a newborn neutron star rather than a black hole. A heavily mass-loaded outflow is launched by a neutrino-driven wind from the hot neutron star surface, leading to a dirty fireball\cite{Zhang&Dai2009}. Finally, a dirty jet could result from an initially unclean funnel along the rotational axis of a star where the jet has to propagate through\cite{Woosley1999}, jet instabilities that cause the orientation of the jet to vary with time\cite{Zhangwq2004}, or the mixing of surrounding materials into the jet during its propagation inside the star\cite{Zhangwq2003}.  
}

{A conservative lower limit of the intrinsic event rate density for EFXTs similar to EP241113a, derived
from the 15 months of operation of EP-WXT, is $\sim 4 \times 10^{-3} \, \rm Gpc^{-3} \, yr^{-1}$\cite{supplementary}.} The true rate should be significantly higher, given that EP has detected many more fast X-ray transients without gamma-ray counterparts but their physical origins remain uncertain owing to the lack of timely follow-up observations and localizations. Continued observations with EP and other similar instruments will reveal the full zoo of these explosions with more dirty fireballs being identified. 


\clearpage

\begin{figure}[htbp]
    \centering
    \begin{overpic}[width=0.45\textwidth]{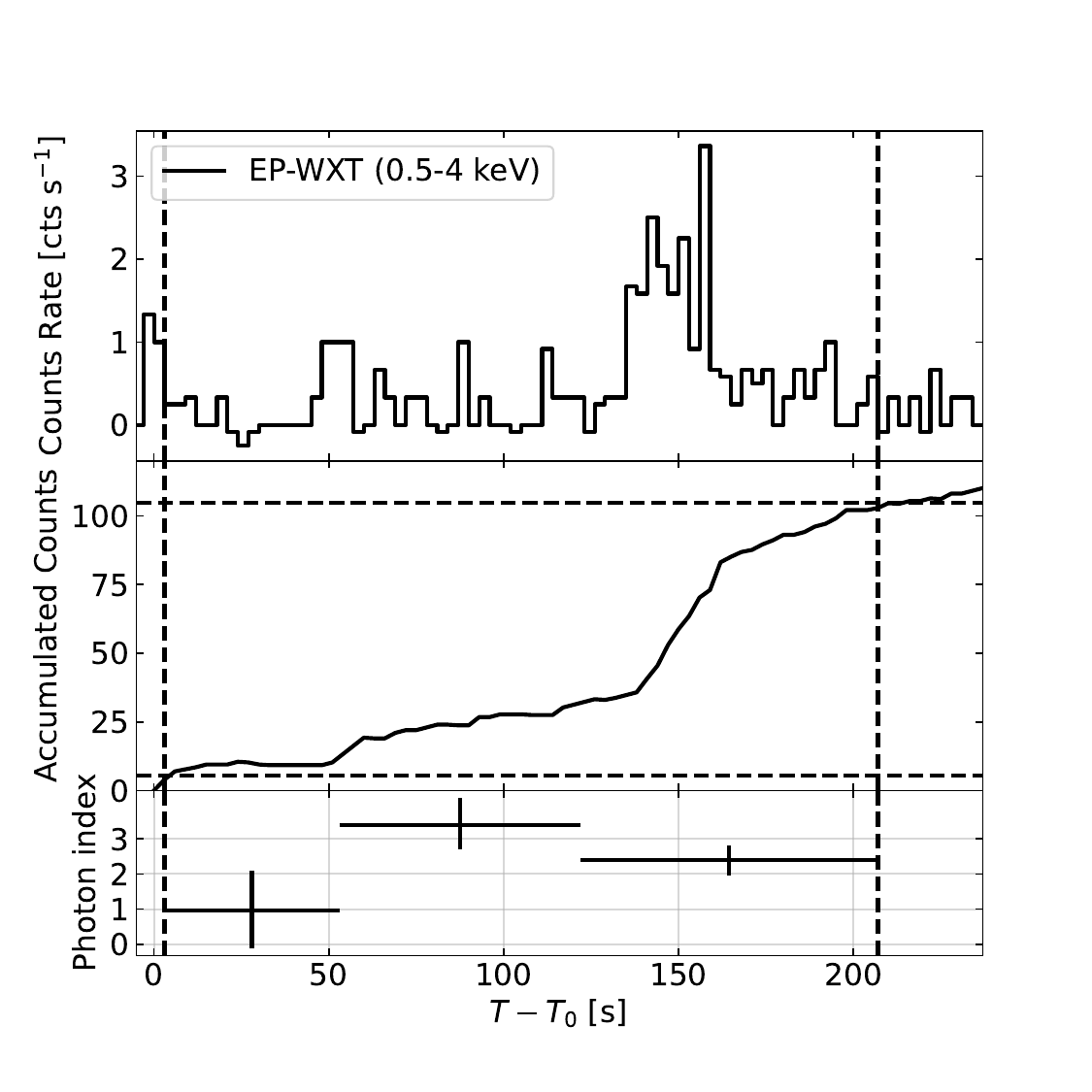}
        \put(2,90){\textbf{(a)}} 
    \end{overpic}
    \hspace{0.05\textwidth} 
    \begin{overpic}[width=0.45\textwidth]{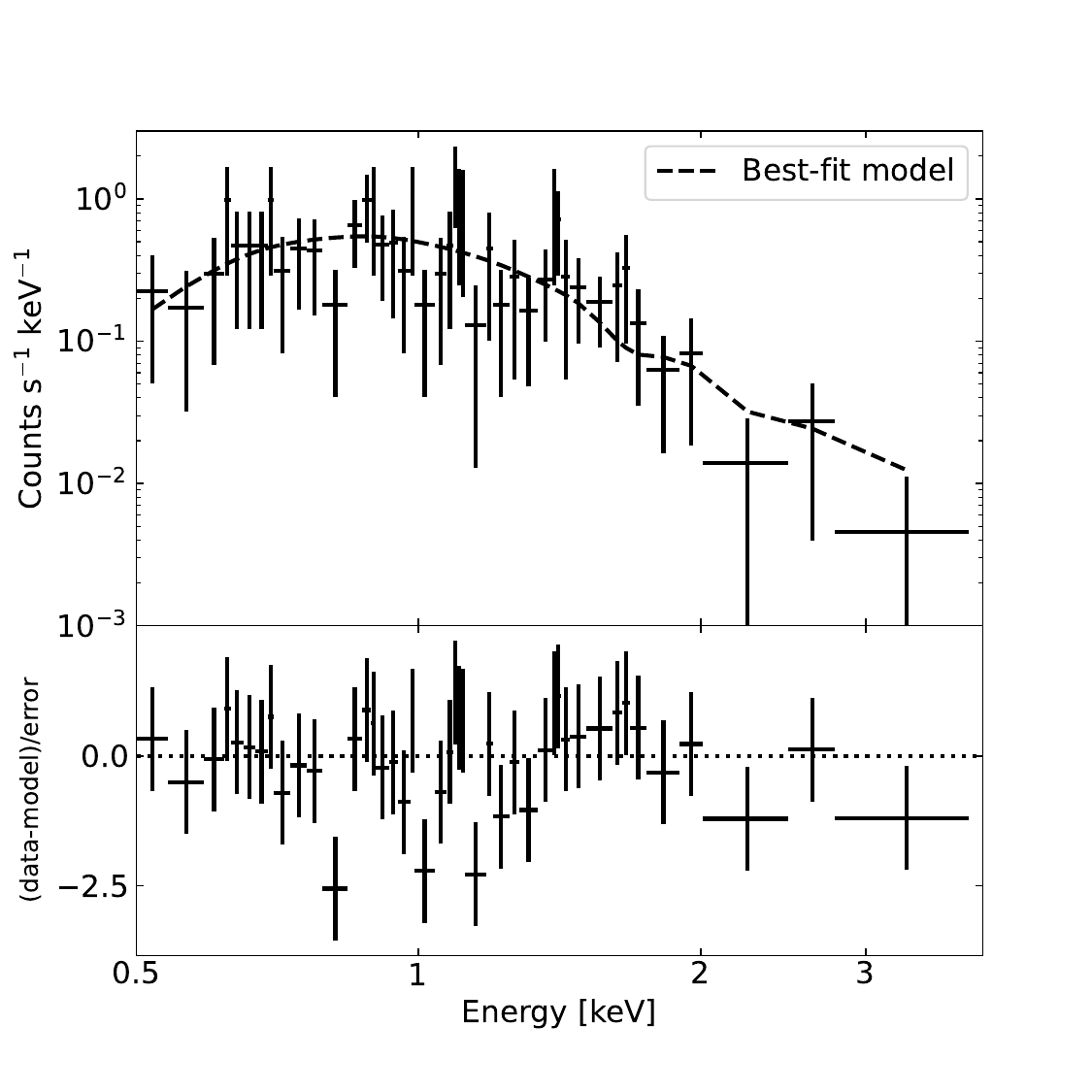}
        \put(2,90){\textbf{(b)}} 
    \end{overpic}

    \vskip 0.3cm

    \begin{overpic}[width=0.40\textwidth]{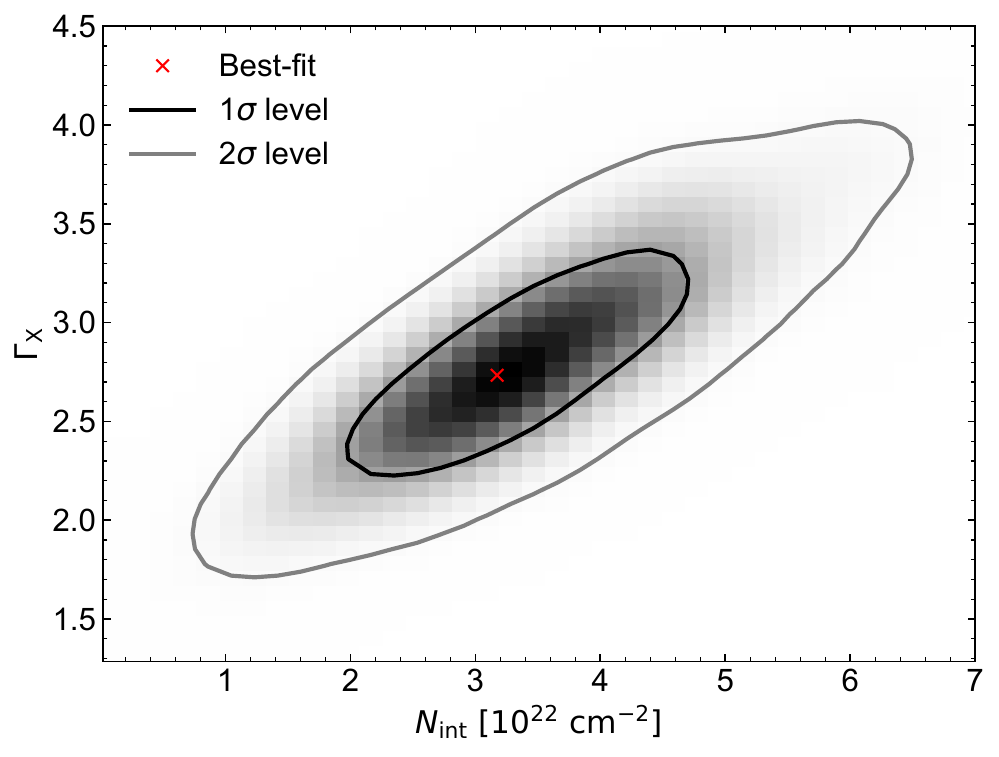}
        \put(-4,76.5){\textbf{(c)}} 
    \end{overpic}
    \hspace{0.05\textwidth} 
    \begin{overpic}[width=0.45\textwidth]{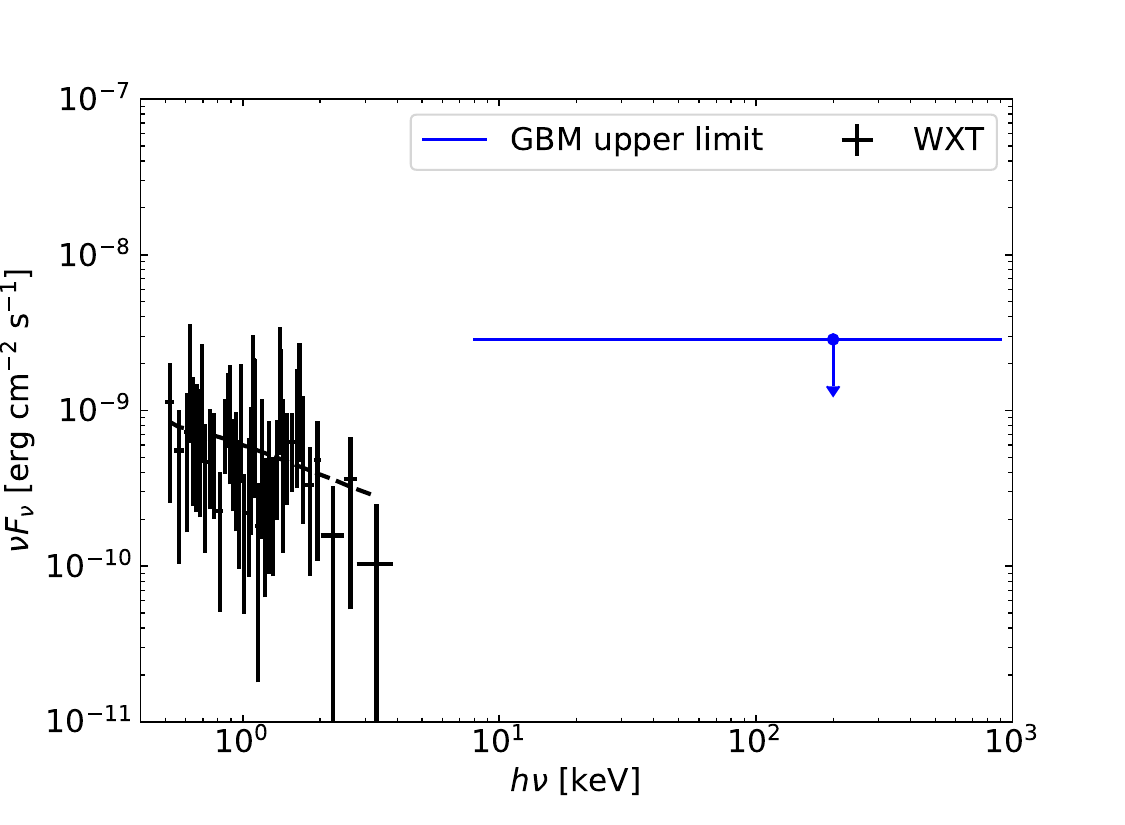}
        \put(7,68){\textbf{(d)}} 
    \end{overpic}

    \caption{\textbf{Characteristics of the  prompt X-ray emission of EP241113a observed by EP-WXT.} (a) Upper panel: The light curve of the net count rate of EP241113a in the $0.5$--$4\, \rm keV$ band, with a time bin size of $3\, \rm s$. Middle panel: Cumulative X-ray photon count distribution as a function of time. Lower panel: The photon index in  the absorbed power-law model for three time bins. The two vertical dashed lines mark the epochs $T_{05} = T_0 + 3 \, \rm s$ and $T_{95} = T_0 + 207 \, \rm s$, corresponding to the times when the cumulative fluence reaches $5\%$ and $95\%$ of the total fluence, respectively. The horizontal dashed line marks these count thresholds. (b) The fitting results of the EP-WXT spectrum in the $0.5$–$4 \, \rm keV$ band during $T_{90}$, assuming an absorbed power-law model\cite{supplementary}. The dashed line in the upper panel represents the best-fit model. (c) The best-fit values of the photon index $\Gamma_X$ and the intrinsic absorption ($N_{\rm{int}}$) for the WXT emission assuming the absorbed power-law model  (indicated by a cross), along with the confidence contours corresponding to the $1\sigma$ and $2\sigma$ levels. (d) The intrinsic SED of EP-WXT during $T_{90}$ (black data point). The dashed line represents the same absorbed power-law model with the best-fit parameters from panel (b), but corrected for absorption. {The blue upper limit represents the differential flux limit  in the energy range $[8,\,900]~\rm{keV}$ derived from the Fermi/GBM integrated flux limit (shown in  the  Table~\ref{tab_GBM_up}), assuming a single power-law spectrum with a photon index of $-2$.}}
    \label{fig_WXT_data}
\end{figure}

\clearpage

\begin{figure}[htbp]
    \centering
    \begin{overpic}[width=0.46\textwidth]{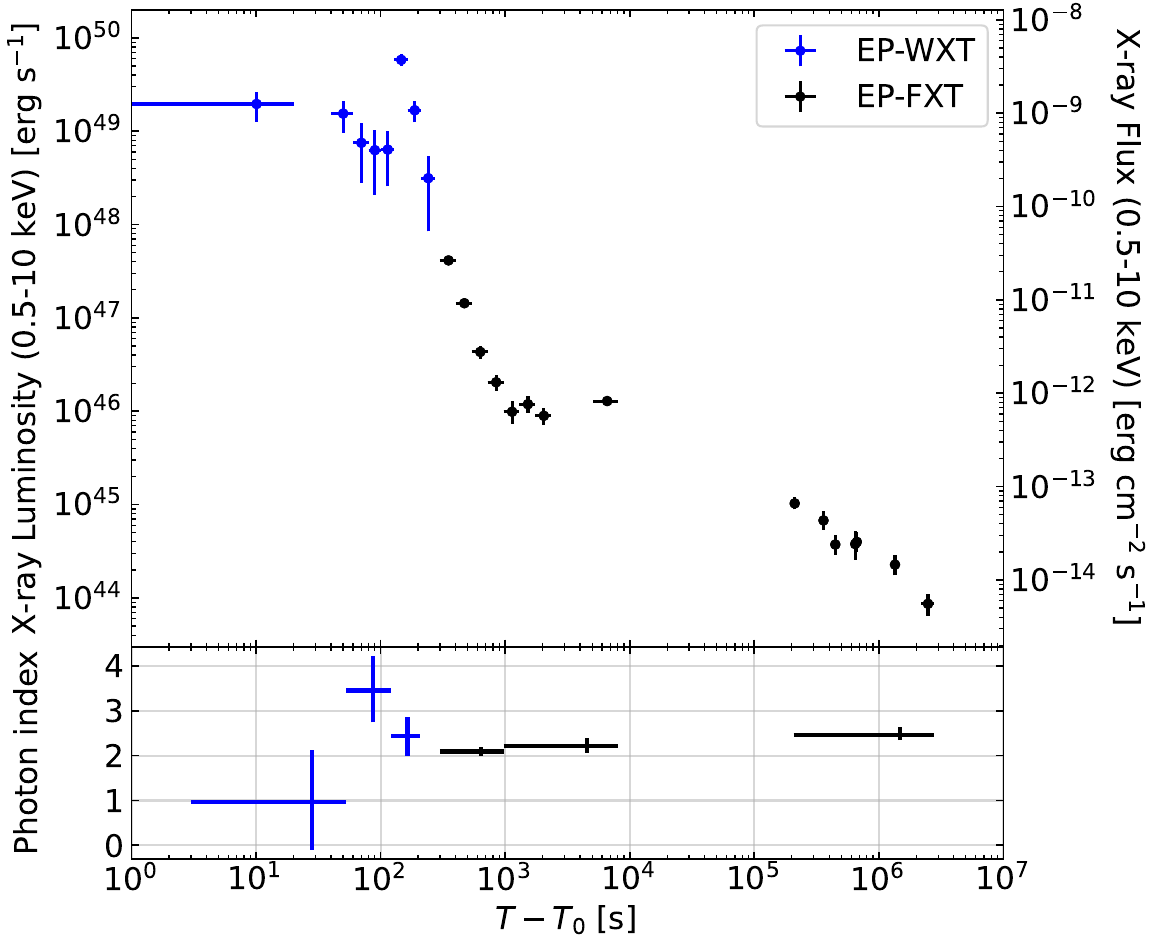}
        \put(2,85){\textbf{(a)}} 
    \end{overpic}
    \hspace{0.05\textwidth} 
    \begin{overpic}[width=0.415\textwidth]{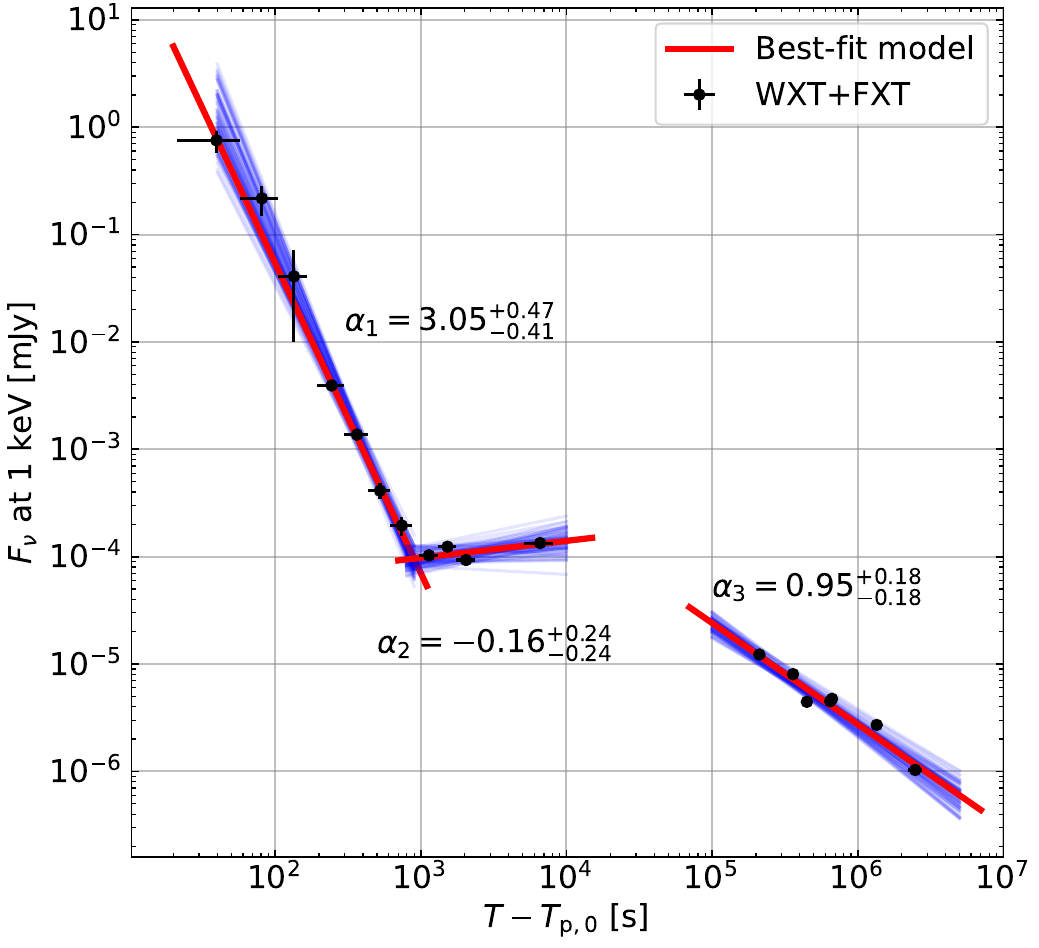}
        \put(2, 94){\textbf{(b)}} 
    \end{overpic}

    \caption{\textbf{X-ray light curve of EP241113a and the photon indices of the spectra for various time intervals.} (a) The upper panel shows the X-ray luminosity and flux observed by EP-WXT (blue) and EP-FXT (black) in the $0.5$–$10 \, \rm keV$ band. 
    For EP-WXT, the flux in the $0.5$--$10\,\rm{keV}$ band is derived by extrapolating the observed count rate in the $0.5$--$4\,\rm{keV}$ band, using a count rate-to-flux conversion factor derived from the best-fit spectral parameters of the absorbed power-law model (see Table~\ref{tab_spectrum_fitting}).
   The data during the steep decay and plateau phases were binned into logarithmically spaced time intervals with a bin size of $\Delta \log{t\,(\rm s)} = 0.1$. 
   The lower panel shows the photon index fitted with an absorbed power-law model during different time intervals, as listed in the  Table~\ref{tab_spectrum_fitting}. All errors are shown at the $1\sigma$ confidence level. (b) MCMC fitting results for the temporal decay indices of the X-ray flux density during the steep decay, plateau, and normal decay phases\cite{supplementary}. The red line denotes the best-fitting model, obtained using the parameter set that yields the minimum reduced $\chi^2$ among all completed runs. 
   The blue shaded region illustrates 70 representative MCMC realizations drawn from the posterior distribution within the $1\sigma$ credible interval. The uncertainties of the temporal indices, $\alpha_1$, $\alpha_2$, and $\alpha_3$, are quoted at the $1\sigma$ confidence level. 
}
    \label{fig_WXT_luminosity}
\end{figure}

\clearpage

\begin{figure}[htbp]
    \centering
    \hspace{0.05\textwidth} 
    \begin{overpic}[width=\textwidth]{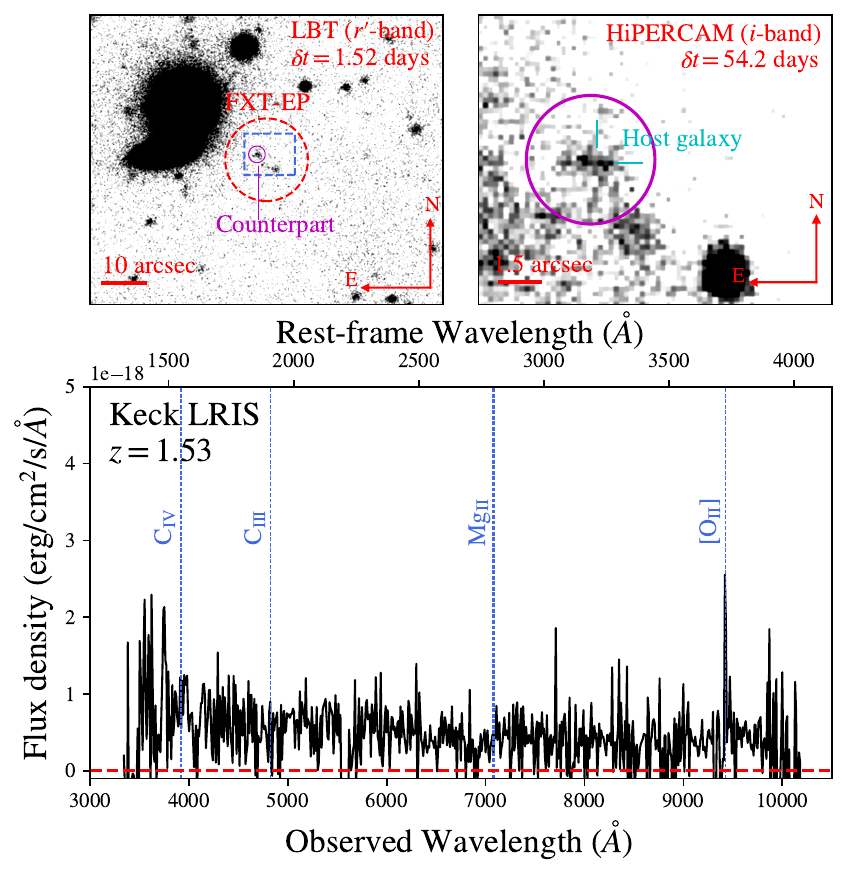}
    \end{overpic}
    \caption{\textbf{Images and host-galaxy spectrum of EP241113a.} 
    The top left panel shows $r'$-band image of the field of EP241113a using the LBT telescope taken 1.52~days after the trigger\cite{supplementary}, where the X-ray position uncertainty of the transient measured by the EP-FXT telescope and its optical counterpart are marked by red-dashed and magenta circles, respectively. 
    The top right panel shows a zoom-in at the position of the transient taken by GTC-HiPERCAM ($i$-band) 54.2~days after the X-ray trigger, with the host galaxy of the transient marked by cyan ticks. 
    The bottom panel shows a Keck-LRIS spectrum taken of the optical transient at 16.78~days after the transient detection. The strong emission line at 9429\AA~is interpreted as $[\rm O_{\rm II}]\lambda3727$ at $z=1.53$.}
    \label{fig_host}
\end{figure}

\clearpage

\begin{figure}[htbp]
    \centering
    \hspace{0.05\textwidth} 
    \begin{overpic}[width=1.0\textwidth]{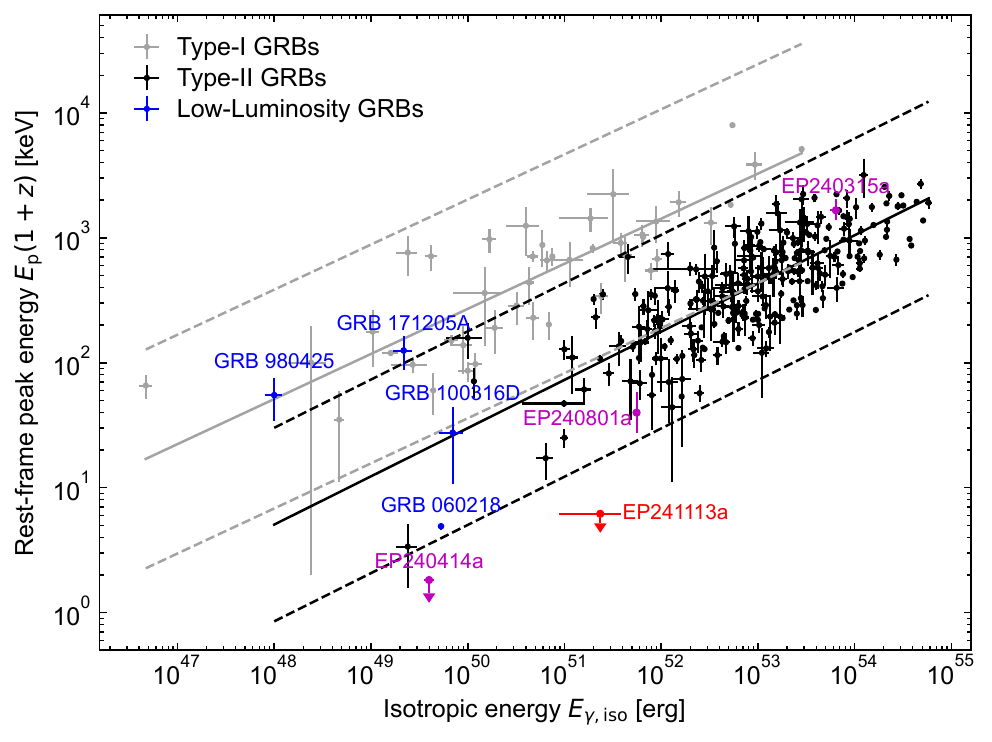}
    \end{overpic}

    \caption{\textbf{The rest-frame peak energy versus isotropic energy for EFXTs detected by EP and GRBs  (the Amati relation).} Type I and Type II GRBs correspond to those originating from compact object mergers and massive star core collapses, respectively. All data points are plotted with $1\sigma$ uncertainties.  $E_{\gamma, \rm iso}$ is calculated over the $1-10^4\,\rm keV$ energy range in the rest frame.  The data for GRB 100816D, EP240414a, and EP240801a are adopted from Refs.\cite{Sun2025,Jiang2025}, while the remaining GRB samples are taken from Ref.\cite{Liu2025}. The solid lines represent the best-fit Amati relations for Type I and Type II GRBs, and the dashed lines indicate the corresponding $3\sigma$ intrinsic scatter regions (see Ref.\cite{supplementary, Liu2025}). As illustrated, EP241113a is located outside the $3\sigma$ scatter region of the Amati relation.}
    \label{fig_Amati}
\end{figure}

\clearpage

\clearpage



\newcommand\aastex{AAS\TeX}
\newcommand\latex{La\TeX}
\newcommand{\apj}{Astrophys. J.}
\newcommand{\procspie}{Proc. SPIE}
\newcommand{\pasp}{Publ. Astron. Soc. Pac.}
\newcommand{\apjs}{Astrophys. J. Supp.}
\newcommand{\araa}{Annu. Rev. Astron. Astrophys.}
\newcommand{\mnras}{Mon. Not. R. Astron. Soc.}
\newcommand{\apjl}{Astrophys. J. Let.}
\newcommand{\aap}{Astron. Astrophys.}
\newcommand{\aj}{Astron. J.}
\newcommand{\nat}{Nature}
\newcommand{\na}{New Astron. Rev.}
\newcommand{\aaps}{A\&AS}
\newcommand{\apss}{Ap\&SS}
\newcommand{\aapr}{A\&ARv}
\newcommand{\solphys}{Solar Phys}
\newcommand{\fcp}{Fundamentals of Cosmic Physics}
\newcommand{\prl}{Phys. Rev. Lett.}
\newcommand{\prd}{Phys. Rev. D}
\newcommand{\pasj}{Publ. Astron. Soc. Jpn.}
\newcommand{\pasa}{Publ. Astron. Soc. Aust.}
\newcommand{\ssr}{Space. Sci. Reviews.}
\newcommand{\actaa}{Acta Astron.}
\newcommand{\jcap}{Journal of Cosmology and Astroparticle Physics}

\bibliography{ref}
\bibliographystyle{Science}

\section*{Acknowledgments}
This work is based on data obtained with the Einstein Probe (EP), a space mission supported by the Strategic Priority Program on Space Science of the Chinese Academy of Sciences, in collaboration with ESA, MPE, and CNES (grant XDA15310000); the Strategic Priority Program on Space Science of the Chinese Academy of Sciences (grant No. E02212A02S) and the Strategic Priority Research Program of the Chinese Academy of Sciences (Grant No. XDB0550200). Data for this paper has in part have been obtained under the International Time Programme of the CCI (International Scientific Committee of the Observatorios de Canarias
of the IAC) with the NOT and GTC operated on the island of La Palma by the Roque de los Muchachos. {The radio reduction in this work used resources from the China SKA Regional Centre (CNSRC). e-MERLIN is a National Facility operated by the University of Manchester at Jodrell Bank Observatory on behalf of STFC, part of UK Research and Innovation.}
{\bf Funding:} We acknowledge the support by  the National Natural Science Foundation of China (NSFC grants 12333006, 12121003, 12203022, U2031105, and 12233002).  This work is also supported by  the  National Key R$\&$D Program of China under grant No. 2022YFF0711404. 
P.G.J. is funded by the European Union (ERC, Starstruck, 101095973) (Views and opinions expressed are however, those of the author(s) only and do not necessarily reflect those of the European Union or the European Research Council Executive Agency. Neither the European Union nor the granting authority can be held responsible for them). 
D.M.S. and M.A.P.T. acknowledge support by the Spanish Ministry of Science via the Plan de Generacion de conocimiento PID2020-120323GB-I00. D.M.S. also acknowledges support via a Ramon y Cajal Fellowship RYC2023-044941. A.R.(INAF) acknowledges support by the project Supporto Arizona \& Italia.
W.X.L. is supported by the National Key R\&D Program of China (grant Nos. 2022YFA1602902, 2023YFA1608100, 2023YFA1607800, 2023YFA1607804), the National Natural Science Foundation of China (NSFC; grant Nos. 12120101003, 12373010, 12173051, and 12233008), and the Strategic Priority Research Program of the Chinese Academy of Sciences with Grant Nos. XDB0550100 and XDB0550000.  V.S.D., S.L. and HiPERCAM are supported by STFC grant  ST/Z000033/1. {T.A. acknowledges the support of the Xinjiang Tianchi Talent Program. T.A. and A.L.W.  are supported by the National SKA Program of China grant numbers 2022SKA0130103 and FAST special funding (NSFC 12041301).} P.O. acknowledges STFC grant ST/W000857/1.




\subsection*{Author contributions}
W.Y. has been leading the Einstein Probe project as Principal Investigator since the mission proposal stage. X.Y.W., B.Z., C.Y.D., and Y.H.W.  initiated the study. X.Y.W., B.Z.,  P.J., Y.L., and X.F.W. coordinated the scientific investigations of the event and led the discussions. C.Y.D., Y.L., and J.Y. processed and analysed the WXT data. C.Y.D. processed and analyzed the  FXT data with the assistance of Y.L. and H.S..  P.J. and J.Q. obtained and reduced the LBT, Keck, and GTC data. H.L.L., Y.L.Q., and L.P.X. obtained and reduced the SVOM/VT data. J.Y. reduced the Fermi/GBM data. Y.P.Y., X.L.C., and X.W.L. obtained and reduced the Mephisto data. D.X. and J.A. obtained and reduced the other optical data.  A.L.W. and T.A.  processed and analyzed radio data. C.Y.W. and J.Y. contributed to the plot of the Amati relation. R.Y.L., Z.G.D., and M. E. R. contributed to the theoretical investigation of the event. X.Y.W., B.Z., C.Y.D., W.Y., P.J., Y.L., J.Q., J.Y., and Y.H.W.  drafted the manuscript with the help of all authors.

\subsection*{Competing interests}
There are no competing interests to declare.

\subsection*{Data and materials availability}
All processed data supporting the findings of this study are included in the figures and tables within the main text. Additional data are available from the corresponding authors upon reasonable request. Publicly available datasets were also utilized, including those obtained from GCN Circulars. The code used to generate the results and figures in this study is available from the authors upon reasonable request.

\clearpage


\setcounter{page}{1}
\renewcommand{\thepage}{S\arabic{page}}

\section*{}
\begin{center}
{\large Supplementary Materials for}

{\bf An energetic dirty fireball detected in soft X-rays}

\noindent C.-Y. Dai$^{1,2}$†, J. Quirola-V\'asquez$^{3}$†, Y.-H. Wang$^{4,5}$†, H.-L. Li$^6$, J. Yang$^{1,2}$,  X.-L. Chen$^{7}$, A.-L. Wang$^{8,9}$, H. Sun$^6$,  X.-Y. Wang$^{1,2}$*, B. Zhang$^{10,4}$*, P. G. Jonker$^3$*, Y. Liu$^6$*, W. Yuan$^{6,11}$*, D. Xu$^{6,12}$, Z.-G. Dai$^{13}$, M.~E. Ravasio$^{3,14}$, L. Piro$^{15}$,  P. O'Brien$^{16}$, D. Stern$^{17}$, H.-M. Zhang$^{18}$, Y.-P. Yang$^7$, T. An$^{8,11}$, Y.-L. Qiu$^6$, L.-P. Xin$^{6}$, W.-X. Li$^6$,   R.-Y. Liu$^{1,2}$,   X.-F. Wu$^{19}$, C.-Y. Wang$^{20}$, D.-M. Wei$^{19}$, Y.-F. Huang$^{1,2}$,  F.~E. Bauer$^{21}$, W.-H. Lei$^{22}$, B.-B. Zhang$^{1,2}$, N.-C. Sun$^{6,11,23}$, H. Gao$^{23,24}$, V. S. Dhillon$^{25,26}$, J. An$^{6}$,   C.-H. Bai$^{27}$,   A. Martin-Carrillo$^{28}$, H.-Q. Cheng$^6$, J.~A. Chacon Chavez$^{14}$, Y. Chen$^9$, G.-W. Du$^7$, J.~N.~D. van Dalen$^3$,  A. Esamdin$^{27}$, Y.-Z. Fan$^{19}$, X. Gao$^{27}$, F. Harrison$^{29}$, J.-W. Hu$^6$, M.-Q. Huang$^{13}$, S.-M. Jia$^9$, A.~J. Levan$^{3,30}$, C.-K. Li$^9$, D.-Y. Li$^6$, E.-W. Liang$^{18}$, S. Littlefair$^{25}$, X.-W. Liu$^7$, Z.-Y. Liu$^{13}$, Z.-X. Ling$^{6,11}$, D.~B. Malesani$^{31,32}$, H.-W. Pan$^6$,  A. Rodriguez$^{29}$, A. Rossi$^{33}$, D. Mata S\'anchez$^{26,34}$, J. Sánchez-Sierras$^3$,  X.-J. Sun$^{35}$, M.~A.~P. Torres$^{26,34}$, A.~P.~C. van Hoof$^3$, X.-F. Wang$^{36}$,  Q.-Y. Wu$^6$, X.-P. Xu$^{6,11}$, Y.-F. Xu$^{6,11}$, Y.-W. Yu$^{37}$, C. Zhang$^6$, M.-H. Zhang$^6$, S.-N. Zhang$^9$, Y. Zhang$^{27}$, Y.-H. Zhang$^{38}$, Z.-P. Zhu$^6$\\

\vskip 1cm

* Corresponding authors:  X.-Y. Wang (xywang@nju.edu.cn), B. Zhang (bing.zhang@unlv.edu), P. G. Jonker (p.jonker@astro.ru.nl), Y. Liu (liuyuan@nao.cas.cn), W. Yuan (wmy@nao.cas.cn)\\

† These authors contributed equally to this work.

\end{center}

\vskip 1cm
\noindent{\bf This PDF file includes:}\\

\noindent\hspace*{1cm} Materials and Methods\\
\noindent\hspace*{1cm} Figures S1 to S9\\
\noindent\hspace*{1cm} Tables S1 to S10\\
\noindent\hspace*{1cm} References \textit{(49--90)}
\clearpage

\setcounter{figure}{0}
\setcounter{table}{0}
\renewcommand{\figurename}{Figure}
\renewcommand{\thefigure}{S\arabic{figure}} 
\renewcommand{\thetable}{S\arabic{table}}   
\setcounter{section}{0}
\renewcommand{\thesection}{S\arabic{section}} 
\setcounter{equation}{0}
\renewcommand{\theequation}{S\arabic{equation}}

\setcounter{figure}{0}
\setcounter{table}{0}
\renewcommand{\figurename}{Figure}
\renewcommand{\thefigure}{S\arabic{figure}} 
\renewcommand{\thetable}{S\arabic{table}}   
\setcounter{section}{0}
\renewcommand{\thesection}{S\arabic{section}} 
\setcounter{equation}{0}
\renewcommand{\theequation}{S\arabic{equation}}

\section{Materials and Methods}

\subsection{Observations and Data Reduction.}

\noindent\textbf{EP-WXT.} 
WXT detected the transient EP241113a onboard at 19:12:53 UTC on November 13, 2024, yielding a signal-to-noise ratio of 9.6 and triggering the onboard processing unit. The acquired X-ray photons were processed using a dedicated data reduction pipeline and the calibration database (CALDB) specifically designed for WXT (Liu et al. in prep.). The CALDB incorporates results from both on-ground calibration experiments and in-orbit calibration observations ({Cheng et al. in prep.}). This procedure closely follows that implemented for the EP-WXT pathfinder, the \textit{Lobster Eye Imager for Astronomy}\cite{ZhangC2022, Cheng2024}. The photon positions were re-projected onto celestial coordinates in the J2000 frame.

To ensure data quality, the Pulse Invariant (PI) value—corresponding to the energy in the channel unit—was calculated for each event, incorporating bias and gain corrections stored in CALDB. Bad or flaring pixels were flagged and removed, while events with grade values between 0 and 12 and without anomalous flags were retained to construct a cleaned events file. The final X-ray image was extracted in the 0.5–4 keV band for further analysis. The light curve and spectra of the source and background were extracted from selected regions: a circular region with a radius of $9'$ for the source, and an annular region with inner and outer radii of $18'$ and $36'$ for the background, respectively. The source spectrum has $110$ net counts in the 0.5–4 keV band during $T_{90}$ after background subtraction.

\noindent\textbf{EP-FXT.} 
Follow-up observations were conducted using the EP Follow-up X-ray Telescope (FXT) \cite{chen2020}. These observations spanned from $\sim 40 \, \rm s$ to 30 days post-trigger (see  Table~\ref{tab_FXT_obs}).

We selected circular regions with radii of $1'$ and $3'$ for the source and background extraction, respectively. Data processing for both source and background regions of the FXT utilized the FXT data analysis software (\texttt{fxtsoftware} v1.20), accessible at \href{http://epfxt.ihep.ac.cn/analysis}{http://epfxt.ihep.ac.cn/analysis}. The reduction pipeline involved particle event identification, PI conversion, grade calculation, and selection (events with grades 0–12), as well as bad pixel flagging. Additionally, housekeeping files were used to filter for intervals with optimal observational conditions to produce high-quality data.

\noindent\textbf{{Fermi}/GBM observations.}
Throughout the duration of the WXT signal of EP241113a, {Fermi}/GBM maintained full temporal and spatial coverage of EP241113a. Specifically, the NaI detectors n9, n6, and n7 remained within an angle of $\lesssim 40^\circ$ relative to the location of EP241113a during this period. However, no onboard GBM trigger was associated with EP241113a. Furthermore, a targeted search\cite{Goldstein2019arXiv} for GRB-like signals was conducted within the time interval encompassing the entire WXT duration of EP241113a, across timescales ranging from 64 ms to 32.768 s. No signal, both temporally and spatially consistent with EP241113a was identified, in agreement with  visual inspection of the data\cite{GCN.38238}. 

The 8.192 s timescale has been utilized to estimate upper limits for short GRB-like signals\cite{Goldstein2019arXiv}, and the longest target search timescale, 32.768 s, is adopted here to estimate upper limits for long GRB-like signals. We then derived the 3$\sigma$ flux upper limit on both the 8.192 s and 32.768 s timescales during the $T_{90}$ duration of the WXT signal of EP241113a, based on data from the n9 detector, which had the smallest viewing angle relative to EP241113a. The adopted spectral model templates include a soft Band function\cite{Connaughton2015ApJS} ($\Gamma_X=1.9$, $\Gamma_{X,2}=3.7$ and $E_{\rm p}=70~{\rm keV}$) and a normal Band function\cite{Poolakkil2021ApJ} ($\Gamma_X=1.1$, $\Gamma_{X,2}=2.2$ and $E_{\rm p}=180~{\rm keV}$). Additionally, we employed power-law model templates with varying photon spectral indices. To calculate the 3$\sigma$ flux upper limit, we adjust the normalization parameter of the spectral model template to ensure that the significance of the model-predicted signal photon counts relative to the background photon counts reaches 3$\sigma$. At this point, the flux derived from the spectral model is then considered the 3$\sigma$ flux upper limit. During this procedure, the significance calculation method is consistent with the case of Poisson measurement and Gaussian background described in Ref.\cite{Vianello2018ApJS}. The resulting flux upper limits in the 8--900 keV energy range are listed in  Table~\ref{tab_GBM_up}.

\noindent\textbf{Optical and NIR  photometric observations.}

{\it LBT observations.} We performed optical observations of the field of the X-ray transient using the Large Binocular Cameras (LBC\cite{Pedichini2003}) mounted on the prime focus swing arms of the Large Binocular Telescope (LBT, Mount Graham, United States), with an approximate mid-time of 2024-11-15 07:45:00 UTC, i.e., $\sim$1.52~days after the burst. Observations were taken in the $r^\prime$ band with a total exposure of 22 minutes and performed under an average seeing of $\sim1.5^{\prime\prime}$. 
Data were reduced using the data reduction pipeline developed at INAF - Osservatorio Astronomico di Roma \cite{Fontana2014}, which includes a standard process for bias subtraction and flat-fielding, bad pixel and cosmic ray masking, astrometric calibration, and stacking. A significant uncataloged optical source was detected at the position RA$_{\rm J2000.0}=$08$^{\rm h}$47$^{\rm m}$59$^{\rm s}$.39,  Dec$_{\rm J2000.0}=$+52$^\circ$22$^\prime$54$^{\prime\prime}$.7, with an apparent magnitude of $r^\prime=23.35\pm0.15$~AB~mag (calibrated against Pan-STARRS field stars).

{\it Keck NIR observations.} On 2024-11-20, the transient was observed by the multi-object spectrometer and imager (MOSFIRE\cite{McLean2012}) instrument mounted on the Keck I telescope, using $J$ ($50\times30$~sec), $H$ ($25\times23$~sec) and $K_s$ ($21\times29$~sec) filters  $\sim$6.83, 6.85, and 6.87 days after the X-ray trigger, respectively. Data were reduced using a standard \texttt{IRAF} process, astrometric corrected using \texttt{Astrometry.net}\cite{Lang2010}, and calibrated against the 2MASS catalog\cite{Skrutskie2006}. No NIR source was identified at the position of the optical transient (GCN Circular 38233\cite{GCN.38233}), yielding a limit of $J>23.35$, $H>22.40$, and $K_s>22.84$~AB mag. Moreover, an extended object was detected in the $J$ and $K_s$ images, which was interpreted as the host galaxy of the transient with $JHK_s$ photometry of $J=22.80\pm0.13$, $H=23.11\pm0.30$, and $K_s=22.06\pm0.27$~AB mag.

{\it SVOM VT observations~.}
The VT (Visible Telescope) is an optical telescope onboard the Chinese-French Space Variable Objects Monitor mission (SVOM\cite{Wei2016}). The effective aperture is 43 cm. The field of view is $26'\times26'$, with a pixel scale of 0.76 arcsecond. VT conducts the observation with two channels, VT\_B and VT\_R, simultaneously, covering the wavelength ranges of 400-650 nm and 650-1000 nm, respectively.
Detailed information on VT is presented in {Qiu et al. (in prep.)}. 
During the commission phase, the transient was observed by VT five times in Target Of Opportunity (ToO) mode, between November 30, 2024 and January 24, 2025.
All data were processed with the standard procedure  with \texttt{IRAF} package, including bias correction, dark correction, and flat-field correction. 
After  pre-processing, the images for each band  were stacked to increase the signal-to-noise ratio. Details of the VT observations are presented in  Table~\ref{tab:vtdata}.

{\it Mephisto observations.}
The simultaneous multiband photometric observations with Multi-channel Photometric Survey Telescope (Mephisto) were conducted starting at 2024-11-16T19:42:59 UTC, 3 days after the EP trigger. 
Mephisto is equipped with two Andor Technology single-chip CCD cameras for the blue ($uv$) and yellow ($gr$) channels. The spatial sampling of the images in these channels is $0^{\prime\prime}.429/\text{pix}$. The red channel ($iz$) utilizes a single-chip camera developed at the National Astronomical Observatories of China (NAOC). The wavelength ranges of the $u$, $v$, $g$, $r$, $i$, and $z$ filters are 320–365, 365–405, 480–580, 580–680, 775–900, and 900–1050 nm, respectively, with central wavelengths at 345, 385, 529, 628, 835, and 944 nm, respectively.

We observed EP241113a with $300 \, \rm s$ exposure times in all bands on November 16, 21, 23, and December 2. For data reduction, the raw frames were processed through a dedicated pre-processing pipeline developed for Mephisto. This pipeline includes bias subtraction, dark subtraction, flat fielding, and cosmic-ray removal. Photometric calibration was performed using  Gaia BP/RP low-resolution spectra (XP spectra) of non-variable stars. Since there were no detections in either single frames or stacked frames for any band, we calculated 3/5-sigma limiting magnitudes for each frame based on the image FWHM and the background fluctuations at the target position. The 3/5-sigma limiting magnitudes are listed in  Table~\ref{tab:limiting_mags_Mephisto}, where  MJD refers to the mid-point of each observation.

{\it TRT/GOT/PAT17 observations.}
We also manually triggered the 70 cm telescopes at Sierra Remote Observatories in the USA, one of the nodes of the worldwide Thai Robotic Telescope (TRT), in the $R$ band within one day of the burst. Four days later, we acquired further observations in the clear and Sloan $r$, $i$ band using the Nanshan 17-inch photometric auxiliary telescope (PAT17) and Gaoyazi-0.5m telescope (GOT), both located in Xinjiang, China. Photometric calibration was performed using the Gaia-DR3\cite{2023A&A...674A...1G} catalog for the clear band and the Pan-STARRS DR2\cite{2016arXiv161205560C} catalog for other bands. Upper limits are tabulated in  Table~\ref{tab:limiting_mags_PAT17S}, in terms of AB magnitudes.

\noindent\textbf{Host-galaxy properties.}
On 2024-11-30 13:49:35.5 UTC (i.e., a mid-time of 16.78~days after the X-ray trigger), spectroscopic observations were taken using the Low Resolution Imaging Spectrometer (LRIS\cite{Oke1995}) mounted at the Keck I telescope with its red and blue cameras, with a total exposure time of $3\times1000$~sec. The data were reduced using the Python Spectroscopic Data Reduction Pipeline (PypeIt\cite{Prochaska2020}). The LRIS spectrum at the position of EP241113a shows a low signal-to-noise trace, likely corresponding to the host galaxy of the transient (Fig.~\ref{fig_host}). A single, weak emission line is detected at 9420\AA. Among the possible interpretations, we favor [O II] at $z = 1.53$, due to the lack of any other lines, which would otherwise be detected if the line were [O III] (at $z = 0.88$) or H$\alpha$ (at $z = 0.44$).

\begin{figure}[htbp]
    \centering
    \hspace{0.05\textwidth} 
    \begin{overpic}[width=\textwidth]{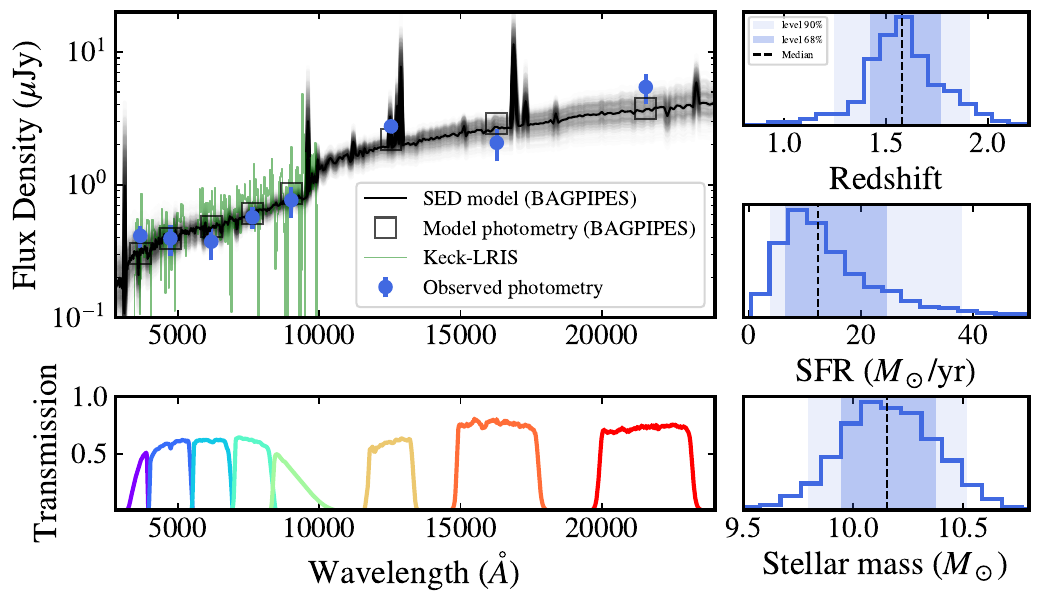}
    \end{overpic}
    \caption{\textbf{The SED of the host galaxy of EP241113a.} 
    Posterior SED models for the host galaxy of EP241113a are obtained from \texttt{Bagpipes}, and the relative transmission functions of the different filters are used in the fitting process. As a comparison, we plot the Keck-LRIS spectrum, which is consistent with the best-fit SED. Moreover, the figure shows the posterior distribution of the photometric redshift, star-formation rate (SFR), and stellar mass, while the color gradient depicts the 68\% and 90\% confidence levels, respectively.}
    \label{fig_SED}
\end{figure}

To detect the host galaxy, the field of the transient was observed by the High PERformance CAMera (HiPERCAM\cite{Dhillon2021}), mounted on the Gran Telescopio Canarias (GTC) at the Roque de los Muchachos observatory on 2025-01-07 01:21:20.6 UTC (i.e., 54.2~ days after the X-ray trigger), when the light of the afterglow counterpart vanished. HiPERCAM obtained photometry of the target in the $ugriz$-bands simultaneously, in $40\times30$~second exposures, and the data were reduced with the HiPERCAM pipeline. The host galaxy of the transient looks slightly extended and faint, with $ugriz$ photometry of $u=24.86\pm0.21$, $g=24.91\pm0.28$, $r=24.97\pm0.30$, $i=24.51\pm0.20$, and $z=24.19\pm0.29$~AB~mag, calibrated against SDSS ($u$ band), Pan-STARRS ($riz$ bands), and Legacy field stars ($g$ band).
We derive the host properties of EP241113a through the SED fitting of the combined photometry (optical + NIR). We consider a star-formation history (SFH) model described by a delayed exponentially declining function using the package \texttt{Bagpipes}\cite{Carnall2018,Carnall2019}. The free parameters of the model and their priors are listed in  Table~\ref{tab:SED_model}. We consider the redshift to be a free parameter to explore the consistency of the spectral interpretation.
The  Fig.~\ref{fig_SED}  depicts the best-fitting SED of the available photometric data derived using \texttt{Bagpipes}. We obtain a photometric redshift of $z=1.58_{-0.16}^{+0.19}$, which is fully consistent with the interpretation of the LRIS spectrum. Moreover, we obtain a stellar mass, star-formation rate and dust attenuation of $\rm log(M_*/M_\odot)=10.15_{-0.21}^{+0.23}$, $\rm SFR=12.33_{-5.85}^{+12.21}$~$M_\odot$~yr$^{-1}$, and $\rm A_V=0.89_{-0.26}^{+0.25}$, respectively.


\noindent\textbf{Radio observations.}
We performed radio  observations using the enhanced Multi Element Remotely Linked Interferometer Network (eMERLIN)  in two epochs: 1) 2024-11-18 20:00 UTC to 2024-11-19 12:21 UTC; and
2) 2024-11-23 08:30 UTC to 2024-11-27 14:49 UTC.  No significant radio emission was detected at 5 GHz in either epoch. We derive $3\sigma$ upper limits of 1.0 mJy for the first epoch and 0.2 mJy for  the second epoch \cite{GCN.38659}.

\subsection{Spectral Analysis.}
\noindent\textbf{EP-WXT.} 
{As the first step, the time-integrated spectrum of the WXT prompt emission during the $T_{90}$ interval (from $T_0 + 3\,\rm{s}$ to $T_0 + 207\,\rm{s}$) is fitted with an absorbed power-law model. Both Galactic and intrinsic absorption are taken into account, parameterized by the equivalent hydrogen column densities $N_{\rm G}$ and $N_{\rm int}$, respectively. The power-law component describes the spectral shape in the observer frame as $N(E) = A \times E^{-\Gamma_X}$, where $A$ is the normalization and $\Gamma_X$ is the photon index. The spectrum is grouped to ensure a minimum of two counts per bin. Given the relatively small number of counts detected by WXT, we employ a Bayesian inference approach as well as the C-statistic, as outlined in Refs.~\cite{Yang2022, Yang2023ApJL}, to model the spectra and explore the parameter space.

The Galactic absorption column density toward EP241113a is fixed at $N_{\rm G} = 2.6 \times 10^{20} \, \rm{cm}^{-2}$ \cite{Dickey1990, Kalberla2005, Willingale2013, HI4PI2016}, and the redshift is fixed at $z = 1.53$. A time-averaged intrinsic absorption of $N_{\rm int} = (3.2^{+1.6}_{-0.9}) \times 10^{22} \, \rm{cm}^{-2}$ and a photon index of $\Gamma_X = 2.7^{+0.7}_{-0.4}$ are obtained with an acceptable fit statistic of $\rm{CSTAT}/(\rm{d.o.f.}) = 0.8$ (see Fig.~\ref{fig_WXT_data}b,c and Table~\ref{tab_spectrum_fitting}), and the corresponding BIC value is 44.46. 
Such a spectral shape with $\Gamma_X = 2.7$ is too soft to represent the low-energy part of a curved $\nu F_\nu$ (i.e., $E^2{\rm d}N/{\rm d}E$) spectrum with a spectral peak. For comparison, the low-energy photon index below the peak energy in GRBs is $\Gamma_X = 1.08^{+0.43}_{-0.44}$\cite{Fermi-catalog}. Therefore, this steep spectrum (with a photon index of $-0.7$ in the $E^2 {\rm d}N/{\rm d}E$ spectrum) implies that the SED of EP241113a should peak somewhere within or below the $0.5$–$4\,\rm{keV}$ WXT band. We note that fitting the spectrum using \textit{XSPEC} with the C-statistic, which is appropriate for low photon counts, yields consistent results.


}

As discussed in the main text, to constrain the spectral peak energy, we fit the WXT spectrum with models that contain a spectral break, including the Band function, a smoothly broken power-law, and a cutoff power-law.
In the fittings, an upper limit is set to the low-energy photon index as $\Gamma_X < 2$ to allow a peak in the $E^2{\rm d}N/{\rm d}E$ spectrum, which is also reasonable in light of the values of GRBs.
These models are defined as follows:
\begin{enumerate}
    \item \textbf{Band function\cite{Band1993}}:
    \begin{equation}
    N(E) =
    \begin{cases}
    A\left(\frac{E}{100\,{\rm keV}}\right)^{-\Gamma_X} \exp\left(-\frac{E}{E_{\rm c}}\right), & E < (-\Gamma_X + \Gamma_{X,2}) E_{\rm c}, \\
    A\left[\frac{(-\Gamma_X + \Gamma_{X,2}) E_{\rm c}}{100\,{\rm keV}}\right]^{-\Gamma_X + \Gamma_{X,2}} \exp(-\Gamma_{X,2} + \Gamma_X) \left(\frac{E}{100\,{\rm keV}}\right)^{-\Gamma_{X,2}}, & E \geqslant (-\Gamma_X + \Gamma_{X,2}) E_{\rm c},
    \end{cases}
    \end{equation}
    where the peak energy is related to the cutoff energy as $E_{\rm p} = (2 - \Gamma_X) E_{\rm c}$.

    \item \textbf{Cutoff power-law model}:
    \begin{equation}
    N(E) = A \left(\frac{E}{100\,{\rm keV}}\right)^{-\Gamma_X} \exp\left(-\frac{E}{E_{\rm c}}\right),
    \end{equation}
    with $E_{\rm p} = (2 - \Gamma_X) E_{\rm c}$.

    \item \textbf{Smoothly broken power-law\cite{Ravasio2018}}:
    \begin{equation}
    N(E) = A E_{\rm j}^{-\Gamma_X} \left[ \left( \frac{E}{E_{\rm j}} \right)^{\Gamma_X n} + \left( \frac{E}{E_{\rm j}} \right)^{\Gamma_{X,2} n} \right]^{-1/n},
    \end{equation}
    where the peak energy is given by: $E_{\rm p} = E_{\rm j} \left( \frac{-\Gamma_X + 2}{-\Gamma_{X,2} + 2} \right)^{-1 / [(-\Gamma_{X,2} + \Gamma_X) n]},$
    and we fix the smoothness parameter to $n = 2$\cite{Ravasio2019AA}.
\end{enumerate}

\begin{figure}[htbp]
    \centering
    \begin{overpic}[width=0.45\textwidth]{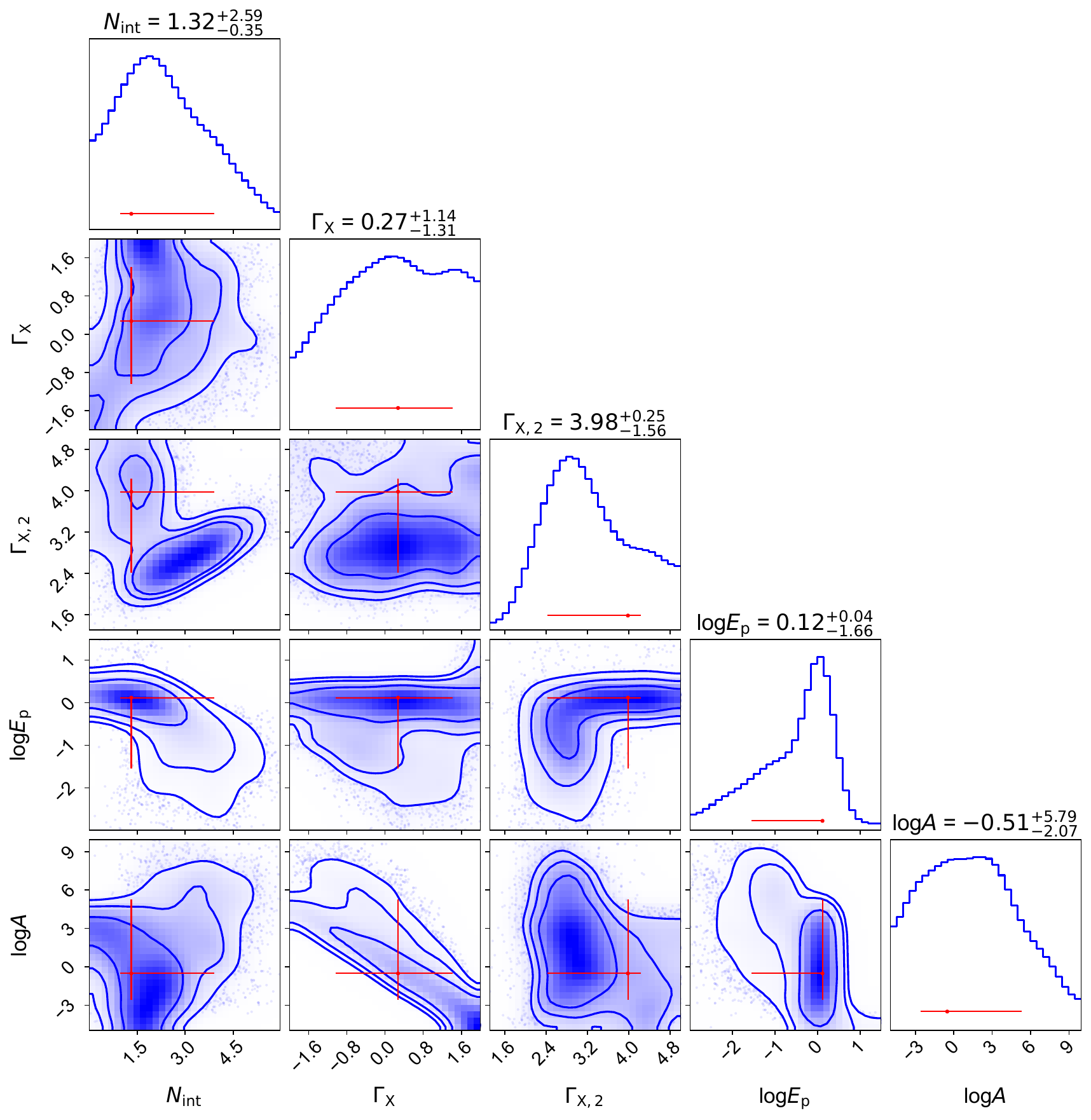}
        \put(-1,100){\textbf{(a)}} 
    \end{overpic}
    \hspace{0.05\textwidth} 
    \begin{overpic}[width=0.45\textwidth]{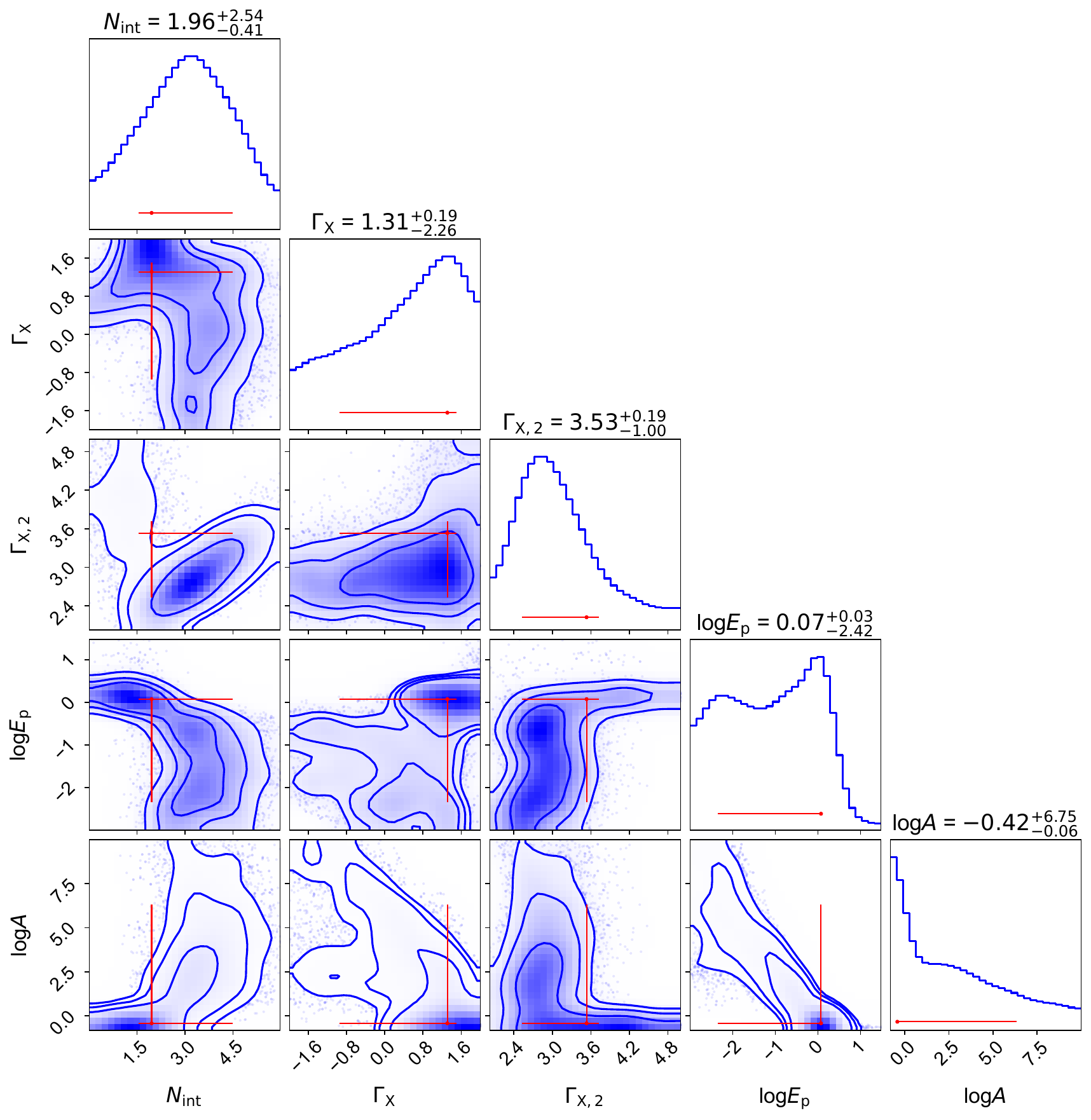}
        \put(-1,100){\textbf{(b)}} 
    \end{overpic}

    \vskip 0.5cm

    \begin{overpic}[width=0.45\textwidth]{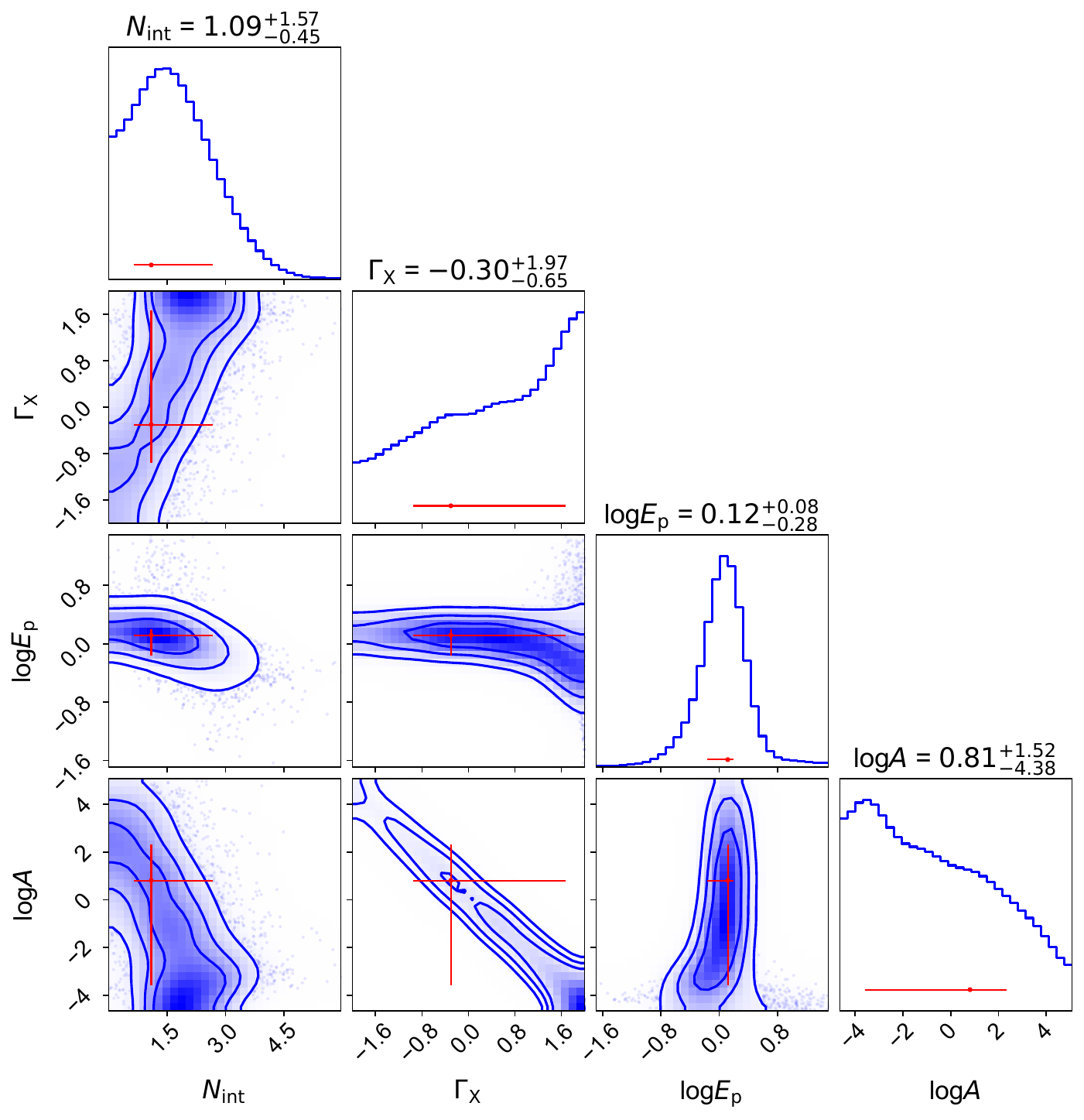}
        \put(-1,100){\textbf{(c)}} 
    \end{overpic}
    \hspace{0.05\textwidth} 
    \begin{overpic}[width=0.45\textwidth]{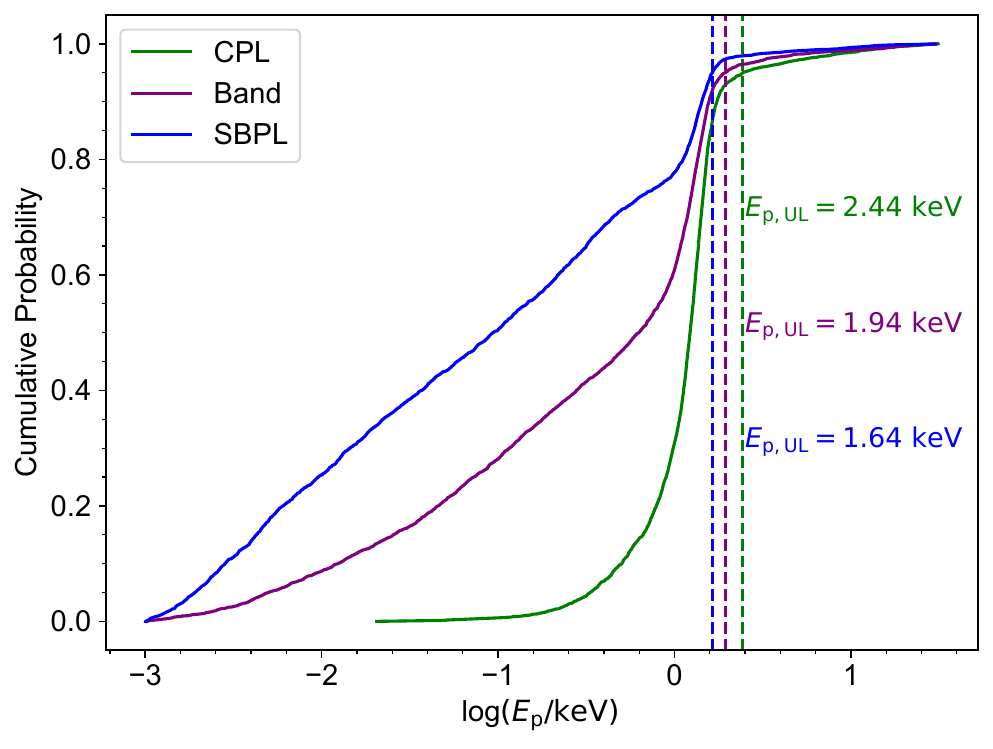}
        \put(-1,100){\textbf{(d)}} 
    \end{overpic}

   \caption{\textbf{WXT spectral fitting results assuming curved models.} 
(a): Corner plot showing the Bayesian posterior distributions assuming the Band function model. (b), (c): Same as (a), but based on the smoothly broken power-law and cutoff power-law models, respectively. (d): Cumulative distribution function (CDF) of $E_{\rm p}$ derived from the Bayesian posterior samples assuming the Band function (purple solid line), smoothly broken power-law (blue solid line), and cutoff power-law (green solid line). The dashed lines indicate the values at which the cumulative probability reaches 95\%.}

    \label{fig_ep_curved_model}
\end{figure}

{For the Band function model, we derive a low-energy photon index of $\Gamma_X = 0.3^{+1.1}_{-1.3}$, a high-energy photon index of $\Gamma_{X,2} = 4.0^{+0.3}_{-1.6}$, an intrinsic column density of $N_{\rm int} = 1.3^{+2.6}_{-0.4} \times 10^{22} \, \rm cm^{-2}$, and a best-fit peak energy of $1.32^{+0.13}_{-1.29} \, \rm keV$, yielding a BIC value of 49.73. The high-energy photon index is consistent with that obtained from the single power-law model fit within the $1\sigma$ uncertainty range. The absorption column density is substantially reduced compared to that obtained from the single power-law fit, due to the intrinsic spectral flattening toward low energies in the curved spectral model. The peak energy $E_{\rm p}$ cannot be well constrained toward its lower bound, primarily because when it drops below the lower bound of the WXT bandpass ($0.5\,\rm keV$), only the high-energy portion of the Band function remains within the WXT band and asymptotically approaches a single power-law.
Nevertheless, an upper limit on $E_{\rm p}$ can still be derived. To do so, we construct the cumulative distribution function (CDF) from the Bayesian posterior samples and define the upper bound, $E_{\rm p, UL}$, as the value of $E_{\rm p}$ corresponding to the 95\% cumulative probability. We find $E_{\rm p, UL} = 1.9\,\rm{keV}$ for the Band function (see Fig.~\ref{fig_ep_curved_model}).

For the smoothly broken power-law model, we obtain $\Gamma_X = 1.3^{+0.2}_{-2.3}$ and $\Gamma_{X,2} = 3.5^{+0.2}_{-1.0}$, an intrinsic column density of $N_{\rm int} = 2.0^{+2.5}_{-0.4} \times 10^{22} \, \rm cm^{-2}$, and a best-fit peak energy of $1.18^{+0.08}_{-1.17} \, \rm keV$, yielding a BIC value of 49.15. Similar to the case above for the Band function model, the peak energy cannot be constrained toward its lower bound. We perform the analysis in the same way as above and derive an upper limit of $E_{\rm p, UL} = 1.6\,\rm{keV}$ (at the $95\%$ confidence level) (see Fig.~\ref{fig_ep_curved_model}). In addition, we explore values of the smoothness parameter $n$ other than the nominal choice of $n = 2$, and find that the effect on the resulting $E_{\rm p}$ value is negligible.


The cutoff power-law model provides a good fit, yielding a BIC value of 46.16. The peak energy can be well constrained, $E_{\rm p} = 1.3^{+0.3}_{-0.6} \, \rm keV$, and the intrinsic absorption column density is further reduced to $1.1^{+1.6}_{-0.5} \times 10^{22} \, \rm cm^{-2}$. This is reasonable, since there must exist a lower bound on $E_{\rm p}$ in order to account for the high-energy part of the WXT spectrum, which is steep enough to be consistent with the rapid exponential decline of the cutoff model. Similarly, we also derive an upper limit of $E_{\rm p, UL} = 2.4\,\rm{keV}$ at the 95\% confidence level by using the CDF (see Fig.~\ref{fig_ep_curved_model}).

}



{The power-law model yields the minimum BIC value. We compare the goodness-of-fit between the power-law and the other models using the BIC comparison approach, adopting the criterion $\Delta {\rm BIC} > 4$ for a positive preference.
\cite{de2016, Kenneth2004}. The Band and smoothly broken power-law models, with $\Delta {\rm BIC} = 5.27$ and $4.69$, respectively, can be statistically ruled out when compared to the power-law model, whereas the cutoff power-law model cannot be statistically distinguished from the power-law model ($\Delta {\rm BIC} < 4$). 
Moreover, we note that in all fits with the curved spectral models, the required excess absorption column density is significantly reduced compared to that of the power-law model, as expected. We therefore adopt $E_{\rm p, UL} = 2.4\,\rm{keV}$, derived from the cutoff power-law model, as a conservative upper limit on the peak energy. To conclude, our spectral analysis argues for the presence of a spectral peak in the SED in $E^{2} {\rm d}N/{\rm d}E$, with the peak energy $E_{\rm p}$ lying at or below approximately 2\,keV.
}


Finally, we consider an absorbed blackbody model, which yields a best-fit temperature of $kT = 0.38^{+0.02}_{-0.07} \, \rm keV$. The spectral peak energy is consistent with the upper limit derived above. 
However, the model does not yield a statistically significant improvement over the power-law model. A summary of the fitting results and corresponding statistical comparisons is provided in  Table~\ref{tab_spectrum_fitting}.

\noindent\textbf{EP-FXT.} An automatic follow-up observation by EP-FXT commenced approximately 2 minutes after the WXT trigger on EP241113a, with a total exposure time of $5076\,\rm{s}$. A bright, decaying source was detected within the WXT positional uncertainty. The source is located at R.A. = $131.9968^\circ$, Dec. = $52.3818^\circ$, with a positional uncertainty of $10''$ (at the $90\%$ confidence level). In addition, EP-FXT performed fourteen follow-up observations of EP241113a, as summarized in Table~\ref{tab_FXT_obs}. During the first five observations (from 2024-11-16T04:38:13 to 2024-11-21T13:31:31 UTC), the source was clearly detected. The remaining nine observations did not yield a detection, but they can be grouped into two epochs since the observation times are quite close with respect to the time since the trigger. The first epoch spans from 2024-11-29T06:28:33.901 to 2024-11-29T13:44:52.795 UTC (ObsIDs: 06800000263 and 08500000219), with a total exposure time of $13134\,\rm{s}$. The second epoch spans from 2024-12-09T06:45:28.875 to 2024-12-15T19:23:45.783 UTC (ObsIDs: 06800000282, 06800000288, 06800000289, 06800000290, 06800000297, 06800000298, and 06800000299), with a total exposure time of $42121\,\rm{s}$. The data from each epoch are combined to assess detection significance. Using the Li-Ma formula \cite{Li-Ma} and photon counts from circular source and background regions (with radii of $1'$ and $3'$, respectively), we find detection significances of $\sigma = 5.1$ and $\sigma = 3.8$ for the first and second epochs, respectively.

The automatic follow-up observation is divided into logarithmically spaced time bins with a bin size of $\Delta \log{t\,(\rm{s})} = 0.1$. The light curve during this period exhibits a steep decay phase from $\sim T_0 + 300\,\rm{s}$ to $\sim T_0 + 1000\,\rm{s}$, followed by a plateau phase extending to $\sim T_0 + 10^4\,\rm{s}$. The subsequent follow-up observation, starting at $\sim T_0 + 2.39 \, \rm{day}$, exhibits a normal decay phase that is shallower than the preceding steep decay phase (see Fig.~\ref{fig_WXT_luminosity}).

There are four time bins in both the steep decay and plateau phases, and seven time bins in the normal decay phase (see Table~\ref{tab_spectrum_fitting}). For the steep decay phase, we perform a joint spectral fit of the four spectra (each grouped to ensure a minimum of two counts per bin) obtained from separate time bins using an absorbed power-law model, \texttt{tbabs} * \texttt{ztbabs} * \texttt{powerlaw}, with textit{XSPEC}. In this fitting, we assume that the photon index and the intrinsic hydrogen column density $N_{\rm int}$ are identical across all four spectra. The redshift and Galactic hydrogen column density are fixed to the same values adopted in the WXT spectral analysis. The same fitting procedure is applied to the spectra during the plateau and normal decay phases. The results of all spectral fittings are summarized in Table~\ref{tab_spectrum_fitting}.

\subsection{The Amati relation.}
We adopt the same GRB samples with well-determined redshifts~\cite{Amati2002, Zhang2009, Zhang2018NatAs, Minaev2020MNRAS} as used in previous work\cite{Liu2025} to examine the location of EP241113a in the Amati relation diagram. The Amati relation, which connects the rest-frame spectral peak energy ($E_{\rm p,z}$) to the isotropic-equivalent gamma-ray energy ($E_{\gamma,\text{iso}}$), is conventionally described by a linear relation in logarithmic space~\cite{Amati2002}:
\begin{equation}
{\rm log}E_{\rm p,z} = u\,{\rm log}E_{\gamma,\rm iso}+v,
\end{equation}
where $u$ and $v$ denote the slope and intercept, respectively.

We employ {\it emcee} to estimate the model parameters. During the fitting procedure, the orthogonal distance regression (ODR) method\cite{Lelli2019MNRAS, Sun2025NSRev} is applied to evaluate the likelihood between the model and the data. Based on this, we calculate the Gaussian intrinsic scatter $\sigma_{\rm int}$ along the orthogonal direction to the model, and successfully reproduce the results reported in Ref.\cite{Liu2025}. 
The best-fit parameters and their associated $1\sigma$ uncertainties are summarized as follows: for Type I GRBs, $u_{\rm I} = 0.35_{-0.05}^{+0.04}$, $v_{\rm I} = -14.9_{-2.2}^{+2.6}$, and ${\rm log}\sigma_{\rm int,I} = -1.29_{-0.13}^{+0.14}$; for Type II GRBs, $u_{\rm II} = 0.39_{-0.02}^{+0.02}$, $v_{\rm II} = -17.9_{-1.0}^{+1.0}$, and ${\rm log}\sigma_{\rm int,II} = -1.42_{-0.05}^{+0.05}$. In Fig.~\ref{fig_Amati}, we show the best-fit  relations for Type I and Type II GRBs, as well as  the corresponding $3\sigma$ intrinsic scatter regions.

For EP240414a, we extrapolate the isotropic energy from $0.5-4 \, \rm keV$ to $1-10^{4} \, \rm keV$ in the rest frame, yielding $E_{\gamma, \rm iso} = (4.0 \pm 0.5) \times 10^{49} \, \rm erg$, with the $1\sigma$ uncertainty of $E_{\gamma, \rm iso}$ calculated by using the error propagation and incorporating the uncertainties of the photon index and the isotropic energy in  $0.5-4 \, \rm keV$ \cite{Sun2025}.

For EP241113a, we calculate the isotropic energy in $1-10^{4} \, \rm keV$  in the rest frame based on the spectral fitting results, yielding $E_{\gamma, \rm iso} = 2.1^{+1.7}_{-0.1} \times 10^{51} \, \rm erg$ and $9.3^{+6.7}_{-0.4} \times 10^{50} \, \rm erg$ for the absorbed power-law and absorbed cutoff power-law models, respectively. Since both models are statistically favored according to the BIC comparison (see discussion above), we adopt the union of their $1\sigma$ uncertainty ranges as the conservative uncertainty range ($[1.0, 3.8] \times 10^{51}$ erg) for $E_{\gamma, \rm iso}$, which is presented in Fig.~\ref{fig_Amati}. As illustrated, EP241113a is located significantly outside the $3\sigma$ scatter region of the Amati relation.

\subsection{Analysis of the X-ray Light Curve.}

The X-ray light curve can be divided into three distinct phases: the steep decay phase (before $\sim 1 \, \rm ks$), the plateau phase ($\sim 1-10\, \rm ks$), and the late afterglow phase (beyond $\sim 100 \, \rm ks$). A full expression for the steep decay phase is given by\cite{ZhangBB2009}:
\begin{equation}
    F_{\nu}(T) = F_{\nu, \rm p} \left( \frac{T - T_{\rm p, 0}}{T_{\rm p} - T_{\rm p, 0}} \right)^{-\alpha_1},
    \label{Eq_temporal}
\end{equation}
where $T_{\rm p, 0}$ denotes the onset time of the final pulse of the prompt emission, $T_{\rm p}=146.6 \, \rm s$ is the peak time of the last pulse, and $F_{\nu, \rm p}$ is the flux at $T_{\rm p}$. The steep decay phase begins immediately after $T_{\rm p}$. We treat $F_{\nu, \rm p}$, $T_{\rm p, 0}$, and $\alpha_1$ as free parameters and perform an MCMC fitting using the Python package {\it emcee}\cite{Foreman-Mackey2013}. The fitting yields $T_{\rm p, 0} = 95.64^{+16.18}_{-20.87}\,\rm{s}$ and $\alpha_1 = 3.05^{+0.47}_{-0.41}$. Subsequently, we employ the same functional form (Eq.~\ref{Eq_temporal}), with $T_{\rm p, 0}$ replaced by the trigger time  to fit the temporal indices for the plateau ($\alpha_2$) and late afterglow ($\alpha_3$) phases. The resulting fit values are $\alpha_2 = -0.16^{+0.24}_{-0.24}$ and $\alpha_3 = 0.95^{+0.18}_{-0.18}$. The fitting results and the corresponding model curves are shown in Fig.~\ref{fig_WXT_luminosity}b.

\subsection{Theoretical Modelling.}
\noindent\textbf {Constraining the viewing angle with the steep decay.\\} 
At the end of the prompt emission, the observer will detect photons emitted from higher latitudes of the jet as the observed time increases. If the jet has a top-hat structure,  the observed flux can be expressed as $F_{\nu} \propto D^2 f'_{\nu'}$, where $D$ is the Doppler factor and $f'_{\nu'}$ represents the intrinsic spectrum of the prompt emission. If the intrinsic spectrum follows a single power law, i.e., $f'_{\nu'} \propto \nu'^{-\beta}$, then the observed flux becomes $F_{\nu} \propto D^2 f'_{\nu'} \propto D^{2+\beta} \propto t^{-(2+\beta)}$.

It has been suggested that GRB jets could have a quasi-universal beaming configuration—
a structured jet with high anisotropy in its angular distribution of the fireball energy about the symmetry axis. 
Usually, two types of jet structure are considered\cite{Rossi2002, Zhang2002}, i.e.,  one  is a power-law structure, given by
\begin{equation}
\begin{cases}
\epsilon(\theta) = \left[ 1 + \left( \frac{\theta}{\theta_{\rm c}} \right)^{k_{\rm c}} \right]^{-1} \\
\Gamma(\theta) = 1 + (\Gamma_{\rm c} - 1) \left[ 1 + \left( \frac{\theta}{\theta_{\rm c}} \right)^{k_{\rm c}} \right]^{-1},
\end{cases}
\label{PL-jet-structure}
\end{equation}
and the other is a Gaussian structure, given by
\begin{equation}
\begin{cases}
\epsilon(\theta) = e^{-(\theta/\theta_{\rm c})^2} \\
\Gamma(\theta) = 1 + (\Gamma_{\rm c} - 1) e^{-(\theta/\theta_{\rm c})^2}.
\end{cases}
\end{equation}

Ref.\cite{Ascenzi2020} investigated the high-latitude emission from a structured jet under the assumption that the prompt emission ceases simultaneously at all latitudes in the source frame, with the emission time given by $t_{\rm em} = \frac{R_{\rm em, 0}}{c \beta_{0}}$, where $R_{\rm em, 0}$ is the radial distance from the central engine to the emission site along the jet axis ($\theta = 0$), and $c \beta_{0}$ is the jet velocity at $\theta = 0$. However, this assumption does not apply to the standard internal shock/dissipation scenario for the prompt emission of GRBs, in which the emission time $t_{\rm em}$ is expected to depend on the latitude angle. Instead, we here model the high-latitude emission within the internal shock framework, assuming that the central engine ceases activity simultaneously across all angles. In this scenario, the emission time in the source frame depends on the latitude angle and is given by $t_{\rm em}(\theta) = \frac{R_{\rm em}(\theta)}{c \, \beta(\theta)}= \frac{t_{\rm v}}{1 - \beta(\theta)}$, where $R_{\rm em}(\theta)=c\beta(\theta) t_{\rm v}/[1-\beta(\theta)]$. Here $t_{\rm v} = T_{\rm p} - T_{\rm p,0}$ represents the variability timescale of the last pulse, where $T_{\rm p}$ and $T_{\rm p,0}$ represent the peak time  and onset time of the last pulse of the prompt emission, respectively. 

\noindent\textbf {1. Calculation of the equal arrival time surface (EATS) for an arbitrarily structured jet.}

\begin{figure}[htbp]
    \centering
    \begin{overpic}[width=0.7\textwidth]{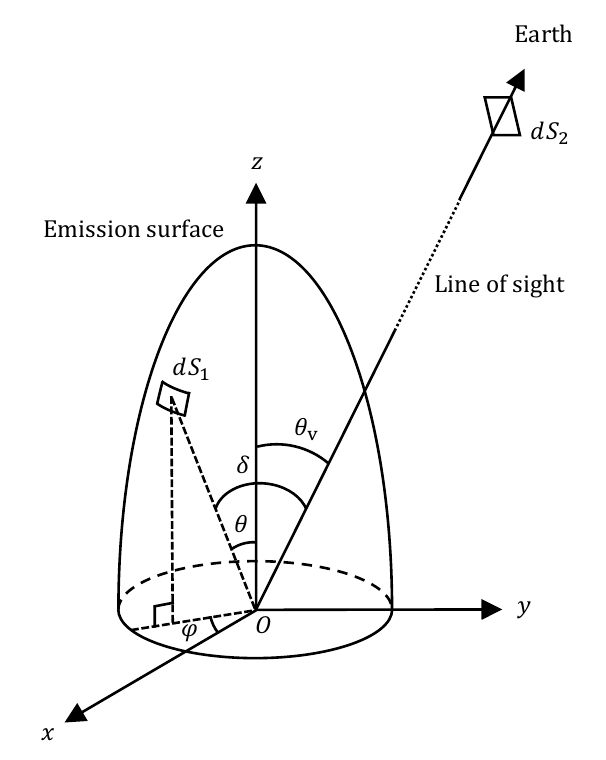}
    \end{overpic}
    
    \caption{\textbf{Schematic illustration of the emission surface.} The figure shows the geometry of the emission surface, where the surface shape (described by $R_{\rm em}(\theta, \phi)$) can be arbitrary and depends on the jet structure. It should be noted that the emission time at any point on the surface does not necessarily remain constant but can vary with the polar angle $\theta$ and azimuthal angle $\phi$. All photons generated from the surface are emitted instantaneously, resembling a "flash". The line of sight lies within the yOz plane, where $\theta_{\rm v}$ denotes the viewing angle. The angle $\delta$ represents the inclination between the radial direction at an arbitrary surface element and the line of sight, and can be expressed by: $\cos \delta = \sin \theta_{\rm v} \sin \theta \sin \phi + \cos \theta_{\rm v} \cos \theta$. $dS_1$ denotes the differential area element of the emission surface, while $dS_2$ represents the area of the receiver.
}

    \label{fig_EATS}
\end{figure}

To account for the EATS effect on a shock surface with an arbitrary emission radius $R_{\rm em}$ and emission time $t_{\rm em}$ (in the source frame), we discretize the shock surface into a series of grids $dS_1(\theta, \phi)$. Provided that the grid size is sufficiently small, photons generated from $dS_1$ can be considered to arrive at the observer simultaneously. We set the time at which the hypothetical photons—emitted from the central engine simultaneously with the final injected wind when the central engine ceases—are received by the observer as $t_{\rm obs} = 0$. For photons generated at an arbitrary surface element $dS_1(\theta, \phi)$ of the shock surface (see  Fig.~\ref{fig_EATS}), the observer time is given by $t_{\rm obs} = t_{\rm em}-\frac{R_{\rm em} }{c}\cos \delta$, where the angle $\delta$ represents the inclination between the radial direction at an arbitrary surface element and the line of sight (see  Fig.~\ref{fig_EATS}). In the internal shock framework, the Lorentz factor of the injected wind can vary significantly.  Once the observer times for photons from all shock surface grids are determined, the observed flux $F_{\nu} \, \rm (erg \, cm^{-2} \, s^{-1} \, Hz^{-1})$ can be calculated by conserving the number of photons as follows:
\begin{equation}
F_{\nu}(t_i) \, d\nu \, \frac{dS_2}{h\nu} \times (t_{i+1} - t_i) = \sum_{t_i < t_{\rm obs}(\theta, \phi) < t_{i+1}} \left[ f_{\nu}(\theta, \phi) \, d\nu \, \frac{d\Omega_{\rm obs}}{h\nu} \right] \times dS_1(\theta, \phi),
\end{equation}
where the observer time is divided into discrete intervals $t_i \, (i = 1,2, ..., i_{\rm max})$. The quantity $f_{\nu}(\theta, \phi) \, \rm (erg\, cm^{-2} \, Hz^{-1} \, Sr^{-1})$ denotes the fluence emitted from the surface element $dS_1(\theta, \phi)$, and the comoving fluence is given by $f'_{\nu'} = \frac{f_{\nu}}{D(\theta, \phi)^2}$, which is assumed to be isotropic in the comoving frame. The Doppler factor is defined as $D(\theta, \phi) = \frac{1}{\Gamma(1 - \beta \cos \delta)}$. The solid angle of the observer area $dS_2$ as seen from the emission point at $dS_1$ is given by $d\Omega_{\rm obs} = \frac{dS_2}{D_{\rm L}^2}$,
where $D_{\rm L}$ represents the luminosity distance to the source. Both the left-hand and right-hand sides of the equation correspond to the number of photons within the frequency range $(\nu, \nu + d\nu)$ that are generated during the last dissipation event and received by the observer within the time interval $(t_i, t_{i+1})$.

\noindent\textbf {2. High-latitude emission from a structured jet and constraint on the viewing angle of EP241113a.}

\begin{figure}[htbp]
    \centering
    \begin{overpic}[width=0.7\textwidth]{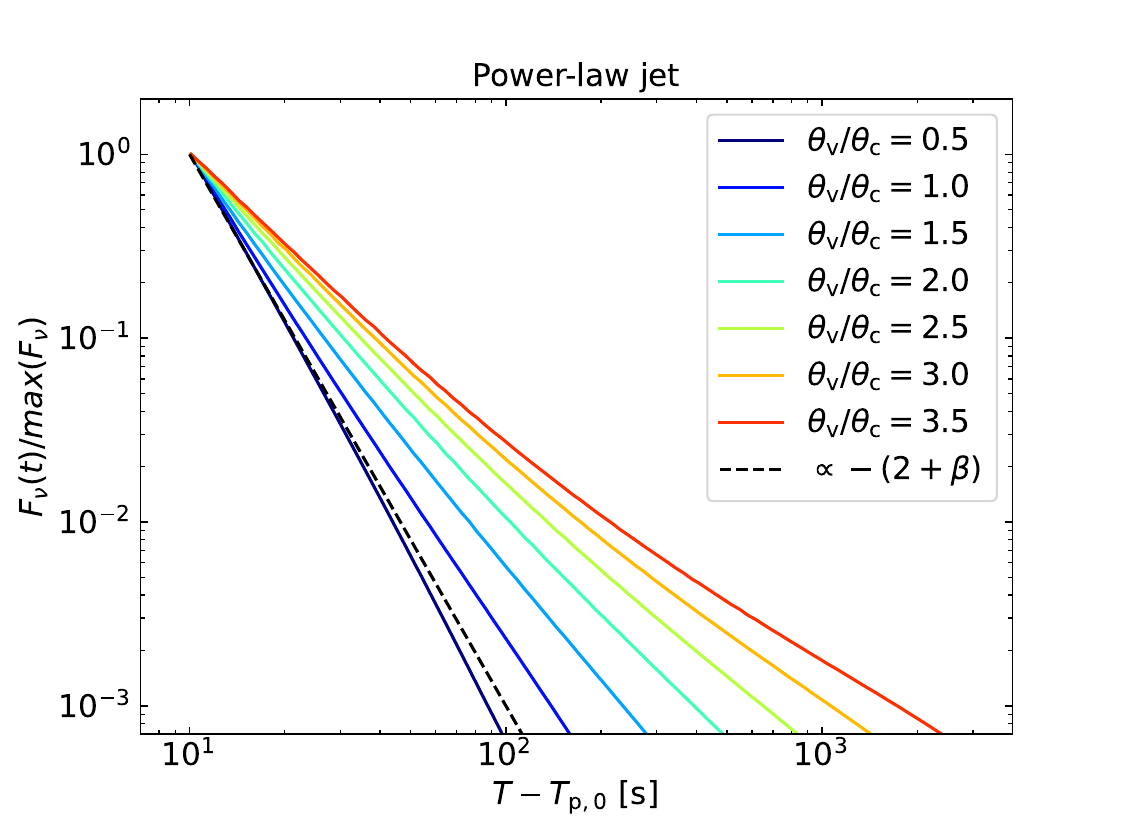}
    \end{overpic}

    \caption{\textbf{High-latitude emission from a power-law jet.} We adopt $\Gamma_{\rm c} = 20$, $\theta_{\rm c} = 20^\circ$, $k_{\rm c} = 2$, $\beta = 1$, and $T_{\rm p}-T_{\rm p,0} = 10 \, \rm s$. The temporal index satisfies $\alpha \simeq 2 + \beta$ as long as the viewing angle $\theta_{\rm v} < \theta_{\rm c}$. If $\theta_{\rm v} > \theta_{\rm c}$, the light curve becomes noticeably shallower than $2 + \beta$.}

    \label{fig_HLE_diff_thetav}
\end{figure}

{Previous simulation studies suggest that structured jets are better described by a power-law jet with an index of $k_{\rm c} \sim 2$  than a Gaussian jet\cite{Gottlieb2021}. For completeness, we consider both power-law and Gaussian jet structures.} Under this framework, we find that the high-latitude emission of a structured jet, when viewed on-axis, closely resembles that of a top-hat jet, and the relation $F_{\nu} \propto t^{-(2+\beta)}$ is preserved, as shown in Fig.~\ref{fig_HLE_diff_thetav}. On the other hand, when the viewing angle is outside of the jet core, $\theta_{\rm v} > \theta_{\rm c}$, the light curve exhibits a much shallower decay. 

By fitting the data during the steep decay phase of EP241113a, we obtain a temporal index of $\alpha_1 = 3.05^{+0.47}_{-0.41}$ (see Fig.~\ref{fig_WXT_luminosity}b), which is in good agreement with the theoretical expectation of $\beta + 2$ (or equivalently, $\Gamma_X + 1=3.1\pm0.1$) for the high-latitude emission viewed on-axis. This agreement  suggests that  EP241113a is an on-axis event.

\begin{figure}[htbp]
    \centering
    \begin{overpic}[width=0.45\textwidth]{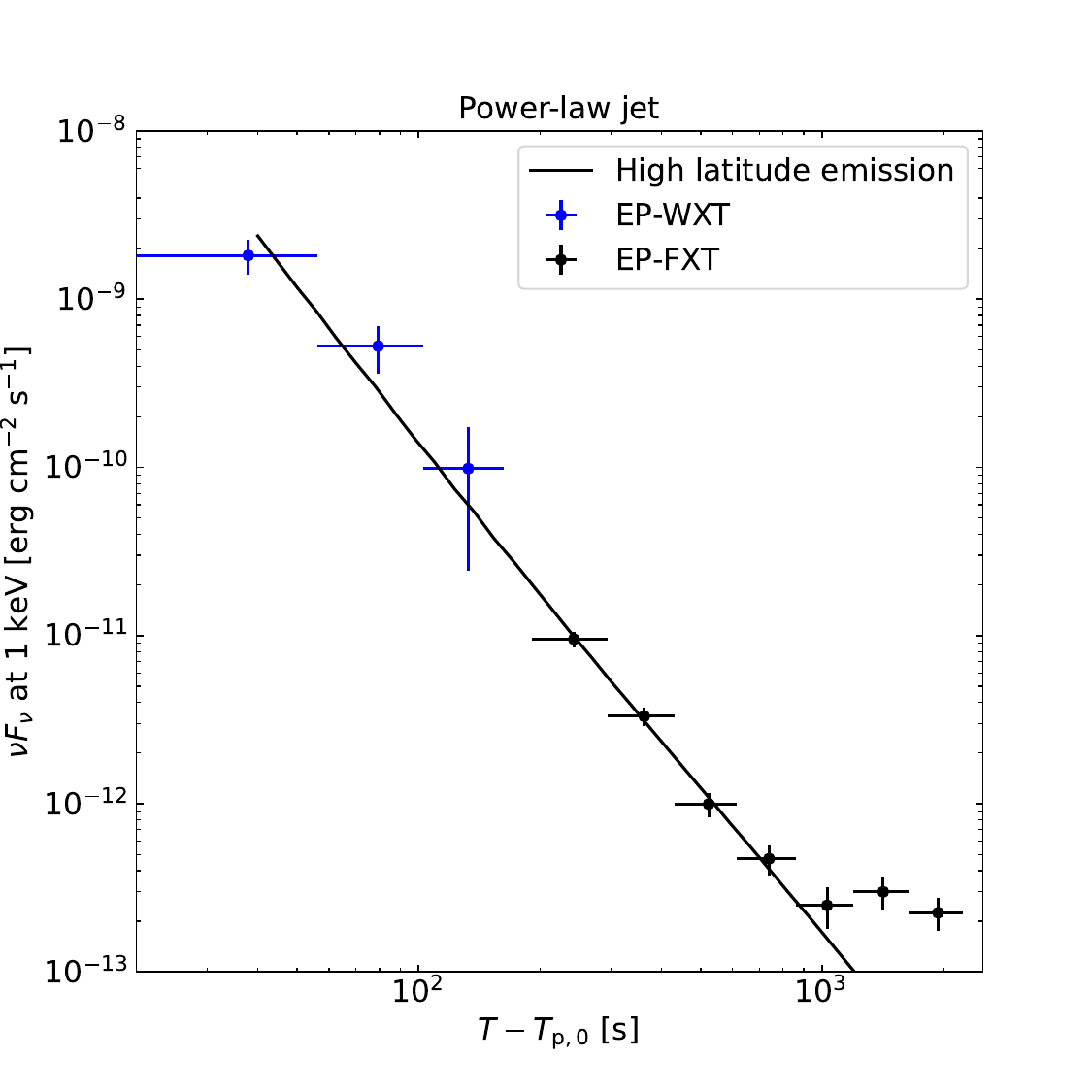}
        \put(2,92){\textbf{(a)}} 
    \end{overpic}
    \hspace{0.05\textwidth} 
    \begin{overpic}[width=0.45\textwidth]{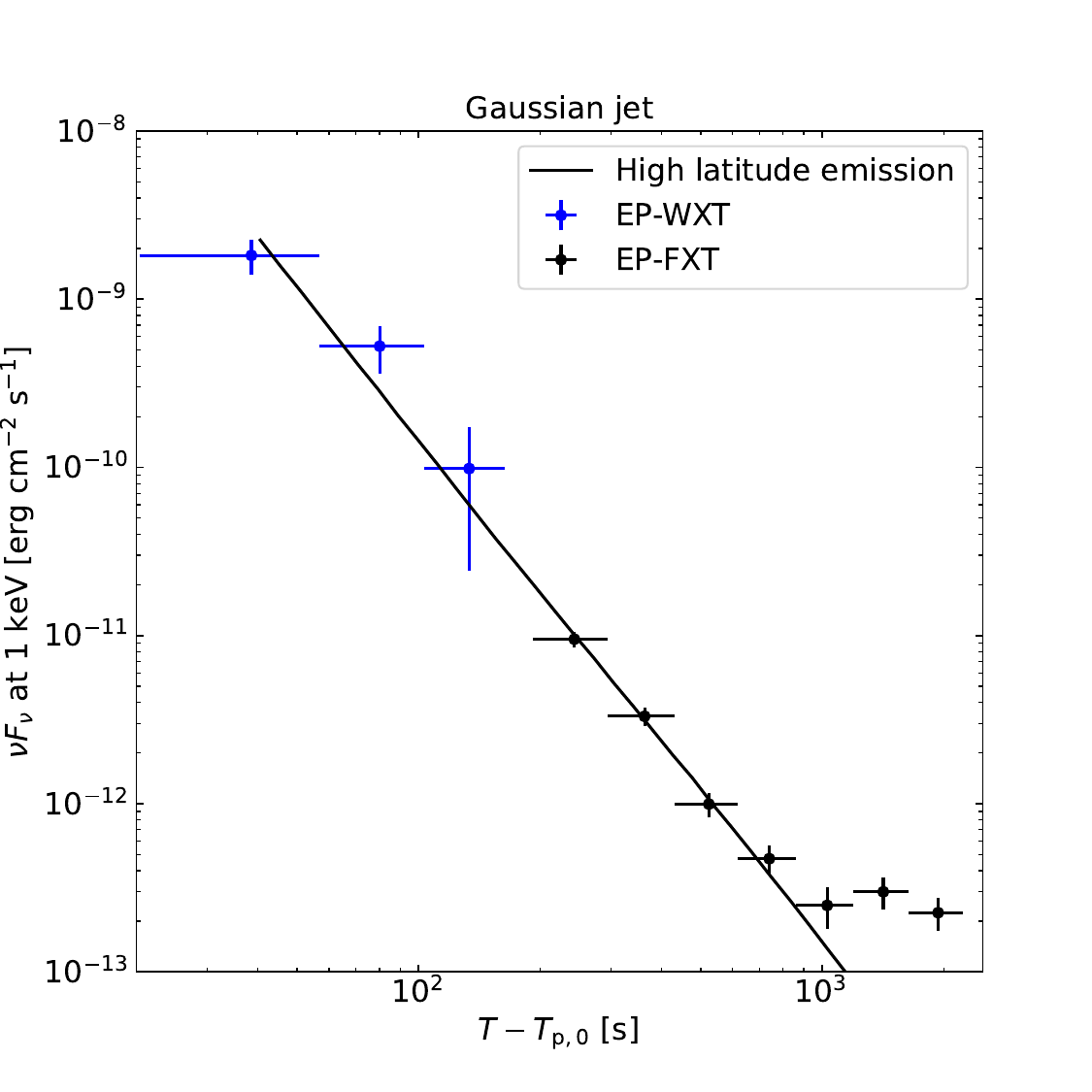}
        \put(2,92){\textbf{(b)}} 
    \end{overpic}

    \vskip 0.3cm

    \begin{overpic}[width=0.45\textwidth]{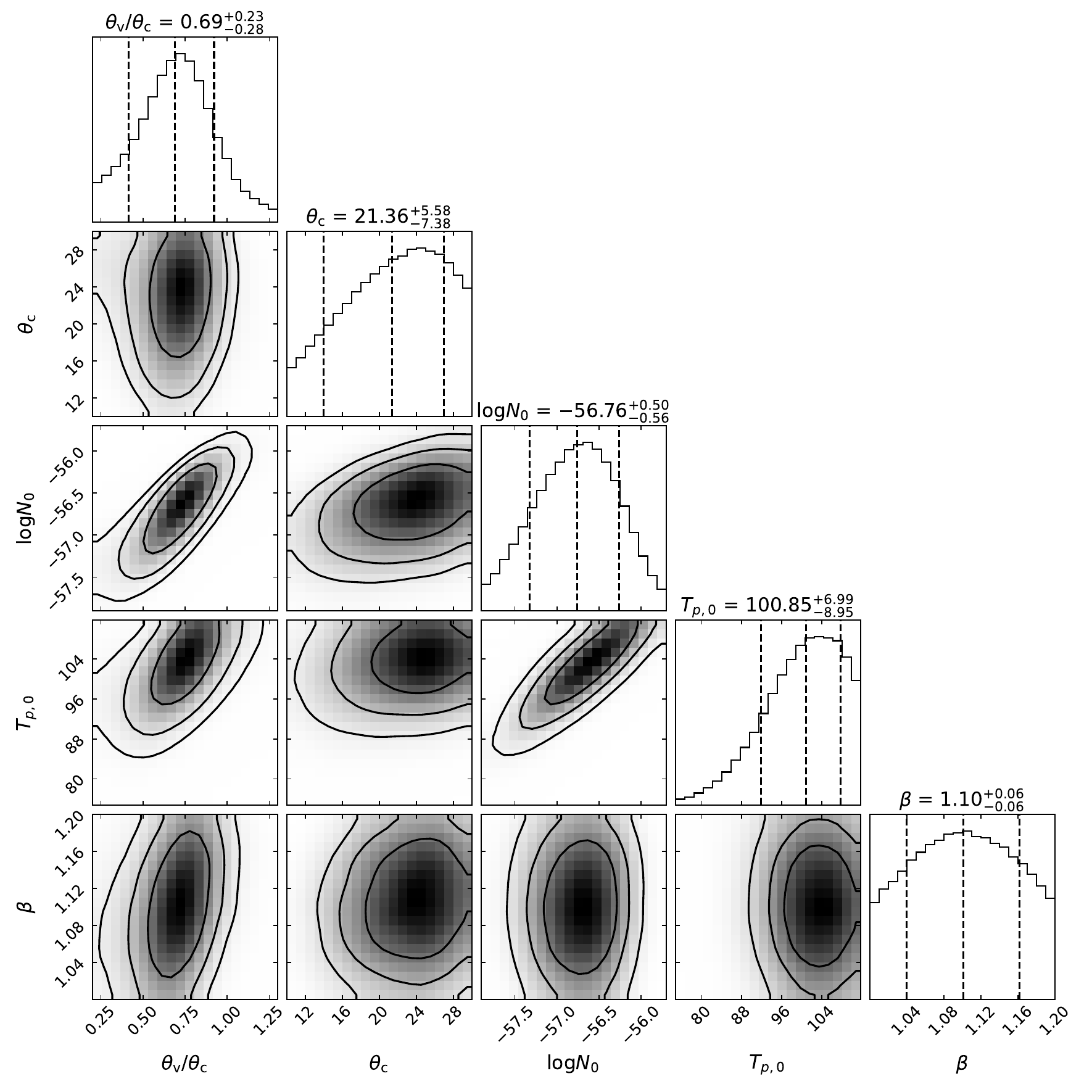}
        \put(2,100){\textbf{(c)}} 
    \end{overpic}
    \hspace{0.05\textwidth} 
    \begin{overpic}[width=0.45\textwidth]{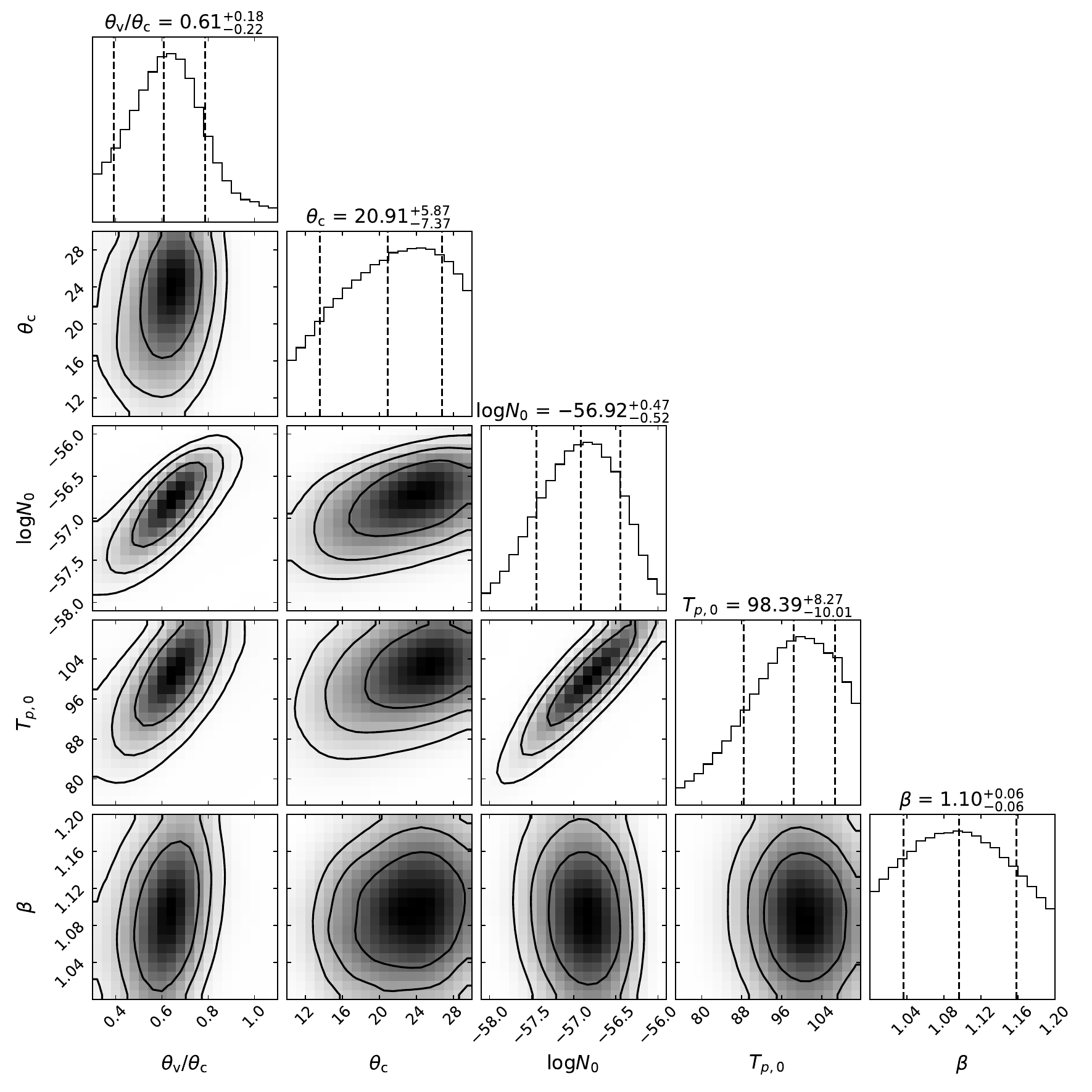}
        \put(2,100){\textbf{(d)}} 
    \end{overpic}

    \caption{\textbf{Fitting results of the steep decay phase based on high-latitude emission.} (a) Best-fit model for the steep decay obtained via MCMC fitting, assuming a power-law jet with $k_{\rm c} = 2$. (b) Same as (a), but for the Gaussian jet model. (c) Corner plot showing the posterior distribution of parameters from the MCMC sampling using the power-law jet model. Dashed lines and labels along the diagonal represent the median values and symmetric $68\%$ credible intervals (corresponding to the 16th and 84th percentiles) of the marginalized distributions. (d) Same as (c), but for the Gaussian jet model. 
The Lorentz factor of the jet core is fixed at $\Gamma_{\rm c} = 20$, and a wide prior range of $\theta_{\rm c} \in [1, 30]\,\rm deg$ is adopted.}

    \label{fig_HLE_fix}
\end{figure}

This decay slope provides a means to constrain the viewing angle under the assumption that the decaying X-ray emission arises from high-latitude emission. We explore the parameter space using MCMC fitting. A uniform prior distribution is adopted for several parameters, including the jet core angle $\theta_{\rm c}$, the ratio of the viewing angle to the core angle $\theta_{\rm v}/\theta_{\rm c}$, the onset time of the final pulse of the prompt emission $T_{\rm p,0}$, and the spectral index $\beta$, while a log-uniform prior is used for the normalization factor $N_0$. We assume $\Gamma_{\rm c} = 20$, consistent with the interpretation of the X-ray plateau (see the following section), and consider a wide range for the jet core angle, with $\theta_{\rm c} \in (1, 30) \, \rm degrees$. The results, presented in  Figure~\ref{fig_HLE_fix} and  Table~\ref{tab_HLE_mcmc_results}, indicate that the steep decay of EP241113a can be interpreted as high-latitude emission from both power-law and Gaussian jets, provided that the viewing angles satisfies $\theta_{\rm v} \leqslant \theta_{\rm c}$. 
{The core opening angle is found to be $\theta_{\rm c} \gtrsim  15^\circ$ at the $1\,\sigma$ confidence level, consistent with the lower bound obtained from analytical estimates:} In the on-axis scenario, the high-latitude emission can be approximated as originating from a uniform jet when $\theta < \theta_{\rm c}$. The minimum value of $\theta_{\rm c}$ can be constrained by the duration of the high-latitude emission \cite{ZhangBB2009}, given by $\theta_{\rm c} \sim \left[ \frac{2c(T_{\rm p,e} - T_{\rm p})}{(1+z)R_{\rm em}} \right]^{0.5}$, where the on-axis emission radius is $R_{\rm em} \sim 2\Gamma_{\rm c}^2 c t_{\rm v}$ and $T_{\rm p,e}$ denotes the end time of the high-latitude emission. Adopting $T_{\rm p,e} \geqslant 1000\,\rm{s}$ and $t_{\rm v} = (T_{\rm p} - T_{\rm p,0}) / (1+z) \simeq 16\,\rm{s}$, we derive $\theta_{\rm c} \geqslant 0.25 \left( \frac{\Gamma_{\rm c}}{20} \right)^{-1}$.

\noindent\textbf {The origin of the X-ray plateau.\\} During $\sim [10^3, 10^4] \, \rm s$, the X-ray light curve of EP241113a exhibits a plateau. The plateau phase could last longer than $10^4{\, \rm s}$, since there is an observation gap between $10^4{\, \rm s}$ and the start time of the normal decay at $2\times 10^5{\, \rm s}$.  The plateau phase  is usually explained by introducing nontrivial modifications to the standard afterglow theory.  One popular model is that the plateau is caused by energy injection into the forward shock that is decelerating in the surrounding medium. Recently, it was shown that the plateau can also arise from the forward shock emission during the free expansion phase if the surrounding medium is a stellar wind \cite{Shen2012,Dereli-Bégué2022}. Furthermore, for a structured jet, it was proposed that the high-latitude emission becomes flat  at late time  because the Doppler boosting becomes  roughly constant when the wing of the structured jet is seen, giving  rise to a plateau in the light curve\cite{Ascenzi2020}. We discuss these three possibilities in the following.

\noindent\textbf {1. Energy injection scenario.}

One of the most commonly used models for the plateau phase is energy injection from the central engine\cite{Dai1998,Zhang2006}. Given that the spectral index in the X-ray plateau is $\beta = \Gamma_X - 1 = 1.2 \pm 0.2$, and assuming $\beta = \frac{p}{2}$, where $p$ is the electron spectral index, we obtain $p = 2.4 \pm 0.4$ if the X-ray frequency satisfies $\nu_X > \nu_{\rm c}$\cite{Sari1998}, where $\nu_{\rm c}$ is the synchrotron cooling frequency. Notably, if $\nu_X < \nu_{\rm c}$, the electron spectral index would be too large ($p = 2\beta+1 = 3.4\pm 0.4$) since modeling of the  GRB afterglows   typically finds $p \sim 2-3$\cite{Panaitescu2001, Aksulu2022}. The plateau of EP241113a lasts up to  at least $\sim  10^4 \, \rm s$ (see Fig.~\ref{fig_WXT_luminosity}b). Assuming $\nu_X > \nu_{\rm c}$, the temporal index of the X-ray light curve is given by $\alpha_{\rm inj} = \frac{(2p-4)+(p+2)q}{4}$ in both wind and ISM environments, assuming the energy injection has the form $L\propto t^{-q}$\cite{Zhang2006}. The temporal index fitted from the plateau is $\alpha_{\rm inj} = -0.16^{+0.24}_{-0.24}$, implying $q = -0.34^{+0.24}_{-0.24}$. Such a form is unusual, since $q=0$, as expected from the  initial spin down from a millisecond pulsar, is the quickest energy injection form and the shallow decay seen in most GRBs requires $q>0$\cite{Zhang2006}. We therefore consider this scenario to be disfavored.

\noindent\textbf {2. Forward shock emission during the free-expansion phase in a wind medium.}

\begin{figure}[htbp]
    \centering
    \begin{overpic}[width=1.0\textwidth]{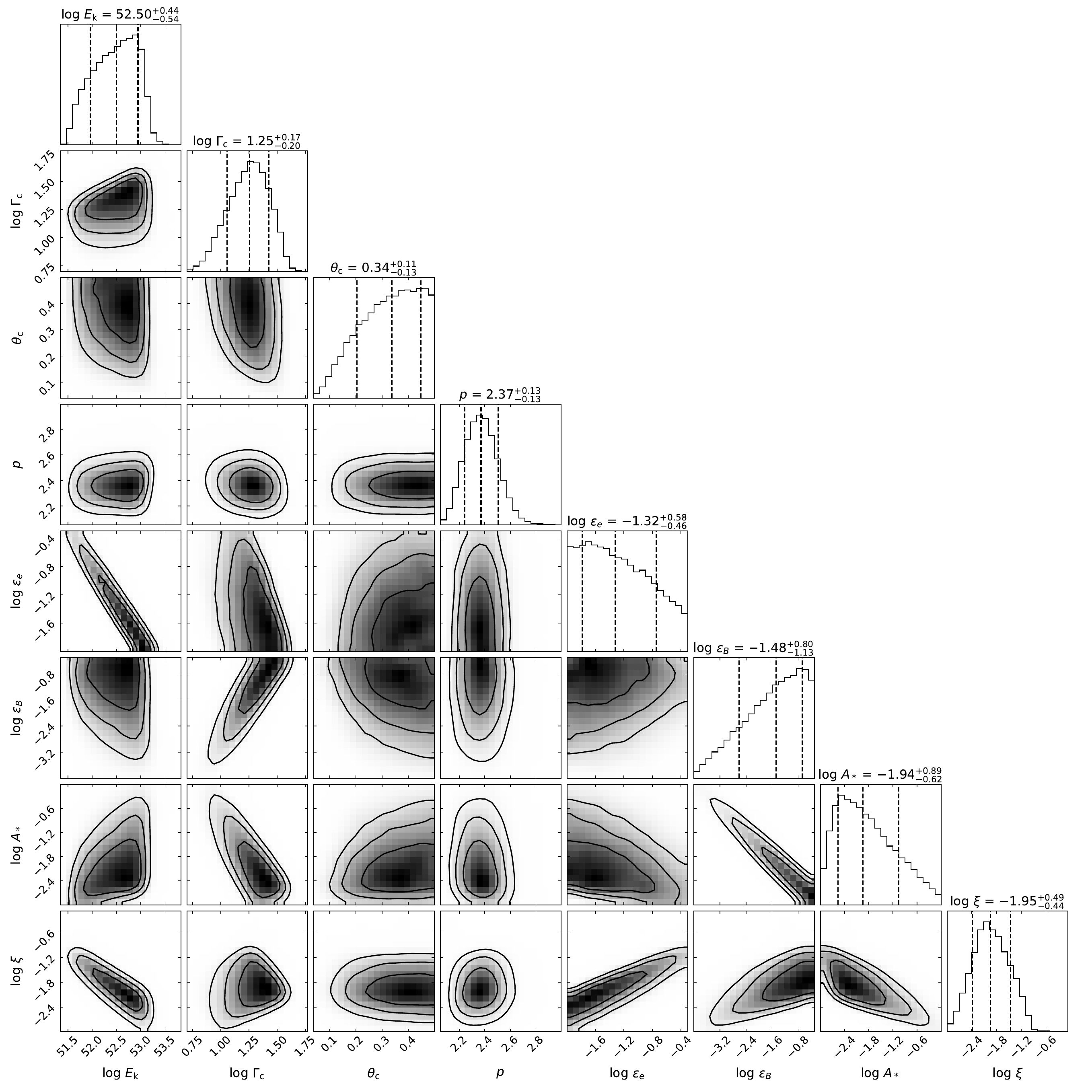}
    \end{overpic}

    \caption{\textbf{MCMC fitting results for multi-band afterglow modeling.} The afterglow emission is modeled as the synchrotron emission from a forward shock expanding in a wind environment, assuming a power-law
structured jet with $k_c = 2$ (see the Methods for details). The diagonal dashed lines and labels indicate the median values along with the symmetric $68\%$ uncertainties (corresponding to the 16th and 84th percentiles) for the marginalized distribution of each parameter.}

    \label{fig_af_contour}
\end{figure}

For the afterglow synchrotron emission from the forward shock expanding in a medium with $n_e\propto R^{-k}$, the temporal and spectral evolution are summarized in  Table~\ref{tab_closure_relation}. As shown, prior to the deceleration, the temporal decay index changes from $F_\nu\propto t^{1/2}$ to $F_\nu\propto t^{(2-p)/2}$ as the spectral regime changes from $\nu_c<\nu_X<\nu_m$ to $\nu_X>\max(\nu_m,\nu_c)$  for a wind medium ($k=2$), where $\nu_m$ is the characteristic frequency corresponding to the minimum Lorentz factor of the injected electrons. Thus, for $p=2.4\pm0.4$, the decay slope changes from  $\alpha=-0.5$ to $\alpha=0.2\pm0.2$ ($F_\nu\propto t^{-\alpha}$), consistent with the plateau of EP241113a. The duration of the plateau can be used to determine an upper limit on the initial Lorentz factor of the jet, since the deceleration time $t_{\rm dec}$ should be longer than the duration of the plateau.  The deceleration time  is determined by the condition that the energy of the swept-up circum-burst medium, given by $E_{\rm sw}(R) \simeq \Gamma^2\int 4\pi n_e(R)m_p c^2 R^2dR$, equals the initial kinetic energy of the shock, $E_{\rm k}$. Based on this, we derive an upper limit on the initial Lorentz factor $\Gamma$ of EP241113a, assuming a density profile of $n_e(R) = (A/m_p)R^{-2}$:
\begin{equation}
    \Gamma \sim \left[ \frac{(1+z)E_{\rm k}}{8 \pi c^3 At_{\rm dec}} \right]^{\frac{1}{4}} = 16 E_{\rm k, 53}^{\frac{1}{4}} A_*^{-\frac{1}{4}} \left( \frac{t_{\rm dec}}{10^4 \, \rm s} \right)^{-\frac{1}{4}}.
\end{equation}
Detailed modeling of the afterglow and MCMC fitting (see Fig.~\ref{fig_afterglow_fitting} and  Fig.~\ref{fig_af_contour} ) gives  $\Gamma\simeq 20$, which is consistent with the above analytical estimate for $E_{\rm k}\simeq 1\times10^{53} {\, \rm erg}$, $A_{*}\simeq 1.5 \times 10^{-2}$ and $t_{\rm dec}\simeq 10^5 {\, \rm s}$ that are obtained in the MCMC fitting.

{We note that  $\Gamma \simeq 20$ is well constrained (see  Fig.~\ref{fig_af_contour}), independent of $E_{\rm k}$ and $A_{*}$. We find that the Lorentz factor $\Gamma $ is primarily constrained by the flux level of the plateau, as we show below. As discussed above, the X-ray plateau should satisfy the condition $\nu_X > \max(\nu_{\rm m}, \nu_{\rm c})$, and the flux density $F_{\nu}$ is given by $F_{\nu} = F_{\rm m} \nu_{\rm m}^{\frac{p-1}{2}} \nu_{\rm c}^{\frac{1}{2}} \nu^{-\frac{p}{2}}$ \cite{Sari1998}. Here, $F_{\rm m} = (1+z) \frac{N_e P_{\rm syn}}{4\pi D_{\rm L}^2}$, where $N_e = \int 4\pi \xi R^2 n_e \, dR$ represents the number of electrons accelerated by the shock, and $\xi$ is the fraction of electrons that are accelerated. The synchrotron radiation power $P_{\rm syn}$ is defined as $P_{\rm syn} = \frac{m_e c^2 \sigma_T}{3e} \Gamma B^\prime$. The break frequencies $\nu_{\rm m}$ and $\nu_{\rm c}$ correspond to the minimum and cooling Lorentz factors ($\gamma_{\rm m}$ and $\gamma_{\rm c}$) of the electron distribution, respectively.
These are given by $\gamma_{\rm m} =\frac{\epsilon_e}{\xi} \frac{m_p}{m_e} \frac{p-2}{p-1}\Gamma$ and $\gamma_{\rm c} = \frac{6\pi m_e c}{\sigma_T \Gamma B^{\prime 2} t}$, where $\epsilon_e$ is the fraction of the shock energy that goes into the electrons and $\sigma_T$ is the Thomson cross-section. The magnetic field in the comoving frame of the shock is given by $B^\prime = \sqrt{32\pi m_p \epsilon_B n_e} \Gamma$, where $\epsilon_B$ is the fraction of energy in the magnetic field, and $n_e$ is the particle number density in the surrounding medium. For a shock expanding freely ($\Gamma = \rm{const.}$) in a stellar wind environment, we have
\begin{equation}
F_{\nu} \propto \left( \frac{p-2}{p-1} \right)^{p-1} \epsilon_B^{\frac{p-2}{4}} \epsilon_e^{p-1} \Gamma^{p+2} \xi^{2-p} A_*^{\frac{p+2}{4}} \nu^{-\frac{p}{2}} t^{\frac{2-p}{2}} D_{\rm L}^{-2},
\end{equation}
for $\nu > \max(\nu_{\rm m}, \nu_{\rm c})$.

Taking $p = 2.4$ and $z = 1.53$, we obtain the following expression for the initial Lorentz factor:
\begin{equation}
\Gamma = 17 \left( \frac{F_{\nu, \rm obs}}{10^{-4} \, \rm mJy} \right)^{0.22} \left( \frac{\epsilon_B}{0.01} \right)^{-0.03} \left( \frac{\epsilon_e}{0.1} \right)^{-0.35} \left( \frac{\xi}{0.1} \right)^{0.13} \left( \frac{A_*}{0.01} \right)^{-\frac{1}{4}} \left( \frac{h\nu_{\rm obs}}{1 \, \rm keV} \right)^{0.28} \left( \frac{t_{\rm obs}}{10^4 \, \rm s} \right)^{0.065},
\end{equation}
which is insensitive to the microphysical parameters of the shock and the wind density. This suggests that the soft, energetic explosion indeed has a Lorentz factor much smaller than that of GRBs.}

\noindent\textbf{3. Late-time high-latitude emission scenario.}

\begin{figure}[htbp]
    \centering
    \begin{overpic}[width=0.45\textwidth]{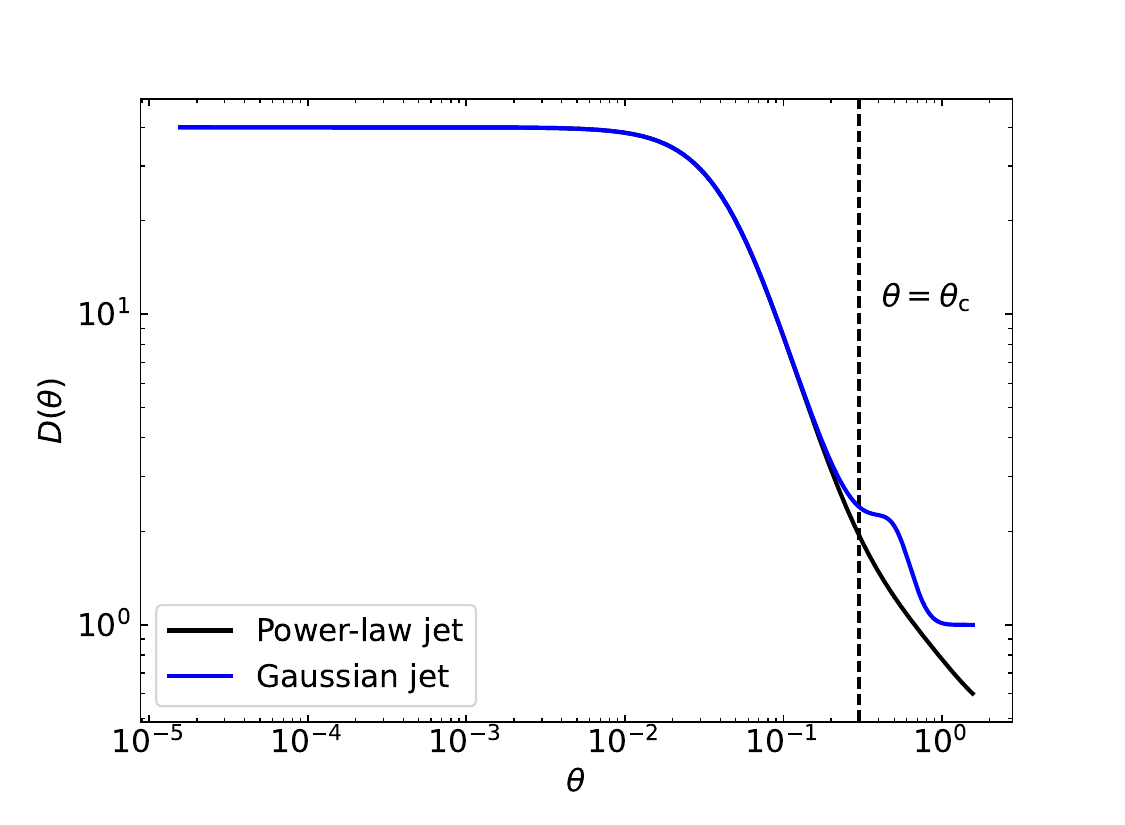}
        \put(2,68){\textbf{(a)}} 
    \end{overpic}
    \hspace{0.05\textwidth} 
    \begin{overpic}[width=0.45\textwidth]{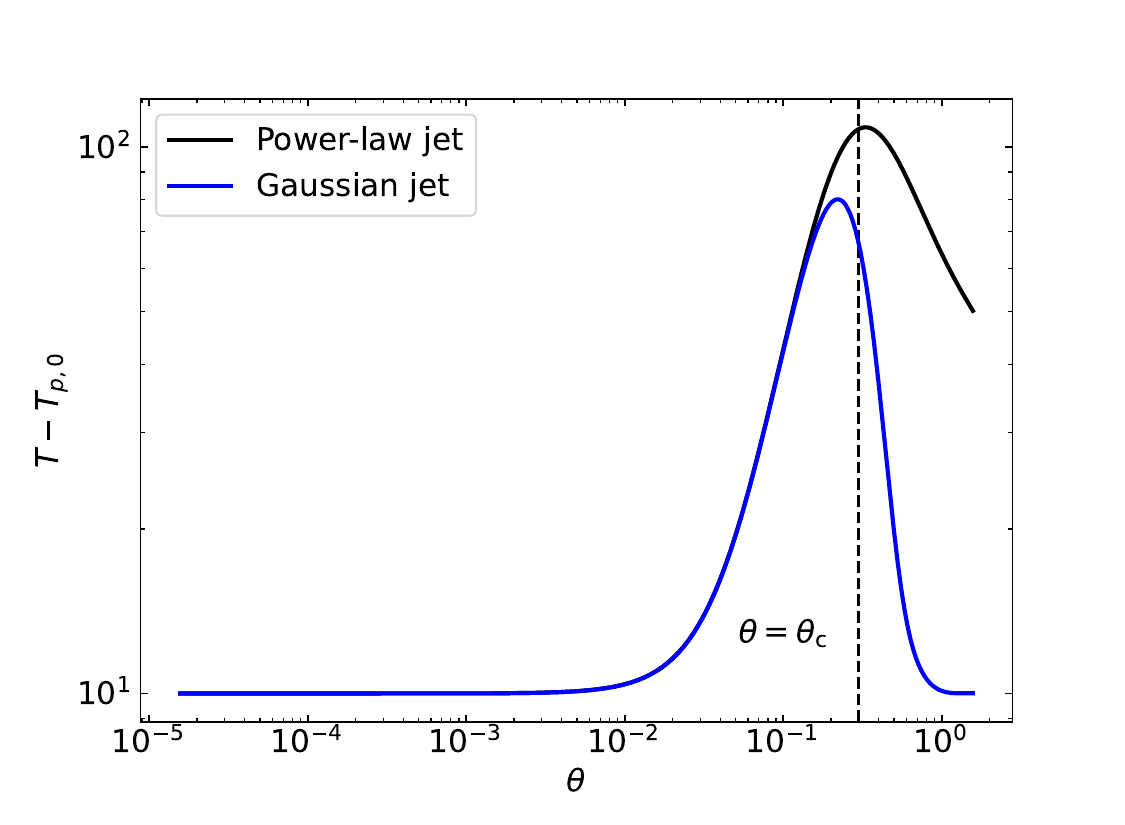}
        \put(2,68){\textbf{(b)}} 
    \end{overpic}

    \caption{\textbf{Doppler factor and observer time of photons from different latitudes of an on-axis jet.} (a) The Doppler factor as a function of latitude angle $\theta$. (b) The observer time of photons from different latitudes $\theta$. We assume $k_{\rm c} = 2$ for the power-law jet and set $\Gamma_{\rm c} = 20$, $\theta_{\rm c} = 0.3$, and $t_{\rm v} = 10 \, \rm s$ for both the power-law and Gaussian jets.}

    \label{fig_HLE_theta}
\end{figure}

When a structured jet is viewed on-axis, the observed flux follows the relation $F_\nu \propto D^{2+\beta} \epsilon(\theta)$, which introduces an additional factor $\epsilon(\theta)$ compared to the top-hat jet. When photons originate from an angle $\theta < \theta_{\rm c}$, both the Lorentz factor $\Gamma(\theta)$ and the energy distribution $\epsilon(\theta)$ can be approximated as constants, yielding $D \propto t^{-1}$. Consequently, the flux scales as $F_\nu \propto t^{-(2+\beta)}$, consistent with the high-latitude emission predicted for a top-hat jet. However, when photons from higher latitudes ($\theta > \theta_{\rm v}$) are observed, both the Lorentz factor $\Gamma(\theta)$ and the jet energy $\epsilon(\theta)$ decrease more rapidly. In this regime, the Doppler factor is approximately given by $D(\theta) = [\Gamma(\theta)(1-\beta(\theta) \cos \theta)]^{-1} \simeq 2\Gamma(\theta)^{-1}\theta^{-2}$. For a power-law jet, this relation becomes $D(\theta) \propto \theta^{k-2}$, while for a Gaussian jet, $D(\theta) \propto \theta^{-2} e^{(\theta/\theta_{\rm c})^2}$ (see  Fig.~\ref{fig_HLE_theta}a). The observed flux then scales as $F_\nu(\theta) \propto D^{2+\beta}\epsilon \propto \theta^{-4+k(1+\beta)-2\beta}$ for a power-law jet, and as $F_\nu \propto \theta^{-4-2\beta} e^{(1+\beta)(\theta/\theta_{\rm c})^2}$ for a Gaussian jet, resulting in a plateau when $\theta \gtrsim \theta_{\rm c}$. However, this plateau cannot persist for more than approximately ten times the duration of its onset (i.e., it cannot last from $\sim 1\, \rm ks$ to $10\, \rm ks$ as observed in the X-ray light curve): the observer-frame arrival times of photons from all latitudes are shown in  Fig.~\ref{fig_HLE_theta}b, which indicates that the last-arriving photons predominantly originate from $\theta \sim \theta_{\rm c}$, corresponding to the onset of the plateau. Therefore, high-latitude emission alone cannot adequately account for the observed X-ray plateau.

\noindent\textbf {Modeling of the multi-band afterglows.\\} 
To place more stringent constraints on the parameter space of the jet, we perform MCMC fitting of the multi-band afterglows under the assumption of a wind-like external medium\cite{Dai1998-2,Chevalier1999} (i.e., $k=2$). The jet structure is assumed to have a power-law form with $k_c=2$ beyond the uniform core (see Eq.~\ref{PL-jet-structure}). We model the multi-band afterglow by considering  synchrotron emission processes. The emission is calculated using {\tt VegasAfterglow}, a high-performance, sophisticated C++ framework designed for modeling GRB afterglows. The source code is available at: \href{https://github.com/YihanWangAstro/VegasAfterglow}{https://github.com/YihanWangAstro/VegasAfterglow}. In this work, we use its modules for forward shock dynamics, synchrotron radiation (including self-absorption), and equal arrival time surface integration. Additional physics, such as reverse shock, inverse Compton scattering, Klein-Nishina corrections, and energy injection, are not included, as they are expected to play a subdominant role in this event.

A log-uniform prior distribution is adopted for several parameters, including the isotropic kinetic energy $E_{\rm k}$, the normalized mass loss rate $A_*$, the fraction of electrons that are accelerated $\xi$, the fraction of shock energy transferred into electrons $\epsilon_e$, the fraction of shock energy transferred into  magnetic field energy $\epsilon_B$, and the initial Lorentz factor of the jet $\Gamma_{\rm c}$ at $\theta = 0$. A uniform prior distribution is applied to the opening angle of the jet core $\theta_{\rm c}$ and the electron spectral index $p$. {The viewing angle is fixed at 0, given that the source is observed on-axis, as argued above.} We use all available X-ray and optical detection data to constrain the model parameters, excluding the contribution from the host galaxy in the optical afterglow. For the $\rm VT\_R$ and $\rm VT\_B$ bands, the observations at 2025-01-24T06:45:25 (UTC) and 2025-01-07T01:21:21 (UTC) are considered to be dominated by host galaxy flux, as the flux has become constant. The marginalized posterior distributions of the model parameters are presented in  Fig.~\ref{fig_af_contour} and  Table~\ref{tab_afterglow_mcmc_results}, while the corresponding multi-band afterglow light curve, computed using the best-fit parameters, is shown in Fig.~\ref{fig_afterglow_fitting}.  The initial Lorentz factor of the jet is well constrained, $\Gamma_{\rm c} \simeq 20$, which is consistent with the analytical derivation. The angle of the jet core ($\theta_c$) cannot be well constrained, as no clear jet break has been detected.
{A relatively low value of $A_*$ may suggest that the properties of the wind ejected by the progenitor prior to its final explosion is very different from the
properties of the wind ejected at earlier times\cite{Dereli-Bégué2022}. }

Thus, from the multi-band afterglow modeling, we find that a low-$\Gamma$ jet with a large isotropic kinetic energy $E_{\rm k}\sim 10^{53}{\, \rm erg}$ is required. This, along with the soft spectrum, the large isotropic X-ray energy $E_{\gamma, \rm iso}$, and the discrepancy with the Amati relation, as well as the on-axis viewing angle derived from the steep decay of the X-ray emission, suggests that EP241113a is produced by a dirty fireball.

\subsection{Event rate.}
According to the sensitivity and detection of EP241113a, we can estimate the event rate $\rho_{\rm DF}$ of such dirty fireball events. Assuming that EP241113a is the only detected dirty fireball event during the operational time of EP-WXT, $T_{\rm OT} \sim 15$ months, we can derive the number of detected dirty fireballs, $N_{\rm DF}$, as follows\cite{Sun2025}:
\begin{equation}
N_{\rm DF} = \frac{\eta_{\rm DC} \Omega_{\rm WXT} T_{\rm OT}}{4\pi} \rho_{\rm DF} V_{\rm max} = 1,
\end{equation}
where $\eta_{\rm DC}$ is the duty cycle of EP-WXT, and $\Omega_{\rm WXT} = 3600$ square degrees represents the field of view of EP-WXT. The sensitivity of EP-WXT with a typical exposure time of $200 \, \rm s$ reaches $\sim 1 \times 10^{-10} \, \rm erg \, cm^{-2} \, s^{-1}$ in the 0.5–4 keV band. The time-averaged luminosity of EP241113a during $T_{\rm 90}$ is $6.7^{+2.4}_{-1.4} \times 10^{48} \, \rm erg \, s^{-1}$. Consequently, the maximum redshift at which the source can be detected is $z_{\rm max} = 4.1^{+0.6}_{-0.3}$.

The effective maximum volume, $V_{\rm max}$, weighted by the density evolution function $f(z)$ (e.g., the star formation history\cite{Yuksel2008}) within $z_{\rm max}$, can be calculated as: 

\begin{equation}
V_{\rm max} = \int_0^{z_{\rm max}} \frac{f(z)}{(1+z)} \frac{dV(z)}{dz} \, dz, 
\end{equation}
where the comoving volume element is given by
\begin{equation}
\frac{dV(z)}{dz} = \frac{c}{H_0} \frac{4\pi D_{\rm L}(z)^2}{(1+z)^2 \left[\Omega_M(1+z)^3 + \Omega_{\Lambda}\right]^{1/2}}.
\end{equation}

From these relations, we can derive the local event rate density as: 
\begin{equation}
\rho_{\rm DF}\simeq (3.9 \pm 0.2) \times 10^{-3} \, \rm Gpc^{-3} \, yr^{-1}.
\end{equation}

However, since some of the sources detected by EP-WXT do not have measured redshifts, the properties of these sources remain unclear. Therefore, the derived value of $\rho_{\rm DF}$ should be considered as a conservative lower limit.

\noindent\textbf{Data Availability\\}
All processed data supporting the findings of this study are included in the figures and tables within the main text. Additional data are available from the corresponding authors upon reasonable request. Publicly available datasets were also utilized, including those obtained from GCN Circulars.

\noindent\textbf{Code Availability\\}
The code used to generate the results and figures in this study is available from the authors upon reasonable request.

\bigskip
\bigskip
\bigskip

\clearpage

\begin{figure}[htbp]
    \centering
    \begin{overpic}[width=0.43\textwidth]{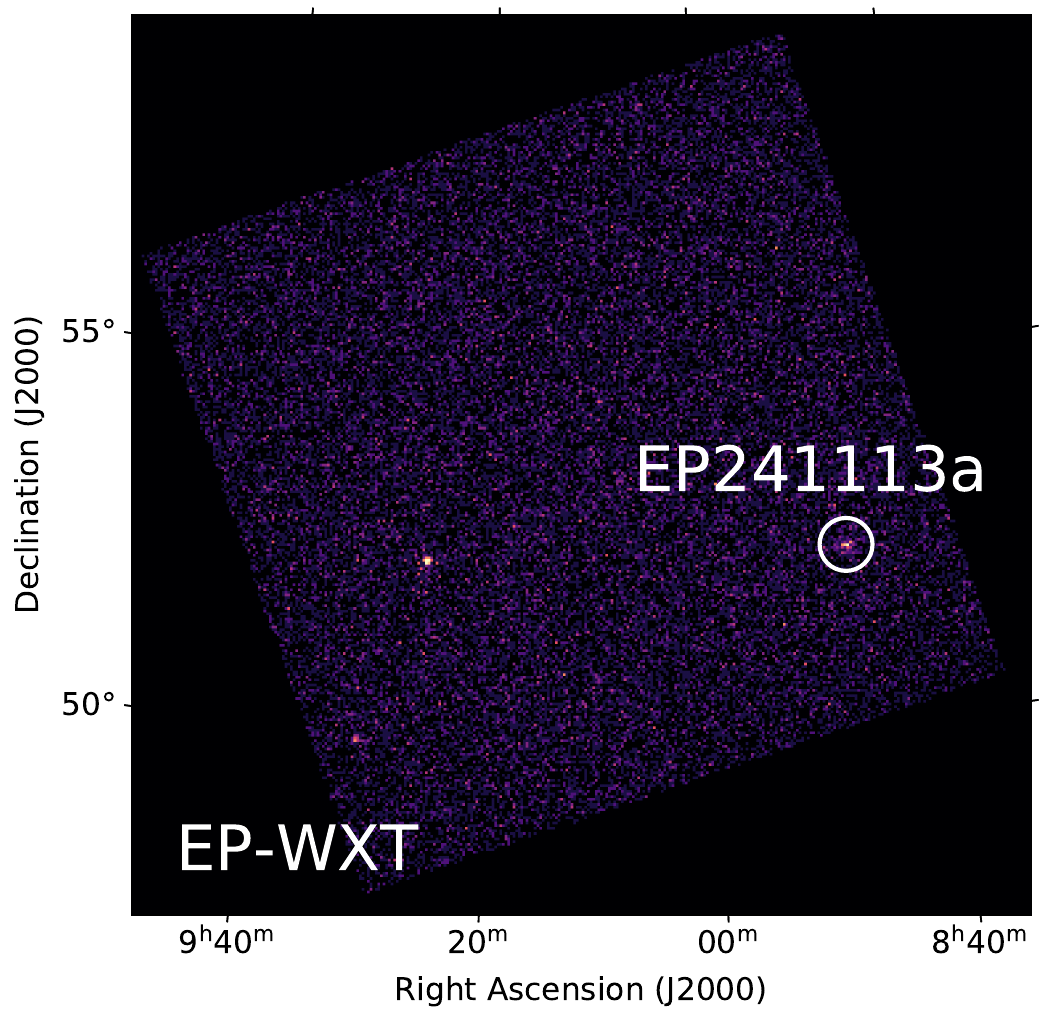}
        \put(2,97){\textbf{(a)}} 
    \end{overpic}
    \hspace{0.05\textwidth} 
    \begin{overpic}[width=0.45\textwidth]{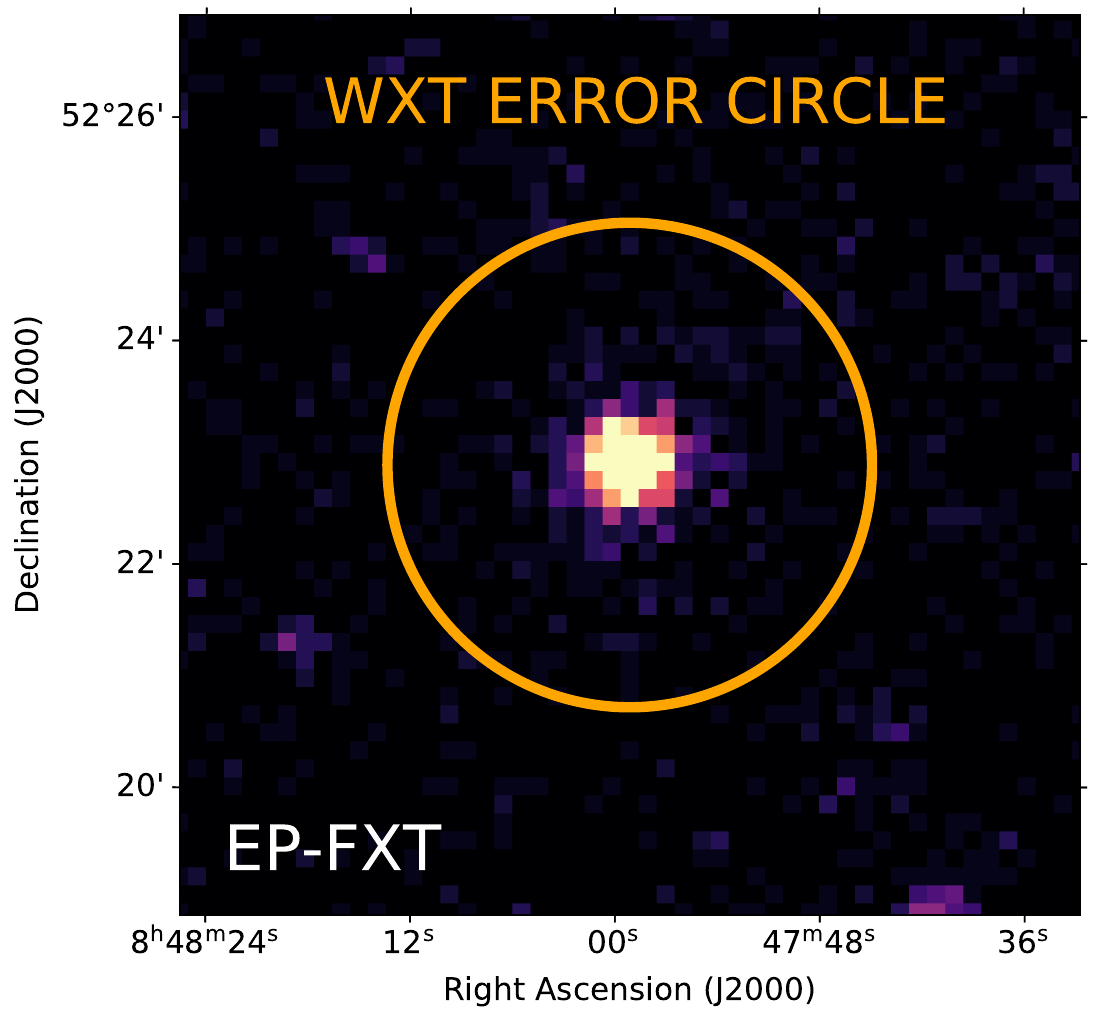}
        \put(2,93){\textbf{(b)}} 
    \end{overpic}
    \caption{\textbf{X-ray image of EP241113a.} (a) The EP-WXT CMOS48 image of EP241113a. (b) The combined EP-FXTA and FXTB image of EP241113a during the time interval 2024-11-13T19:15:12 to 2024-11-13T21:25:43 (UTC). The outer yellow circle represents the positional uncertainty of the WXT with a radius of $2.4'$ (at the $90\%$ confidence level).}
    \label{fig_WXT_FXT_image}
\end{figure}

\clearpage

\clearpage

    


\clearpage

\clearpage

\clearpage

\clearpage

\begin{figure}[htbp]
    \centering
    \begin{overpic}[width=0.7\textwidth]{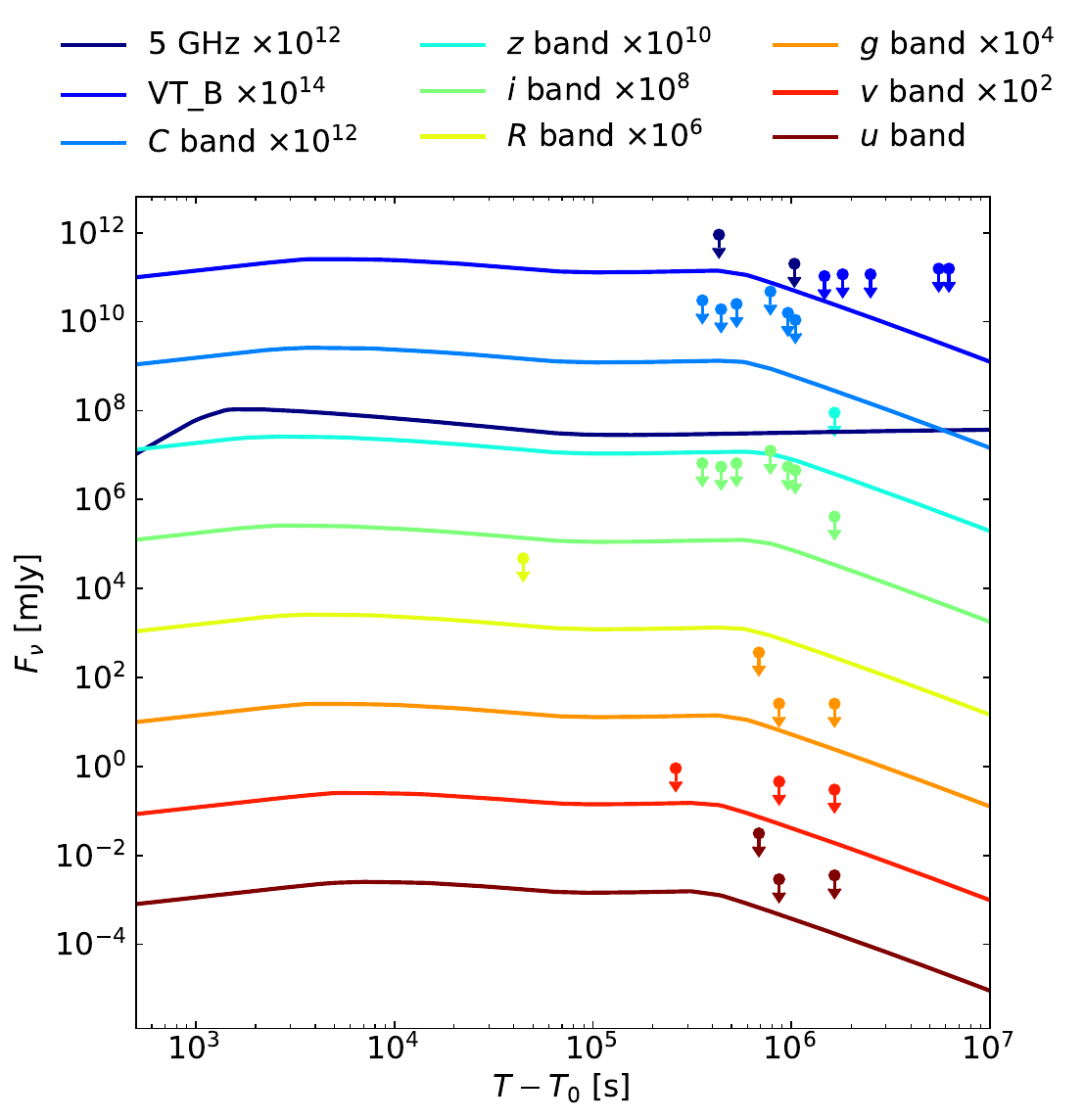}
    \end{overpic}

    \caption{\textbf{The optical and radio upper limits of the afterglow of EP241113a.} The solid line represents the predicted light curve derived from the same model and parameter values as in Fig.~\ref{fig_afterglow_fitting}.}

    \label{fig_af_up}
\end{figure}

\clearpage

\clearpage

\clearpage

\begin{table*}
\centering
\scriptsize
\begin{threeparttable}
\small
\caption{Log of the EP-FXT follow-up observations.}\label{tab_FXT_obs}
\begin{tabular}{ccccc}
\toprule
ObsID & Start$\;$Time & End$\;$Time&Duration&Exposure\\
 & (UTC) &(UTC)&(second)&(second)\\
 \hline
01709120178 & 2024-11-13T19:15:12 & 2024-11-13T21:25:43 & 7831 & 5076 \\
06800000234 & 2024-11-16T04:38:13 & 2024-11-16T07:05:04 & 8811 & 6058 \\
06800000242 & 2024-11-17T22:16:27 & 2024-11-17T23:59:12 & 6165 & 3434 \\
06800000247 & 2024-11-18T22:17:42 & 2024-11-19T00:44:41 & 8819 & 6073 \\
06800000255 & 2024-11-21T06:20:27 & 2024-11-21T07:11:32 & 3065 & 3046 \\
06800000256 & 2024-11-21T11:08:40 & 2024-11-21T13:31:31 & 8571 & 5403 \\
06800000263 & 2024-11-29T06:28:34 & 2024-11-29T08:42:17 & 8023 & 4358 \\
06800000219 & 2024-11-29T09:40:56 & 2024-11-29T13:44:53 & 14637 & 8776 \\
06800000282 & 2024-12-09T06:45:29 & 2024-12-09T11:04:31 & 15542 & 8253 \\
06800000288 & 2024-12-11T16:27:02 & 2024-12-11T19:15:14 & 10092 & 8529 \\
06800000289 & 2024-12-12T02:04:03 & 2024-12-12T03:16:14 & 4331 & 2207 \\
06800000290 & 2024-12-12T11:41:06 & 2024-12-12T14:29:32 & 10106 & 7229 \\
06800000297 & 2024-12-14T15:16:01 & 2024-12-14T17:46:03 & 9002 & 6396 \\
06800000298 & 2024-12-14T20:04:31 & 2024-12-14T20:58:16 & 3225 & 3191 \\
06800000299 & 2024-12-15T16:54:30 & 2024-12-15T19:23:46 & 8956 & 6316 \\

\bottomrule
\end{tabular}
\end{threeparttable}
\end{table*}

\clearpage

\begin{table*}
\centering
\scriptsize
\begin{threeparttable}
\small
\caption{Upper limit of {Fermi}/GBM observation}\label{tab_GBM_up}
\begin{tabular}{cccccc}
\toprule
Model\tnote{*} & $\Gamma_X$ & $\Gamma_{X,2}$ & $E_{\rm p}$  & Energy Range  & $3\sigma$ Upper Limit \\
 &&&(keV)&(keV)&($10^{-8} \, \rm erg \, cm^{-2} \, s^{-1}$)
\\
\midrule
{\bf Time scale = 8.129\,s} & & & & & \\
Band & 1.1 & 2.2 & 180.0 & 8--900 & 3.28 \\
Band & 1.9 & 3.7 & 70.0 & 8--900 & 2.49 \\
PL & 1.0 & -- & -- & 8--900 & 8.51 \\
PL & 2.0 & -- & -- & 8--900 & 2.69 \\
PL & 3.0 & -- & -- & 8--900 & 1.68 \\
\midrule
{\bf Time scale = 32.786\,s} & & & & & \\
Band & 1.1 & 2.2 & 180.0 & 8--900 & 1.91 \\
Band & 1.9 & 3.7 & 70.0 & 8--900 & 1.24 \\
PL & 1.0 & -- & -- & 8--900 & 4.28 \\
PL & 2.0 & -- & -- & 8--900 & 1.35 \\
PL & 3.0 & -- & -- & 8--900 & 0.84 \\
\bottomrule
\end{tabular}
\begin{tablenotes}
\footnotesize
\item[*] PL and Band represent the power-law and Band function model, respectively.

\end{tablenotes}
\end{threeparttable}
\end{table*}

\clearpage

\begin{table*}[ht]
   \caption{Observation and measurement by SVOM/VT. The 3$\sigma$ upper limit is given if the source was not detected.}
   \centering
   \begin{tabular}{cccccccc}
   \hline\hline
 Start Time & End Time   &   Band     & Magnitude &    Exposure   \\
     UT       &     UT       &            &    (AB)         &     second       \\
   \hline                          
    2024-11-30T08:43:32   &  2024-11-30T12:33:17    &   VT$\_$R   &  $22.96\pm0.25$                & $30\times216$  \\
    2024-12-04T14:42:52   &  2024-12-04T18:33:37    &   VT$\_$R   &  $23.20\pm0.20$                & $30\times208$  \\
    2024-12-12T15:19:56   &  2024-12-12T19:14:10    &   VT$\_$R   &  $23.34\pm0.20$                & $50\times136$  \\
    2025-01-16T14:16:02   &  2025-01-16T19:47:48    &   VT$\_$R   &  $23.65\pm0.32$                & $60\times147$  \\
    2025-01-24T06:45:25   &  2025-01-24T15:35:54    &   VT$\_$R   &  $23.60\pm0.32$                & $60\times276$  \\
    2024-11-30T08:43:02   &  2024-11-30T12:33:17    &   VT$\_$B   &  $>23.9$ & $30\times216$  \\
    2024-12-04T14:44:52   &  2024-12-04T18:33:37    &   VT$\_$B   &  $>23.8$  & $30\times199$  \\
    2024-12-12T15:19:56   &  2024-12-12T19:14:10    &   VT$\_$B   &  $>23.8$  & $50\times144$  \\
    2025-01-16T14:15:02   &  2025-01-16T19:47:48    &   VT$\_$B   &  $23.48\pm0.22$                & $60\times148$  \\
    2025-01-24T06:45:25   &  2025-01-24T15:35:54    &   VT$\_$B   &  $23.48\pm0.22$                & $60\times282$  \\
    \hline
    \end{tabular}
   \label{tab:vtdata}
   \end{table*}

\clearpage

\begin{table}[htbp]
\centering
\caption{Limiting magnitudes from Mephisto}
\label{tab:limiting_mags_Mephisto}
\begin{tabular}{@{\extracolsep\fill}lcccccc} 
\toprule
\multirow{2}{*}{\centering Band} & \multirow{2}{*}{\centering MJD} & \multirow{2}{*}{\centering Exposure (s)} & \multicolumn{2}{c}{\centering Limiting magnitude} \\
\cmidrule(lr){4-5}
 & & & \multicolumn{1}{c}{$3\sigma$} & \multicolumn{1}{c}{$5\sigma$} \\
\midrule

\multirow{3}{*}{\centering $u$} 
 & 60635.74372 & 1$\times$300  & 20.31 & 19.75 \\
 & 60637.83503 & 3$\times$300  & 22.89 & 22.34 \\
 & 60646.90450 & 2$\times$300  & 22.67 & 22.11 \\
\midrule

\multirow{3}{*}{\centering $v$}
 & 60630.82561 & 2$\times$300  & 21.61 & 21.06 \\
 & 60637.84581 & 2$\times$300  & 22.36 & 21.81 \\
 & 60646.89661 & 2$\times$300  & 22.81 & 22.26 \\
\midrule

\multirow{3}{*}{\centering $g$}
 & 60635.74370 & 1$\times$300  & 20.08 & 19.53 \\
 & 60637.83504 & 3$\times$300  & 22.95 & 22.40 \\
 & 60646.90451 & 2$\times$300  & 22.96 & 22.40 \\
\midrule

\multirow{3}{*}{\centering $r$}
 & 60630.82563 & 2$\times$300  & 22.29 & 21.73 \\
 & 60637.84582 & 2$\times$300  & 22.53 & 21.98 \\
 & 60646.89663 & 2$\times$300  & 23.11 & 22.56 \\
\midrule

\centering $i$ & 60646.90451 & 2$\times$300 & 22.40 & 21.85 \\
\midrule

\centering $z$ & 60646.89666 & 2$\times$300 & 21.53 & 20.97 \\
\bottomrule
\end{tabular}
\end{table}

\clearpage

\begin{table*}[ht]
   \caption{Limiting magnitudes from TRT/GOT/PAT17}
   \label{tab:limiting_mags_PAT17S}
   \centering
   \begin{tabular}{ccccc}
   \hline\hline
    Mid Time&Band&Exposure&Magnitude Upper limit&Telescope  \\
    (UT) & & (second) & (AB, 5$\sigma$) & \\
   \hline
    2024-11-14T07:32:14.976&$R$&5x180&$>$19.5&TRT-SRO \\
    2024-11-21T16:17:14.000&$r$&60x60&$>$19.4&GOT \\
    2024-11-17T22:10:11.500&Clear&14x120&$>$20&PAT17 \\
    2024-11-17T22:07:37.580&$i$&14x120&$>$19.4&PAT17 \\
    2024-11-17T22:09:02.680&$r$&15x120&$>$19.9&PAT17 \\
    2024-11-18T22:08:03.845&Clear&14x120&$>$20.5&PAT17 \\
    2024-11-18T22:09:13.165&$i$&15x120&$>$19.6&PAT17 \\
    2024-11-18T22:10:38.739&$r$&14x120&$>$20.1&PAT17 \\
    2024-11-19T22:20:07.180&Clear&12x120&$>$20.2&PAT17 \\
    2024-11-19T22:10:12.415&$i$&14x120&$>$19.4&PAT17 \\
    2024-11-19T22:07:55.680&$r$&14x120&$>$19.8&PAT17 \\
    2024-11-22T21:04:51.105&Clear&3x120&$>$19.5&PAT17 \\
    2024-11-22T21:06:00.360&$i$&4x120&$>$18.7&PAT17 \\
    2024-11-22T21:03:43.480&$r$&4x120&$>$19.2&PAT17 \\
    2024-11-24T22:09:19.780&Clear&15x120&$>$20.7&PAT17 \\
    2024-11-24T22:10:28.419&$i$&14x120&$>$19.6&PAT17 \\
    2024-11-24T22:08:12.035&$r$&14x120&$>$20.3&PAT17 \\
    2024-11-25T22:09:55.755&Clear&14x120&$>$21.1&PAT17 \\
    2024-11-25T22:07:21.935&$i$&14x120&$>$19.8&PAT17 \\
    2024-11-25T22:08:48.114&$r$&15x120&$>$20.5&PAT17 \\
    \hline
    \end{tabular}
   \end{table*}

\clearpage

\begin{table*}
    \centering
    \hspace{0.6cm}
    \caption{Parameters and prior distributions considered for fitting the SED of the hosts of EP241113a. }
    \scalebox{1.0}{
    \begin{tabular}{llll}
    \hline \hline
    Parameter & Definition & Prior \\ 
     \hline
    $z$ & Redshift & $\mathcal{U}{\sim}[0.1,3.5]$  \\
    $t_{\rm age}$~(Gyr) & Age of the galaxy at the time of observation & $\mathcal{U}{\sim}[0.1,15.0]$ \\
    $\tau$~(Gyr) & e-folding time of exponential SFH & $\mathcal{U}{\sim}[0.1,10.0]$ \\
    $\log(M_F/M_\odot)$ & Total mass formed & $\mathcal{U}{\sim}[6.0,13.0]$ \\
    $\log(Z/Z_\odot)$ & Stellar metallicity & $\mathcal{U}{\sim}[-1.0,0.5]$ \\
    $\log(U)$ & Nebular component & $\mathcal{U}{\sim}[-4.0,-1.0]$ \\
    $A_{\rm V,dust}$ & Dust attenuation & $\mathcal{U}{\sim}[0.0,2.0]$ \\    
    \hline 
    \end{tabular}
    }
    \label{tab:SED_model}
\end{table*}

\clearpage

\begin{table*}
\centering
\scriptsize
\begin{threeparttable}
\small
\caption{MCMC parameter estimate (median) for the high-latitude emission model with 1$\sigma$ uncertainties.}\label{tab_HLE_mcmc_results}
\begin{tabular}{cccc}
\toprule
Parameter & Prior & Posterior & Posterior \\
 & & (Power-law jet) & (Gaussian jet)  \\
\hline
$\Gamma_{\rm c}$ & 20 (Fixed) & -- & -- \\
$\theta_{\rm c} \, \rm (deg)$ & $\mathcal{U}{\sim}$[0, 30] & $21.36^{+5.58}_{-7.38}$ & $20.91^{+5.87}_{-7.37}$ \\
$\theta_{\rm v}/\theta_{\rm c}$ & $\mathcal{U}{\sim}$[0, 10] & $0.69^{+0.23}_{-0.28}$ & 
$0.61^{+0.18}_{-0.22}$ \\
$k_{\rm c}$ &--&--& 2 (Fixed) \\
log $N_0$ & $\mathcal{U}{\sim}$[-50, -70] & $-56.76^{+0.50}_{-0.56}$ & $-56.92^{+0.47}_{-0.52}$\\
$T_{\rm p,0} \, \rm (s)$ & $\mathcal{U}{\sim}$[74.8, 111.8] & $100.85^{+6.99}_{-8.95}$ & $98.39^{+8.27}_{-10.01}$ \\
$\beta$ & $\mathcal{U}{\sim}$[1.0, 1.2] & $1.10^{+0.06}_{-0.06}$ & $1.10^{+0.06}_{-0.06}$ \\
$\chi^2_{\rm min}/\rm d.o.f$  & -- & 3.5/4 & 3.9/4 \\
\bottomrule
\end{tabular}
\end{threeparttable}
\end{table*}

\clearpage

\begin{table*}
\setlength{\tabcolsep}{4pt} 
\centering
\scriptsize
\begin{threeparttable}
\caption{Spectral results of X-ray observations of EP241113a. Errors represent the 1$\sigma$ uncertainties. The upper limit is quoted at the 95\% confidence level.}
\label{tab_spectrum_fitting}
\begin{tabular}{llcccccccc}
\toprule
Instrument & Time Interval &  Model\tnote{*} & $\Gamma_X$ & $\Gamma_{X,2}$ & $E_{\rm p}$ & $ kT_{\rm BB}$ & $ N_{\rm int}$ &   Flux\tnote{‡}  & CSTAT/(d.o.f.) \\
& &  & & & (keV) & (keV) & ($ 10^{22} \, \rm{cm^{-2}}$)  & ($\rm erg\,cm^{-2}\,s^{-1}$) &  \\
\hline
 
\multirow{8}{*}{EP-WXT} & [3\,s, 207\,s] &  PL  &  $2.7_{-0.4}^{+0.7},$ & -- & -- & -- & $3.2^{+1.6}_{-0.9}$ & $1.1^{+0.5}_{-0.2} \times 10^{-9}$ & $33.10/41$ \\
& [3\,s, 207\,s] &  Band  & $0.3^{+1.1}_{-1.3}$ & $4.0^{+0.3}_{-1.6}$ & $<1.9$ & -- & $1.3^{+2.6}_{-0.4}$ & $7.6^{+5.4}_{-0.5}\times 10^{-10}$  & $31.26/39$ \\
& [3\,s, 207\,s] &  SBPL  & $1.3^{+0.2}_{-2.3}$ & $3.5_{-1.0}^{+0.2}$ & $<1.6$ & --  & $2.0^{+2.5}_{-0.4}$ & $8.8^{+6.1}_{-0.8} \times 10^{-10}$ & $30.00/39$ \\
& [3\,s, 207\,s] &  CPL  & $-0.3^{+2.0}_{-0.7}$ & -- & $1.3^{+0.3}_{-0.6}$ & --  & $1.1^{+1.6}_{-0.5}$ & $6.8^{+3.1}_{-0.3} \times 10^{-10}$ & $31.02/40$ \\
& [3\,s, 207\,s] &  BB  &  -- & -- & -- & $0.38^{+0.02}_{-0.07}$ & $<2.1$ & $6.2^{+1.4}_{-0.6} \times 10^{-10}$ & $30.60/41$ \\
& [3\,s, 53\,s] &  PL  &  $1.0^{+1.2}_{-1.1}$ & -- & -- & -- & $3.2$ (Fixed) & $4.3^{+2.6}_{-1.9} \times 10^{-10}$ & $7.74/6$ \\
& [53\,s, 122\,s] &  PL  &  $3.5^{+0.8}_{-0.7}$ & -- & -- & -- & $3.2$ (Fixed) & $6.4^{+1.7}_{-1.5} \times 10^{-10}$ & $10.82/16$ \\
& [122\,s, 207\,s] &  PL  &  $2.4^{+0.4}_{-0.4}$ & -- & -- & -- & $3.2$ (Fixed) & $2.3^{+0.4}_{-0.3} \times 10^{-9}$ & $57.94/63$ \\
\hline
\multirow{15}{*}{EP-FXT}& [300\,s, 402\,s] & PL & \multirow{4}{*}{$2.1^{+0.1}_{-0.1}$ \tnote{†}} & -- & -- & -- &\multirow{4}{*}{$<0.03$}  & $2.7^{+0.2}_{-0.2} \times 10^{-11}$ & \multirow{4}{*}{287.90/324}  \\
& [402\,s, 540\,s] & PL &  & -- & -- & -- &   & $9.2^{+1.1}_{-1.0} \times 10^{-12}$ &  \\
& [540\,s, 724\,s] & PL &  & -- & -- & -- &   & $2.8^{+0.5}_{-0.4} \times 10^{-12}$ &  \\
& [724\,s, 971\,s] & PL &  & -- & -- & -- &   & $1.3^{+0.3}_{-0.3} \times 10^{-12}$ &  \\ 
\cline{2-10}
& [971\,s, 1303\,s] & PL & \multirow{4}{*}{$2.2^{+0.2}_{-0.2}$ \tnote{†}} & -- & -- & -- & \multirow{4}{*}{$<0.3$} & $6.3^{+1.8}_{-1.6} \times 10^{-13}$ & \multirow{4}{*}{248.04/273} \\
& [1303\,s, 1748\,s] & PL &  & -- & -- & -- &   & $7.6^{+1.6}_{-1.4} \times 10^{-13}$ &  \\
& [1748\,s, 5085\,s] & PL &  & -- & -- & -- &   & $5.7^{+1.2}_{-1.1} \times 10^{-13}$ &  \\
& [5085\,s, 8112\,s] & PL &  & -- & -- & -- &   & $8.3^{+0.9}_{-0.8} \times 10^{-13}$ &  \\
\cline{2-10}
& [2.39\,d, 2.50\,d] & PL & \multirow{7}{*}{$2.5^{+0.2}_{-0.1}$ \tnote{†}}  & -- & -- & -- & \multirow{7}{*}{$<0.2$} & $6.6^{+1.2}_{-0.9} \times 10^{-14}$ & \multirow{7}{*}{321.11/269} \\
& [4.13\,d, 4.20\,d] & PL & & -- & -- & -- & & $4.3^{+1.1}_{-0.9} \times 10^{-14}$ & \\
& [5.13\,d, 5.23\,d] & PL & & -- & -- & -- & & $2.4^{+0.7}_{-0.5} \times 10^{-14}$ & \\
& [7.47\,d, 7.50\,d] & PL & & -- & -- & -- & & $2.4^{+0.9}_{-0.8} \times 10^{-14}$ & \\
& [7.67\,d, 7.77\,d] & PL & & -- & -- & -- & & $2.6^{+0.7}_{-0.6} \times 10^{-14}$ & \\
& [15.47\,d, 15.77\,d] & PL & & -- & -- & -- & & $1.5^{+0.4}_{-0.3} \times 10^{-14}$ & \\
& [25.48\,d, 32.01\,d] & PL & & -- & -- & -- & & $5.6^{+1.6}_{-1.4} \times 10^{-15}$ & \\
\bottomrule
\end{tabular}

\begin{tablenotes}
\footnotesize
\item[*] PL, Band, SBPL, CPL, and BB represent the power-law, Band function, smoothly broken power-law, cutoff power-law, and blackbody models, respectively.
\item[†] The photon indices of the spectra are assumed to remain constant within each of the following temporal phases: the steep decay phase ([300\,s, 971\,s]), the plateau phase ([971\,s, 8112\,s]), and the late afterglow phase ([2.39\,d, 32.01\,d]).
\item[‡] The unabsorbed flux is measured in the $[0.5, 4.0]\,\rm{keV}$ energy range for WXT and in the $[0.5, 10.0]\,\rm{keV}$ range for FXT.

\end{tablenotes}
\end{threeparttable}

\end{table*}

\clearpage

\begin{table*}
\centering
\scriptsize
\begin{threeparttable}
\small
\caption{Closure relations for synchrotron emission in different spectral regimes.}\label{tab_closure_relation}
\begin{tabular}{lll}
\toprule
Spectral Regime & $F_{\nu}$ & $F_{\nu}(k = 2, p = 2.4) \tnote{*}$ \\
\midrule
\textbf{Coasting phase} ($t \leqslant t_{\rm dec}$) & & \\
$\nu < \nu_{\rm m} < \nu_{\rm c}$ & $\propto \nu^{1/3} t^{3 - \frac{4k}{3}}$ & $\propto \nu^{1/3} t^{1/3}$ \\
$\nu < \nu_{\rm c} < \nu_{\rm m}$ & $\propto \nu^{1/3} t^{\frac{11}{3} - 2k}$ & $\propto \nu^{1/3} t^{-1/3}$ \\
$\nu_{\rm m} < \nu < \nu_{\rm c}$ & $\propto \nu^{\frac{1 - p}{2}} t^{3 - \frac{k(5 + p)}{4}}$ & $\propto \nu^{-0.7} t^{-0.7}$ \\
$\nu_{\rm c} < \nu < \nu_{\rm m}$ & $\propto \nu^{-1/2} t^{2 - \frac{3k}{4}}$ & $\propto \nu^{-1/2} t^{1/2}$ \\
$\max(\nu_{\rm m}, \nu_{\rm c}) < \nu$ & $\propto \nu^{-p/2} t^{2 - \frac{k(2 + p)}{4}}$ & $\propto \nu^{-1.2} t^{-0.2}$ \\
\midrule
\textbf{Deceleration phase} ($t > t_{\rm dec}$) & & \\
$\nu < \nu_{\rm m} < \nu_{\rm c}$ & $\propto \nu^{1/3} t^{1 + \frac{2}{-4 + k}}$ & $\propto \nu^{1/3} t^{0}$ \\
$\nu < \nu_{\rm c} < \nu_{\rm m}$ & $\propto \nu^{1/3} t^{1 + \frac{10}{3(-4 + k)}}$ & $\propto \nu^{1/3} t^{-2/3}$ \\
$\nu_{\rm m} < \nu < \nu_{\rm c}$ & $\propto \nu^{\frac{1 - p}{2}} t^{\frac{5}{4} + \frac{2}{-4 + k} - \frac{3p}{4}}$ & $\propto \nu^{-0.7} t^{-1.6}$ \\
$\nu_{\rm c} < \nu < \nu_{\rm m}$ & $\propto \nu^{-1/2} t^{-1/4}$ & $\propto \nu^{-1/2} t^{-1/4}$ \\
$\max(\nu_{\rm m}, \nu_{\rm c}) < \nu$ & $\propto \nu^{-p/2} t^{1/2 - \frac{3p}{4}}$ & $\propto \nu^{-1.2} t^{-1.3}$ \\
\bottomrule
\end{tabular}
\begin{tablenotes}
\footnotesize
\item[*] We adopt $p = 2.4$, assuming the electron spectral index follows the relation $p = 2\beta$, where $\beta = \Gamma_X - 1 = 1.2^{+0.2}_{-0.2}$ was obtained by fitting the X-ray spectrum during the plateau phase (see Table~\ref{tab_spectrum_fitting}).
\end{tablenotes}
\end{threeparttable}
\end{table*}

\clearpage

\begin{table*}
\centering
\scriptsize
\begin{threeparttable}
\small
\caption{MCMC parameter estimate (median) for the multi-band afterglow model with 1$\sigma$ uncertainties.}\label{tab_afterglow_mcmc_results}
\begin{tabular}{ccc}
\toprule
Parameter & Prior & Posterior\\
\hline
log $E_{\rm k} \, \rm (erg)$ & $\mathcal{U}{\sim}$[51, 54] & $52.50^{+0.44}_{-0.54}$ \\
log $\Gamma_{\rm c}$ & $\mathcal{U}{\sim}$[0.7, 3] & $1.25^{+0.17}_{-0.20}$ \\
$\theta_{\rm c} \, \rm (rad)$ & $\mathcal{U}{\sim}$[0, 0.5] & $0.34^{+0.11}_{-0.13}$ \\
$p$ & $\mathcal{U}{\sim}$[2.01, 3] & $2.37^{+0.13}_{-0.13}$ \\
log $\epsilon_e$ & $\mathcal{U}{\sim}$[-2, -0.3] & $-1.32^{+0.58}_{-0.46}$ \\
log $\epsilon_B$ & $\mathcal{U}{\sim}$[-4, -0.3] & $-1.48^{+0.80}_{-1.13}$ \\
log $A_*$ & $\mathcal{U}{\sim}$[-3, 1] & $-1.94^{+0.89}_{-0.62}$ \\
log $\xi_e$ & $\mathcal{U}{\sim}$[-3, 0] & $-1.95^{+0.49}_{-0.44}$ \\
\hline
$\chi^2_{\rm min}/\rm d.o.f.$  & -- & 10.88/11 \\
\bottomrule
\end{tabular}
\end{threeparttable}
\end{table*}

\clearpage

\end{document}